\documentstyle[emulateapj]{article}
\begin{document}

\slugcomment{To appear in the September 2000 Astrophysical Journal Supplement Series}

\title{A Uniform Analysis of the Ly-$\alpha$ forest at z=0 - 5: \\ 
I. The sample and distribution of clouds at z $>$ 1.7}

\author{Jennifer Scott\altaffilmark{1}, Jill Bechtold\altaffilmark{1}, 
 \& Adam Dobrzycki\altaffilmark{2}}

\altaffiltext{1}{Steward Observatory, University of Arizona, Tucson,
AZ 85721, USA\\
e-mail: [jscott,jbechtold]@as.arizona.edu}

\altaffiltext{2}{Harvard-Smithsonian Center for Astrophysics,
60 Garden Street, Cambridge, MA 02138, USA\\ e-mail:
adobrzycki@cfa.harvard.edu}

\begin{abstract}
We present moderate resolution data for 39 quasi-stellar objects (QSOs)
at $z \approx 2$ obtained
at the Multiple Mirror Telescope.  These data are combined with 
spectra of comparable resolution of 60 QSOs with redshifts
greater than $1.7$ found in the literature   
to investigate the
distribution of Ly$\alpha$ forest lines in redshift and equivalent width. 
We find a value for $\gamma$, the parameter describing the number distribution
of Ly$\alpha$ forest lines in redshift, of $1.88\pm0.22$ for lines stronger than
a rest equivalent width of
0.32 $\AA$, in good agreement with some previous studies.  The
Kolmogorov-Smirnov test was applied to the data and it is found that this 
single power law is a good fit over the relevant redshift
ranges. Simulations of the Lyman $\alpha$ forest were performed
to determine the completeness of the line lists and 
to test how well the analysis recovers the underlying line statistics, 
given this level of completeness. 

\end{abstract}

\keywords{quasars: absorption lines}

\section{Introduction}
The spectra of quasi-stellar objects (QSOs) 
blueward of Ly$\alpha$ emission show a large number
of absorption lines primarily due to Ly$\alpha$ absorption by intervening 
neutral hydrogen along the line of sight to the QSO (Lynds, 1971; Sargent
et al. 1980; Weymann, Carswell, \& Smith, 1981).
The evolution of the number density of these lines can be described by
a power law in $1+z$ with index $\gamma$.  It has been debated whether 
the number densities of 
absorbers of different column densities, Lyman $\alpha$ forest, Lyman
limit systems, and damped Lyman $\alpha$ systems, evolve differently
with redshift which would indicate that these absorbing structures represent
different populations of objects (Lanzetta 1991,
Storrie-Lombardi 1994,1995, Giallongo et al. 1996).

Several authors have carried out the analysis of the statistics of
the Lyman $\alpha$ forest on other data sets
(Lu et al. 1991, hereafter LWT; Bechtold 1994, hereafter B94; 
Williger et al. 1994;
Cristiani et al. 1995; Giallongo et al. 1996; Kim et al. 1997; Weymann et al.
1998.)
At high redshift, LWT find 
$\gamma= 2.75 \pm 0.29$; and B94, whose resolution class M sample is
a subsample of our total sample, finds
$\gamma=1.85 \pm 0.27$.
These results were obtained using a fixed equivalent width threshold of
0.36 $\AA$ and using only Ly$\alpha$ lines outside the immediate 
vicinity of the QSO.
Work with high resolution spectra has yielded results similar to
those of both LWT and B94.
Cristiani et al. (1995) find $\gamma= 1.86\pm 0.21$ for Lyman $\alpha$
forest lines with log(N$_{HI}) \geq 13.8$ from R $\sim$ 22,000 
spectra of 3 QSOs; and
Giallongo et al. (1996) find $\gamma= 2.7$ for lines
with log(N$_{HI}) \gtrsim 14$ from R $\sim$ 25,000 spectra of 10 QSOs. 
Similarly, Kim et al. (1997)  report $\gamma= 2.78$ from a sample of 
R $\sim$ 36,000 spectra of 5 QSOs taken using the HIRES Spectrograph on
the Keck Telescope.

The evolution is significantly flatter at redshifts less than 1.7. 
The first Hubble Space Telescope Quasar
Absorption Line Key Project paper (Bahcall et al. 1993) reported 
$\gamma = 0.50 \pm 0.77$
for lines from 13 objects with W $>$ 0.32 $\AA$.  The addition of four more
quasars at z $<$ 1.3 yielded $\gamma= 0.58 \pm 0.50$ for lines
with W $>$ 0.24 $\AA$ (Bahcall et al. 1996).
Weymann et al. (1998) have analyzed the complete HST Key Project 
Ly$\alpha$-absorption line sample
presented by Jannuzi et al. (1998).
They find $\gamma= 0.15 \pm 0.23$ for a fixed equivalent width
threshold of 0.24 $\AA$.    From a large sample of archival
HST/Faint Object Spectrograph (FOS)
spectra which includes the public data from the Key Project, 
Dobrzycki \& Bechtold (1997) find $\gamma =
0.87 \pm 0.27$ for lines with W $>$ 0.36 $\AA$ at z $<$ 1.7, and notes that this
value flattens to $\gamma = 0.14 \pm 0.45$ for Ly$\alpha$ lines with
equivalent widths between 0.24 and 0.36 $\AA$.  This is supported 
by Weymann et al. (1998) who also find a trend of increasing $\gamma$ with
increasing line strength.

In this paper, a homogeneous sample of moderate resolution 
QSO spectra is used to investigate the number density evolution of
Ly$\alpha$ systems and how this changes with redshift and with 
varying equivalent width thresholds.
Specifically, the Ly$\alpha$ forest in the redshift range
between 1.6 and 2.0  was targeted because few lines of sight in the
literature cover this range,
as it extends down to 
wavelengths of $\sim$3200 $\AA$.
Improvements in CCD technology allowed
us to obtain data in this spectral region.
We present new data for 39 objects and supplement
this sample with 60 objects from the literature. 
Metal line systems in these spectra
have been identified and removed from the final analysis of the Ly$\alpha$
forest (Murdoch et al. 1986, hereafter MHPB).  
The resulting Ly$\alpha$ absorption line sample 
is comprised of 2079 lines in the range 1.6 $<$ z $<$ 4.1
when a variable equivalent width threshold is used,
or 1131 lines using a fixed rest equivalent width threshold of 0.32 $\AA$.
Future papers will extend this work to the local universe using 
a sample of low-redshift spectra from the HST/Faint Object Spectrograph
(FOS) archives.
Much recent work has promoted treating the Ly$\alpha$ forest
as density fluctuations in the continuous field of the intergalactic medium
(Bi 1993; Reisenegger \& Miralda-Escud\'{e} 1995;
Hernquist et al. 1996; Miralda-Escud\'{e} et al. 1996;
Bi \& Davidsen 1997; Croft et al. 1998,1999; Weinberg et al. 1999).
However, in this paper, we
will continue to interpret the Ly$\alpha$ forest as a series of discrete
lines for comparison to previous work.
The spectra presented in this data set will be of particular 
use to those interested in using the
Ly$\alpha$ forest to recover the power spectrum of mass fluctuations in
the universe (Croft et al. 1998,1999) and are available on the web.
For the spectra used in this paper, 
see the authors' homepages, which are
accessible from either http://www.as.arizona.edu or http://cfa-www.harvard.edu.

In Section~\ref{sec-absdata}, 
the details of the observations are outlined and the absorption line
sample is presented.
Notes on individual sample QSOs and their corresponding absorption line systems
are given in Section~\ref{sec-notes}.
In Section~\ref{sec-results}
the statistics of the Ly$\alpha$ forest are derived and discussed.
Lyman $\alpha$ forest simulations
based on our sample of QSOs were performed to test the ability of 
our analysis to recover the underlying line statistics in our moderate
resolution data.
These are also presented and discussed in Section~\ref{sec-results}. 
In a subsequent paper, Paper II in this series (Scott, Bechtold, Dobrzycki, \& Kulkarni 2000), 
we will investigate the measurement 
of the mean  intensity of the ionizing background radiation using this 
Ly$\alpha$ forest sample.

\section{Data}
\label{sec-absdata}

\subsection{Observations and Data Reduction}
A sample of 39 QSOs was observed using the Multiple Mirror Telescope and
Blue Channel Spectrograph.  The observations are summarized in 
Table~\ref{table-mmtobj}.  Each object's redshift is given in column (3)
and the reference for that redshift is given in column (4).

\begin{figure}
\epsscale{1.00}
\plotone{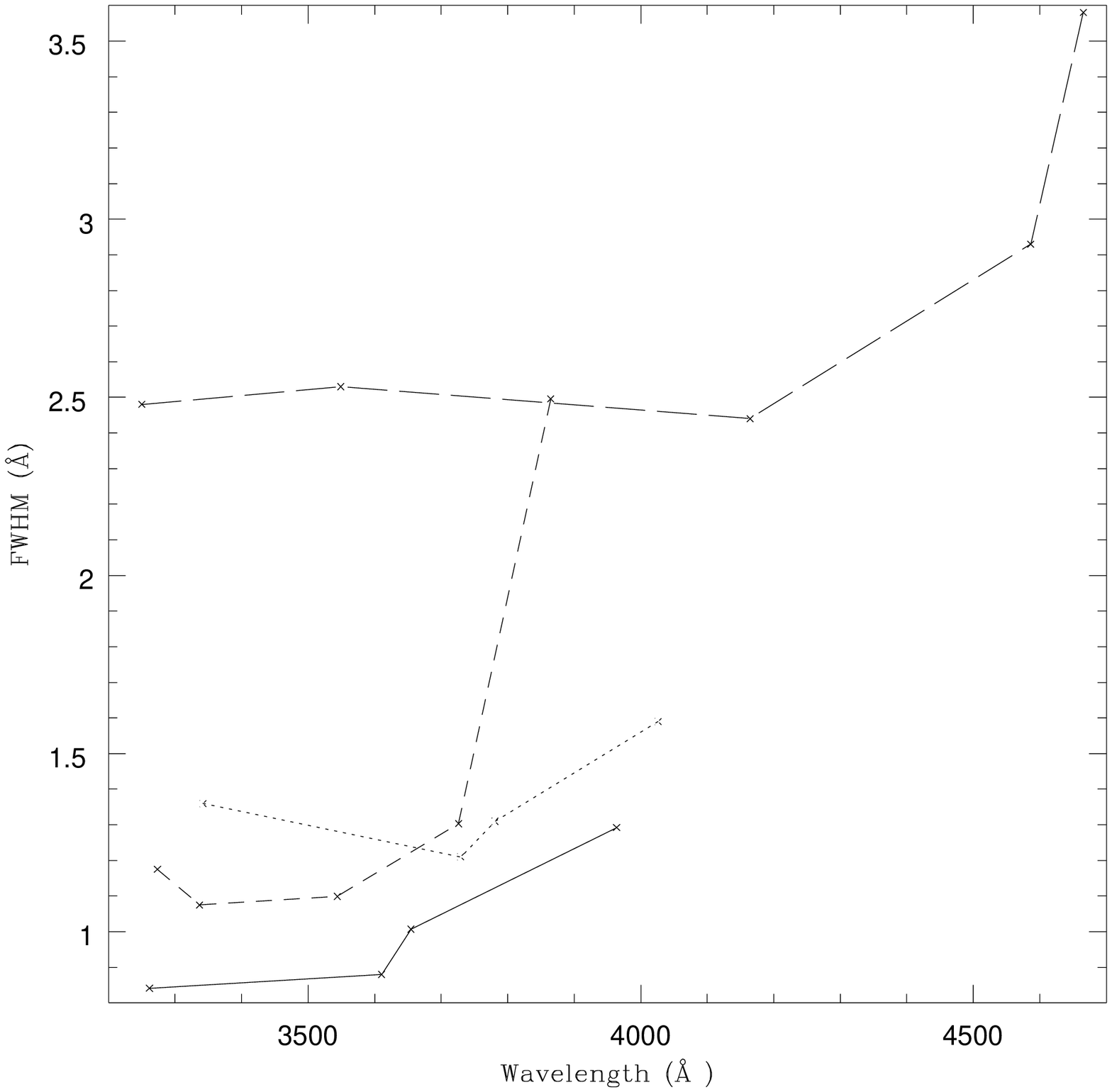}
\caption{
FWHM of comparison lines versus wavelength
for four separate
instrumental setups listed in Table~\ref{table-mmtobj}: 
solid line- (1) Big Blue Reticon,  832 l mm$^{-1}$ 2$^{nd}$ order, 1$\arcsec$x3$\arcsec$ slit;
short dashed line- (2) 3Kx1K CCD, 832 l mm$^{-1}$ 2$^{nd}$ order, 1$\arcsec$x180$\arcsec$ slit;
dotted line- Same as previous setup but with improved field flattener (see text);
long dashed line- (3) 3Kx1K CCD, 800 lmm$^{-1}$  1$^{st}$ order, 1$\arcsec$x180$\arcsec$ slit
\label{fig:fwhm} }
\end{figure}

The three instrumental setups used are as follows: (1) the $``$Big Blue" 
image tube and photon 
counting Reticon detector, a
832 l mm$^{-1}$ grating blazed at 3900 $\AA$ in the
second order with a CuSO$_{4}$ red blocking filter, and a
1$\arcsec$ $\times$ 3$\arcsec$ slit; (2) the 3K $\times$ 1K CCD, the 832 l mm$^{-1}$ grating
blazed at 3900 $\AA$
in the second order with a CuSO$_{4}$ order blocking filter, and a
1$\arcsec$ $\times$ 180$\arcsec$ slit; and (3) the 3K $\times$ 1K CCD, 800 l mm$^{-1}$ 
grating blazed at 4050 $\AA$ in the first order, and a 1$\arcsec$ $\times$ 
180$\arcsec$ slit.
All these spectra have a spectral resolution  of $\sim$1 $\AA$ with the
exception of the spectra of 1207+399 and 1408+009 taken with the 800 l 
mm$^{-1}$ grating, which have a resolution of $\sim$2.5 $\AA$.
Thinning and backside illumination of a Loral CCD along with the
use of antireflection coatings and backside surface charging 
(Lesser 1994) improved  the
quantum efficiency of the 3K $\times$ 1K CCD used to over 80\% at
3200 $\AA$.
The exposures from the first runs using the improved CCD at the MMT
suffer from a variable 
focus across the chip due to problems with the original field flatteners used.
Figure~\ref{fig:fwhm}
shows the FWHM of the comparison lamp lines as a function of 
wavelength.  The July 1993 data was taken on the first
run with this CCD detector; and a number of problems were encountered,
including poor charge transfer efficiency and  a jump in the bias level  
of $\sim$8 ADU in the center of the chip.  On this run, the FWHM rises
to $\sim$2.5 $\AA$ at the red end of the spectrum (short-dashed line in 
Figure~\ref{fig:fwhm}).

Wavelength calibration was performed using He-Ne-Ar-Hg-Cd lamp exposures; and
domeflats or quartz exposurers were used to correct for pixel-to-pixel 
variations.
When available, a few half-hour exposures of each object are combined; 
and the total integration time is listed in  Table~\ref{table-mmtobj}.
The QSO spectra are available in HTML and Postscript format at
http://qso.as.arizona.edu/$\tilde{\;\;}$jscott/Spectra/index.html.

Cosmic rays were removed from the data during the reduction process.
Bad columns on the CCD were left in the spectrum in order to keep track 
of their positions. The flux
in these regions was set to a value of -1000.; 
and they were excluded from the
analysis. In some spectra, some clearly non-Gaussian features are
present at the red end, mainly redward of Lyman $\alpha$ emission.
Because these features occur
at the same pixel in each of the spectra in which they are visible, they
are identified as traps in the CCD.  They
are discussed individually in Section~\ref{sec-notes} below.

\subsection{Line Identification Process}
\label{sec-lineid}
The continuum was fit iteratively to each spectrum and significant (3$\sigma$
or greater) 
absorption lines were found by measuring the equivalent width
in bins of size equal to 2.46 times the FWHM of the 
comparison lines in pixels, the point at which a Gaussian
is 1.5\% of its peak value (B94, Young et al. 1979).  
Lines of 3$\sigma$ significance and
above
were used to help identify metal line systems, but only lines of greater
than 5$\sigma$ significance were used in the analysis of the Lyman $\alpha$
forest statistics.  

Using the technique described in Dobrzycki and Bechtold (1996), 
we produced a set of 30 simulated z=2.48 pure Ly$\alpha$ forest spectra
in order to determine how reliably our program for finding significant lines,
FINDSL,
recovers those generated by the simulations.
We use values of 1.82 and 1.46 for  Ly$\alpha$ forest statistics
$\gamma$ and $\beta$, but the results
of this analysis should not be sensitive to the value of $\gamma$ as the
redshift path covered in each spectrum is small.  The lower and upper 
column density limits chosen were 1x10$^{12}$ and 7x10$^{14}$ cm$^{-2}$
respectively; and the mean Doppler parameter and width of the
Doppler parameter distribution used were 28 km s$^{-1}$ and 10 km s$^{-1}$.
The column density limits were chosen to give the same total
absorption in the simulated spectra as is seen in the spectrum
of 0955+472, the object spectrum which served as the template for this series
of simulations.

We determine matches between the simulation line list output and  the
FINDSL line lists on the basis  of the best wavelength match between
simulated and recovered lines.
At 5$\sigma$ significance,
the line lists are 55\% complete.  When blending is accounted for by
matching all simulated lines within 2.46 resolution elements of each
recovered line to that recovered line, 99\% of the lines  in the simulation
are recovered. 
These completeness values for 3$\sigma$ lines
are 49\% and 98\% respectively.
Obviously, FINDSL can do nothing to help us
overcome the finite resolution of the data, but when this is
taken into consideration, this test indicates that it  does a good job of
recovering the lines it is capable of recovering. 

Our simulations also revealed another interesting point.  Of the 3$\sigma$ lines
$``$recovered" by FINDSL, a small percentage, $\sim$0.25\%,
were not generated by the simulation program.
In other words, FINDSL found some 
lines in the noise.  This was not true of the 5$\sigma$ lines, however,
so we expect no spurious lines to be present in the line lists used 
for the analysis of Lyman $\alpha$ forest statistics.  We do use
lines with significance levels between 3$\sigma$ and 5$\sigma$ for metal line
identification purposes; but expect that any low occurrence of spurious
lines would have no effect on those identifications due
to the all the constraints
that were placed upon metal line matches to qualify as true metal line systems,
which are discussed in more detail below.

The July 1993 CCD data suffers from a gradient in the FWHM across the 
spectrum as discussed in  Section~\ref{sec-absdata}, 
rising from $\sim$1.1 $\AA$ in the
blue end to $\sim$2.5 $\AA$ in the red (Figure~\ref{fig:fwhm}).  
This variation has 
some impact on how FINDSL identifies significant lines.  Using a
FWHM of 2.5 $\AA$ for $\lambda >$ 3700 $\AA$, results in fewer
significant lines identified relative to the case where a FWHM of 
1.1 $\AA$ is used over the full spectrum.   Inspection of the
fits for these two cases for several objects in our sample
leads us to conclude that the two cases give
consistent total equivalent widths for absorption features, but that
using a search window based on a FWHM of 1.1 $\AA$, 
even at the red ends of these spectra,
gives the most reasonable line identifications, as the larger window
tended to blend distinct features together. 
Table 2 gives a
list of the vacuum, heliocentric wavelengths of all lines identified along
with the equivalent width of each line as determined by a Gaussian fit
to the line.  

We generated additional
synthetic Ly$\alpha$ forest spectra with no metal lines in 
order to determine
the maximum number of metal line identifications that our software will
identify spuriously in the Ly$\alpha$ forest, or equivalently, the minimum 
number of metal line identifications needed to qualify as a metal line system,
cf. Dobrzycki and Bechtold (1996). 
The simulation parameters used in this case were 
$\gamma$=1.5, $\beta$=1.46, N$_{lower}$=2x10$^{12}$ cm$^{-2}$, 
N$_{upper}$=10$^{16}$ cm$^{-2}$, $<$b$>$=28 km s$^{-1}$, and $\sigma_{b}$=
10 km s$^{-1}$.
We find that our program will find metal line systems in the
Lyman $\alpha$ forest that may
appear to be reasonable based on the species present and doublet ratios,
if the number of required matches between the
data and a table of possible metal lines is set to a number less than four
if there are less than $\sim$100 lines in the spectrum, and less than five
if there are more than $\sim$100 identified lines in the spectrum.  If
a system shows lines redward of Ly$\alpha$ emission, this requirement is
relaxed since this spectral region is free of Ly$\alpha$ forest absorption
lines.

The search list of metal lines, their wavelengths, and their f values   
was taken from Table 4 of Morton et al. (1988) supplemented with Fe II
$\lambda$1143 and $\lambda$1145 and N I $\lambda$1135 from their Table 3.
Redshift systems were identified by first running our metal line searching
program to find systems with our prescribed number of matches.  Metal
line matches within 3$\sigma$ of an observed significant line are counted.
The output of this program was 
analyzed for consistency with required doublet ratios and f values. Lines
found by this program were rejected if a) the weaker line of a doublet
is detected while the stronger is not or b) a weak line of a species is
detected while a stronger line of the same species and ionization state is not 
(eg. Si II $\lambda$1304 is detected but Si II $\lambda$1260 is not).
Next, lines with rest equivalent width greater than about 1 $\AA$ were
tentatively identified as Ly$\alpha$ for a metal line system.  
The resulting redshift was used as a trial redshift and the matches with
metal lines were noted and critiqued as above.

A metal line system identification is considered a strong one 
if it is corroborated by a spectrum from the literature
that extends redward of Ly$\alpha$ emission.
A system is considered reasonable if it consists of at least 
the minimum number of lines and the strengths of those lines  are in 
agreement with the expected f values and range in doublet ratios.

An identification is marked as a possible identification if either the doublet
ratio gives a value less than one or greater than two, ie. one of the doublet 
lines is a blend if it is present,  or if the separation between that line 
and another line in the redshift system (excluding doublet pairs) 
is greater than $\sim$200 km s$^{-1}$ but less than $\sim$300 km s$^{-1}$.
Once metal lines were identified, they were removed from the line list used
for the Ly$\alpha$ forest analysis.  Also, the redshift path covered by
each line was removed from the analysis by removing a region of width
2.5$\sigma$ centered on the wavelength centroid of the line.   The line $\sigma$
and line centroid were taken from the Gaussian fit. 

The redshift of any  spurious line in our 3$\sigma$ 
line lists identified as a metal line
would also have to match with other metal lines in our
line list, specifically to a strong Lyman $\alpha$ line if it is observable in
the spectrum.  For this reason, we expect that the possible low occurrence of
false lines of less than 5$\sigma$ significance in our line lists has 
no effect on the metal line systems identified below.

\section{Notes on Individual Objects}
\label{sec-notes}

\subsection{Q 0006+020  \hspace{0.5in} $z_{em}=$2.340}

This QSO was identified by \markcite{foltz1989} 
Foltz  et al. (1989). \markcite{tyt1993} 
Tytler et al. (1993), hereafter T93, 
discuss the redshift systems they find in their red (4312 $\AA$ - 7059 $\AA$),
low resolution (8.6 $\AA$ FWHM) spectrum of this object.   We do not confirm the
first system they find at $z_{abs}=$1.131. 
This identification was based on the detection of 
Mg II $\lambda\lambda$2796, 2803 at 5960 $\AA$ and 5975 $\AA$ respectively
which we do not detect in our red spectrum of this object, which is 
presented in Paper II of this series. 
The second system \markcite{tyt1993} T93 find is at 
$z_{abs}=$2.034 for which they 
identify the C IV doublet at
4700 $\AA$ and Al II $\lambda$1670 at 5073 $\AA$.  The positions of
Ly $\alpha$ and Si  III $\lambda$1206  for this redshift lie on 
bad columns in the data, but we identify N II $\lambda$1083
at 3289 $\AA$, a possible N V doublet at 3757 $\AA$ and 3770 $\AA$, 
Si II $\lambda$
1260 at 3825 $\AA$, and C II $\lambda$1334 at 4050 $\AA$.
In addition, our red spectrum of this QSO confirms the C IV doublet
and Al II identifications of T93 while also revealing the Si IV doublet
at 4227 $\AA$ and 4252 $\AA$ and a possible Si II $\lambda$1526 line
at 4632 $\AA$.
Identifying the 4700 $\AA$ line in the spectrum of \markcite{tyt1993} 
T93 as
Si IV $\lambda$1393 reveals the 
third system, at  $z_{abs}=$2.374. We identify Ly$\beta$ at  3460 $\AA$,
O VI $\lambda$1031 and $\lambda$1037 at 3482 $\AA$
and 3501 $\AA$, and N I $\lambda$1200 at 4050 $\AA$.   Our red data 
confirm the 4700 $\AA$ feature as well as the C IV doublet at $\sim$5222 $\AA$
for this redshift. This system is consistent with an
associated absorber as proposed by \markcite{foltz1989}
Foltz et al. (1989).

We also detect several other systems  using the methods and criteria 
described above:  

$z_{abs}=$1.6094-  This is a system showing  Si II $\lambda$1260 at 3289 $\AA$,
C II $\lambda$1334 at 3482 $\AA$, Si IV $\lambda$1393 at 3637 $\AA$
(the position of the $\lambda$1402 component 
lies on a bad column but there is an absorption
feature at this wavelength in our red spectrum), and Si II $\lambda$1526
at 3984 $\AA$ (which is blended with Ly$\alpha$ at $z_{abs}=$2.2775.)
In addition, our red spectrum (see Paper II) shows a line at 4362 $\AA$, 
consistent with
Al II $\lambda$1670 for this redshift.

$z_{abs}=$1.8189-  At this redshift, we identify Ly$\alpha$ at 3427 $\AA$,
N I $\lambda$1200 at 3383 $\AA$,
a tentative N V  doublet at 3491 $\AA$ and 3501 $\AA$ (where the $\lambda$1242
component must be blended with Ly$\alpha$ at $z_{abs}=$1.880 and/or
O VI $\lambda$1037 at $z_{abs}=$2.375),
and C II $\lambda$1334 at 3762 $\AA$.  The C IV doublet at this redshift 
is visible  in our red spectrum at a wavelength of 4367 $\AA$.

$z_{abs}=$1.8409-   For this system, we detect Ly$\alpha$ 3454 $\AA$,
Si III $\lambda$1206 at 3427  $\AA$, Si II $\lambda$
1260 at 3579 $\AA$, and C II $\lambda$1334 at 3791 $\AA$.  Our red spectrum
does not show any lines redward of Ly$\alpha$ consistent with this 
redshift.

$z_{abs}=$1.8802-   This system is composed of Ly$\alpha$ at 3501 $\AA$, a
N V doublet at 3568 $\AA$ and 3579 $\AA$, C II $\lambda$1334 at 3845 $\AA$,
and a possible weak Si IV $\lambda$1393 line at 4015 $\AA$ (no $\lambda$1402 is
detected.)  No lines redward of Ly$\alpha$ emission are detected in the red spectrum.

$z_{abs}=$2.2775-   This is a system showing Ly$\alpha$ at 3984 $\AA$, 
Ly$\beta$ at 3363 $\AA$, Si III $\lambda$1206 at 3955 $\AA$, and the 
N V doublet at 4060 $\AA$ and 4072 $\AA$.  (The position of Fe II $\lambda$1145 
falls on a bad column for this redshift.)  A possible C IV doublet 
identification is made from the red spectrum at 5076 $\AA$.

Lastly, we find a possible Mg II absorber at
$z_{abs}=$0.448.  However, the implied Fe II lines are not consistent
with line ratios.  Therefore, since only two lines
are found, this system cannot qualify as a metal line system by our criteria.

\subsection{Q 0027+014 \hspace{0.5in} $z_{em}=$2.310}
\label{sec-q0027}

\markcite{ss1992} Steidel \& Sargent (1992), hereafter SS92,
find a single Mg II 
system for this object at $z_{abs}=$1.2664 using 
their red setup
(5128-8947 $\AA$) with 4-6 $\AA$ resolution.  In addition to Mg II
$\lambda\lambda$2796, 2803 (at 6336 $\AA$ and 6352 $\AA$ respectively), 
they identify Fe II $\lambda$2382 at 5400 $\AA$.  We
confirm this system with our detection of the C IV  doublet at 3508 $\AA$
and 3513 $\AA$ as well as Al II
$\lambda$1670 at 3786 $\AA$.  Our red spectrum of this object 
(see Paper II)
shows the Fe II line found by SS92, but shows only marginal evidence
for the Mg II doublet.

We also identify two other redshift systems in our spectrum: 

$z_{abs}=$1.8415-  We find
Ly$\alpha$ at 3454 $\AA$, N I $\lambda$1200 at 3411 $\AA$, 
Si III $\lambda$1206 at 3428 $\AA$, Si II $\lambda$1260 at 3582 $\AA$, 
a possible, blended C II $\lambda$1334 line 3793 $\AA$, and the 
Si IV doublet at 3960 $\AA$ and 3986 $\AA$.  However, the doublet ratio
for the Si IV doublet is greater than two; therefore, the $\lambda$1393
component must be blended.  Our red spectrum shows Si II $\lambda$1526 at
4337 $\AA$, the C IV doublet at 4403 $\AA$, Fe II $\lambda$1608 at
4572 $\AA$, and Al II $\lambda$1670 at 4748 $\AA$.

$z_{abs}=$1.9859-  Ly$\alpha$ for this possible system is found at 3630 $\AA$.  
At this redshift, we also identify N II  $\lambda$1083,
Fe II $\lambda$1145, Si II $\lambda$1193 and $\lambda$1260
lines at 3237 $\AA$, 3419 $\AA$,
3563 $\AA$, and 3763 $\AA$.  
The equivalent widths relative to Ly$\alpha$ indicate each 
of these must be blended.  A  Si III $\lambda$1206 line is found at
3603 $\AA$.  The red spectrum shows no lines for this redshift redward of
Ly$\alpha$ emission.  

\subsection{Q 0037-018 \hspace{0.5in} $z_{em}=$2.341}
\label{sec-q0037}

\markcite{wolfe1986} Wolfe et al. (1986), hereafter W86, 
find a candidate damped Ly$\alpha$ system 
present in the spectrum of this object at 3602 $\AA$ ($z_{abs}=$1.962) with
an observed equivalent width of 15.5 $\AA$.
They also note an absorption feature at 3832 $\AA$ ($z_{abs}=$2.152).  
However, since their objective was to search for and characterize damped 
Ly$\alpha$ systems only, they do not produce detailed 
line lists for their spectra. 
These lines are not confirmed by
our data.  We find no  significant absorption feature at 3602 $\AA$; but we do
find a line at 3604 $\AA$. 
We also find no significant line at 3832 $\AA$.  Due to the low signal-to-noise
at the blue end of our spectrum, we truncated the spectrum for the 
purposes of our line searches. The usable portion of our 
spectrum therefore extends from $\sim$3542 $\AA$  to $\sim$4110 $\AA$.
The features at 3998 $\AA$, 4003 $\AA$, 4007 $\AA$, and 4011 $\AA$ are
identified as traps in the CCD, as they appear in many other object spectra.

\subsection{Q 0049+007 \hspace{0.5in} $z_{em}=$2.279}
We find a system consistent with Ly$\alpha$ at 3540 $\AA$.
\markcite{ss1992} SS92 (cf. Section~\ref{sec-q0027}) identify 
this metal line system
at $z_{abs}=$1.9115 on the basis of weak Al III $\lambda$1854 and
$\lambda$1862 lines and a weak Mg II doublet.  
Further corroboration of this system comes from a possible  N V $\lambda$1238
line at 3607 $\AA$ (no $\lambda$1242 is detected) and the Si IV doublet at
4057 $\AA$ and  4084 $\AA$ respectively in our data.  Our red spectrum
of this object (see Paper II) also shows Si II $\lambda$1526 at
4445 $\AA$, and the C IV doublet at 4507 $\AA$ and 4515 $\AA$, consistent
with this system.

In addition, we find five other systems or possible systems from our data:

$z_{abs}=$1.3865- We identify this system based on the
C IV doublet at 3695 $\AA$ and 3701 $\AA$.  We also find 
Si II $\lambda$1526 at 3643$\AA$, and 
Al II $\lambda$1670 at 3987 $\AA$. \markcite{ss1992} SS92 
do not find a Mg II doublet nor do they find any Fe II lines at 
this redshift.  Our red spectrum shows possible Al III $\lambda$1854
and $\lambda$1862 lines at 4426 $\AA$ and 4445 $\AA$.   However, the
feature at 4445 $\AA$ is more likely Si IV $\lambda$1393 at $z_{abs}=$2.1908.

$z_{abs}=$1.5226- This system is composed of O I $\lambda$1302
at 3285 $\AA$, Si IV $\lambda$1393 at 3515 $\AA$ and
$\lambda$1402 at 3540 $\AA$ (blended with Ly$\alpha$ at $z_{abs}=$1.9123), 
a possible identification of  Si II $\lambda$1526 
at 3850 $\AA$, and the C IV doublet at 3905 $\AA$ and 3912 $\AA$.
\markcite{ss1992} SS92 do not detect a Mg II 
doublet or any Fe II lines at this redshift, nor do we find any matching 
lines in our red spectrum.

$z_{abs}=$2.1168-  This is a relatively insecure identification based upon 
Ly$\alpha$ at 3789 $\AA$, a possible O I $\lambda$1302 at 4057 $\AA$ and
possible Si II $\lambda$1193 and $\lambda$1260 lines at 3720 $\AA$ and 3927 
$\AA$.  No lines are found redward of Ly$\alpha$ emission.

$z_{abs}=$2.1667-  This system shows  Ly$\alpha$ at 3850 $\AA$, Ly$\beta$
at 3248 $\AA$,  and a very tentative N V doublet both components of which
must be blends at 3927 $\AA$ and 3935 $\AA$.  We find no lines at this 
redshift in our red spectrum

$z_{abs}=$2.1918-  For this system, we find Ly$\alpha$ at 3880 $\AA$, 
Ly$\beta$ at 3274 $\AA$, Si III $\lambda$1206  at 3850  $\AA$,
and a possible, blended N II $\lambda$1083 at 3458 $\AA$.  In addition,
our red spectrum shows the Si IV doublet at 4447 $\AA$ and 4476 $\AA$ and
the C IV doublet at 4944 $\AA$.

\subsection{Q 0123+257 \hspace{0.5in} $z_{em}=$2.370}

The absorption spectrum of this QSO has been observed by \markcite{so1968}
Schmidt \& Olsen (1968) (SO68), \markcite{ol1975} Oemler \& Lynds (1975) 
(OL75), and
\markcite{wolfe1986} W86 (cf. Section~\ref{sec-q0037}).

We confirm the absorption  features seen by \markcite{so1968} SO68 
at 3900 $\AA$, 4013 $\AA$, 4057 $\AA$, and 4065 $\AA$. 
The remainder of their features lie outside the wavelength
range of our spectrum.  They report an absorption system at $z_{abs}=$2.3683,
an associated absorber, 
from the identification of Ly$\alpha$ and the C IV doublet, as well as a 
possible identification of Si III $\lambda$1206.   
\markcite{ol1975} OL75 discuss several possible redshift
systems.  The only system they find compelling, however, is the 
$z_{abs}=$2.3683 system  of \markcite{so1968} SO68.
We confirm several lines possibly associated with this system:   Ly$\beta$ 
at 3456 $\AA$, O VI 
$\lambda\lambda$1031,1037 at 3473 $\AA$ and 3496 $\AA$,  N II 
$\lambda$1083 at  3645 $\AA$, and Si III $\lambda$1206 at 4064 $\AA$ which
is blended with Ly$\alpha$ at $z_{abs}=$2.3433.
Our red spectrum of this object (see Paper II) shows the C IV doublet
at 5216 $\AA$ and 5226 $\AA$.  We also
confirm the absence of any marked
damped Ly$\alpha$ absorption, as reported by \markcite{wolfe1986} W86. 

We tested all of the possible redshift systems 
proposed by \markcite{ol1975} OL75 and used our usual
methods for finding additional metal line systems.  As a result, we identify 
three  other systems:

$z_{abs}=$0.3207-  This system consists of Fe II $\lambda$2600 at
3433 $\AA$, a Mg II doublet at 3693 $\AA$ and 3702 $\AA$, and Mg I
$\lambda$2753 at 3767 $\AA$.

$z_{abs}=$1.8427-  This system consists
of Ly$\alpha$ at 3456 $\AA$, Si III $\lambda$1206 at 3430 $\AA$, O I 
$\lambda$1302 at 3702 $\AA$, and Si IV $\lambda$1393 at 3962 $\AA$, with a 
possible identification of $\lambda$1402 blended with a feature at 3989 $\AA$.

$z_{abs}=$2.0379-  For this system, we find Ly$\alpha$ at 3693 $\AA$,
Fe II $\lambda$1143 and a blended $\lambda$1145 at 3473 $\AA$ and 3478 $\AA$, 
and N I $\lambda$1200 at 3645 $\AA$.  Neither \markcite{so1968} SO68 nor
\markcite{ol1975} OL75 note any absorption features at the position of Fe II
$\lambda$1608 for this redshift; and our red spectrum shows no lines at this
redshift.

\subsection{Q 0150-203 \hspace{0.5in} $z_{em}=$2.148}
\label{sec-q0150}

The absorption spectrum of 0150-203 (UM675) is first discussed in detail by
\markcite{sbs1988} Sargent et al. (1988), hereafter SBS88. 
Their data provide 
coverage from 3815 $\AA$ to 5038 $\AA$ with 1.5 $\AA$ resolution.
They report several absorption systems from their spectrum:

$z_{abs}=$0.3892-  A Mg II doublet is identified at this redshift. 
\markcite{sbs1988} SBS88 report the possible blending of the
Mg II doublet at $z_{abs}=$0.3892
with a second component at $z_{abs}=$0.3882.  Our spectrum does show
two prominent absorption features at 3883 and 3896 $\AA$.  If these lines
are interpreted as the Mg II doublet the resulting redshifts are
$z_{abs}=$0.38869 for the $\lambda$2796 line and $z_{abs}=$0.38977 for the
$\lambda$2803 line, an unacceptable separation of 233 km s$^{-1}$.  
It is possible to identify three
Fe II lines at this redshift, $\lambda$2344 at 3253 $\AA$,
$\lambda$2382 at 3308 $\AA$, and $\lambda$2586 at 3590 $\AA$.  However,
no Fe II $\lambda$2600 line is found which calls the identification of
the $\lambda$2344 and the $\lambda$2586 lines into question.  Given these
arguments and
the more compelling identification of the 3883 $\AA$ and 3896 $\AA$  features
as the N V doublet at $z_{abs}=$2.134, we consider this system improbable. 

$z_{abs}=$0.7800-  A Mg II system showing Fe II $\lambda$2382 is 
reported by \markcite{sbs1988} SBS88.
The only lines in our search list that fall within the wavelength range of our
data for $z_{abs}=$0.7800 are Al III $\lambda$1854 and $\lambda$1862, but
we detect neither of these, and thus cannot confirm this system.

$z_{abs}=$1.7666-  \markcite{sbs1988} SBS88 detect a weak C IV doublet at 
this redshift.  We confirm this system
from our detection of Ly$\alpha$ at 3363 $\AA$
and possible identifications of 
O I $\lambda$1302 at 3604 $\AA$, and C II
$\lambda$1334 at 3693 $\AA$. 

$z_{abs}=$1.9287- \markcite{sbs1988} SBS88 regard this weak C IV doublet as
a probable system.  We confirm this system through our identifications of
Ly$\alpha$ at 3560 $\AA$ and tentative Si II $\lambda$1193,
$\lambda$1260, and $\lambda$1304 lines at 3494 $\AA$,  3690 $\AA$ and 
3821 $\AA$ respectively. 

$z_{abs}=$2.0083,2.0097- \markcite{sbs1988} SBS88 regard this C IV complex 
as almost certain due to the good redshift agreement between the putative
doublet lines.  The Si IV $\lambda$1393
line is also identified for the $z_{abs}=$2.0097 component of this complex.
We confirm the $z_{abs}=$2.0083 system.
At this redshift, we identify lines of Ly$\alpha$
at 3657 $\AA$, Fe II $\lambda$1145 at 3444 $\AA$, Si II $\lambda$1193 at 3590
$\AA$ (possible), Si II $\lambda$1260 at 3792 $\AA$, N V $\lambda$1238  at
3726 $\AA$ (possible), and C II $\lambda$1334 at 4014 $\AA$. \markcite{y1991}
York et al. (1991) give this component a B rating, as \markcite{sbs1988}
SBS88 only identified the  C IV doublet.
For the  $z_{abs}=$2.0097 component, we confirm Ly$\alpha$ absorption at
3659 $\AA$, or  $z_{abs}=$2.0101.  We also find Fe II $\lambda$1145 at 3446
$\AA$, Si III $\lambda$1206 at 3632 $\AA$, and O I $\lambda$1302 at 3918 $\AA$. 
\markcite{y1991} York et al.
(1991) assign this system an A rating since \markcite{sbs1988}  SBS88
identified both C IV and Si IV $\lambda$1393 at this redshift.  In our spectrum,
the Ly$\alpha$ lines for the components of this complex are within 5 $\AA$ of
a third line, which, if identified as Ly$\alpha$ as well, gives
$z_{abs}=$2.0060.  However, we detect only one other line (Si II $\lambda$1260
at 3788 $\AA$) for this redshift.  This, and the fact that\markcite{sbs1988}
SBS88  find no C IV at $z_{abs}=$2.0060 lead us to regard this 
additional identification as extremely uncertain.

\markcite{ss1992} SS92 (cf. Section~\ref{sec-q0027}) find 
no Mg II systems in their spectrum of this
object although they note that for the SBS88 systems at $z_{abs}=$1.7666,
1.9287, 2.0083, and 2.0097, these lines would have been visible in their 
spectrum if present.  In fact, SS92 find no absorption features in their
spectrum at all.

\markcite{b1991} Beaver et al. (1991), hereafter B91, observed the far-UV 
spectrum of this object using the FOS on HST.
The spectra range from 1630 $\AA$ to 2428 $\AA$ and
were taken using two different apertures each resulting in $\sim$8.0 $\AA$ 
resolution. In addition, optical spectra were obtained with the Lick telescope.
These spectra cover 3250-6350 $\AA$ at 15 $\AA$ resolution and 3540-4120 $\AA$ 
at 1.8 $\AA$ resolution. \markcite{b1991} B91 confirm 
the $z_{abs}=$0.7800 system of SBS88 
with their identification of Ly$\alpha$ at 2161 $\AA$ and Ly$\beta$ at 1836 
$\AA$ as well as their tentative identifications of C II $\lambda$1334 at 2370
$\AA$, and Si III $\lambda$1206 at 2148 $\AA$ and their possible identification
of C III $\lambda$977 at 1736 $\AA$.  They also report one other system:

$z_{abs}=$2.1348-  The optical spectra of \markcite{b1991} B91 show strong 
absorption at 3810 $\AA$ which is identified as Ly$\alpha$. This
identification results in the coincidence of the N V doublet at this redshift 
with the Mg II doublet at $z_{abs}=$0.3892 identified by \markcite{sbs1988}
SBS88.  This system is corroborated by the tentative identification of Ne VIII
$\lambda$770 at 2417 $\AA$ and the uncertain identification of He I 
$\lambda$584 at 1836 $\AA$.
We identify several lines for this associated absorber, including 
Ly$\alpha$ at 3810 $\AA$, O VI
$\lambda\lambda$1031, 1037 at 3234 $\AA$ and 3253 $\AA$, N II
$\lambda$1083 at 3397 $\AA$,
and N V $\lambda\lambda$1238,1242 at 3883 $\AA$ and 3896 $\AA$.
Also, as noted by \markcite{b1991} B91, the spectrum of \markcite{sbs1988}
SBS88 shows some absorption near the position of the C IV
doublet at this redshift ($\sim$4860 $\AA$) but they do not identify this
feature.

We identify one additional system in our data: 

$z_{abs}=$0.3628-  This system consists of Fe II $\lambda$2382,
$\lambda$2586, and $\lambda$2600 at 3249 $\AA$, 3525 $\AA$, and
3542 $\AA$ respectively, as well as Mg II $\lambda\lambda$2796,2803
at 3810 $\AA$ and 3821 $\AA$.  \markcite{b1991} B91 do not detect Ly$\alpha$
for this system in their FOS spectrum, however, its position at 1657 $\AA$
would place it at the blue edge of their data where the signal-to-noise
ratio is poor.

\subsection{Q 0153+744 \hspace{0.5in} $z_{em}=$2.341}

According to our searches, there is no previously published spectrum
of this QSO.  In our spectrum, we find only one possible metal line
system, an associated absorber at $z_{abs}=$2.3456.  We consider this
identification tentative, however, due to the fact that the Ly$\beta$ line
for this system is separated from the position of Ly$\alpha$ by 6$\sigma$.
The other species detected are O VI $\lambda$1031 and $\lambda$1037
at 3453 $\AA$ and 3472 $\AA$, Fe II $\lambda$1143 and Fe II $\lambda$1145
at 3826 $\AA$ and 3831 $\AA$, and Si III $\lambda$1206 at 4037 $\AA$.
In addition, our red spectrum (see Paper II)
does show  a possible C IV doublet at 5179 
$\AA$ and 5188 $\AA$, but the doublet ratio is less than one.

\subsection{Q 0226-038 \hspace{0.5in} $z_{em}=$2.073}
\label{sec-q0226}

The absorption line spectrum of this QSO has been studied by many authors.
The first such investigation was undertaken by \markcite{c1976} Carswell
et al. (1976) using spectrograms 
spanning a wavelength range from 3200 $\AA$ to 6000 $\AA$.
\markcite{ysb1982} Young, Sargent, and Boksenberg (1982b), YSB82 hereafter, 
obtained spectra from 3530 $\AA$ to 5070 $\AA$ with 2.2
$\AA$ resolution.  In addition, \markcite{ltw1987} Lanzetta, Turnshek, \&
Wolfe (1987), LTW87 hereafter, obtained spectra from 6271 $\AA$
to 8766 $\AA$ with 4.5 $\AA$ resolution and a signal-to-noise ratio between
18 and 32.  This object was also observed by \markcite{ss1992} 
SS92 with their red setup and by \markcite{sbs1988} SBS88
(cf. Section~\ref{sec-q0027} and Section~\ref{sec-q0150}). 

The spectrum we obtained for this object is, unfortunately, riddled with
bad columns from the CCD.  Therefore, we find no absorption systems from our
data alone; instead, we use our spectrum to attempt to confirm the
systems found by other authors:

$z_{abs}=$1.3284- \markcite{ss1992} SS92 confirm the Mg II identification
for this system which was found by \markcite{ltw1987} LTW87. \markcite{ss1992} 
SS92  also identify Fe II 
$\lambda$2344, $\lambda$2382, and $\lambda$2600 in their red spectrum.
They further corroborate this system by noting that lines found by 
\markcite{ysb1982} YSB82 at 3606 $\AA$ and 3611 $\AA$ can be identified as the 
C IV doublet and that an unidentified line found by \markcite{sbs1988} SBS88
at 3890 $\AA$ can be identified as Al II $\lambda$1670.
Our data show  the C IV $\lambda$1548 line at 3604 $\AA$, but we find only a 
weak feature at the expected position of $\lambda$1550.  
The position of Al II $\lambda$1670 falls on a bad column in our data.  
There is a feature
at 3555 $\AA$, the expected position of Si II $\lambda$1526; but it is not
identified as a significant line as it falls on another of the many bad columns.

$z_{abs}=$1.3558- \markcite{ysb1982} YSB82 propose the identification of two 
lines, at 3647 $\AA$ and 3654 $\AA$, in the Lyman $\alpha$ forest region of
their spectrum with the C IV doublet at this redshift.
We confirm the presence of these lines; 
however, given the lack of any other lines to strengthen this
identification, the does not meet our criteria for a true metal line system.

$z_{abs}=$2.0435- \markcite{sbs1988} SBS88 identify this system based on the
identification of the Si IV and C IV doublets.   The expected position of
Ly$\alpha$ for this redshift falls on a bad column in our data; and we find only
one other possible line for this redshift, Si III $\lambda$1206 at 3672 $\AA$.

We do not confirm the absorption line at 3703 $\AA$ reported by Carswell
et al. (1976)

\subsection{Q 0348+061 \hspace{0.5in} $z_{em}=$2.056}

\markcite{sbs1988} SBS88 (cf. Section~\ref{sec-q0150}) 
find several absorption systems in their spectrum 
of this QSO (3880 $\AA$ - 5060 $\AA$): 

$z_{abs}=$0.3997-  This system is a single Mg II doublet according to
\markcite{sbs1988} SBS88.  We find only marginal evidence for a Mg II doublet
at 3912 $\AA$ and 3921 $\AA$ from our red spectrum of this object (see Paper 
II).

$z_{abs}=$1.7975-  \markcite{sbs1988} SBS88 find a C IV doublet  at this
redshift.  We verify Ly$\alpha$ absorption at 3400 $\AA$;  we
detect a possible Si III $\lambda$1206 line at 3374 $\AA$; and our
red spectrum shows the C IV doublet identified by \markcite{sbs1988} SBS88
at 4328 $\AA$ and 4336 $\AA$.

$z_{abs}=$1.8409- \markcite{sbs1988} SBS88 find another C IV doublet at
this redshift.  We detect Ly$\alpha$ absorption at 3453 $\AA$, in 
agreement with this system.  A possible Si II $\lambda$1260 line
at 3581 $\AA$ is found for this redshift; and our red spectrum corroborates the
C IV doublet found by \markcite{sbs1988} SBS88 as well as showing Si IV 
$\lambda$1393 at 3958 $\AA$ (but no $\lambda$1402) and a possible
Si II $\lambda$1526 line at 4335 $\AA$.

$z_{abs}=$1.9681- \markcite{sbs1988} SBS88 find a C IV doublet along
with C II $\lambda$1334 and a possible Si IV $\lambda$1393 line at this
redshift.  We find Ly$\alpha$ 3608 $\AA$, 
Si III $\lambda$1206 at 3581 $\AA$,  and a very tentative N V
doublet at 3676 $\AA$ and 3687 $\AA$, which, if present, is highly blended
with Ly$\alpha$ at $z_{abs}=$2.0238 and $z_{abs}=$2.0331.  Our red spectrum
verifies the identifications of  \markcite{sbs1988} SBS88 listed above and 
also shows Fe II $\lambda$1608 at 4775 $\AA$.

$z_{abs}=$2.0237- \markcite{sbs1988} SBS88 find a C IV doublet and possible 
Si IV $\lambda$1393 at this redshift.
\markcite{ss1992} SS92 (cf. Section~\ref{sec-q0027}) confirm this system
in their red spectrum (5128 $\AA$ - 8947 $\AA$) of this object with the 
detection of a Mg II doublet at this redshift.   They do not detect Mg II for
any of the other SBS88 redshifts to which their spectrum is sensitive 
($z_{abs}>$0.83.)  We detect Ly$\alpha$ absorption at 3676 $\AA$,  in 
agreement with this system.  In addition, we identify a possible blended
Si II $\lambda$1193 line at 3608 $\AA$, N I $\lambda$1200 at 3628 $\AA$,
Si III $\lambda$1206 at 3648 $\AA$, Si II $\lambda$1260 at 3812 $\AA$, and C II
$\lambda$1334 at 4035 $\AA$.  Our red spectrum exhibits the features found
by \markcite{sbs1988} SBS88 listed above as well as C II $\lambda$1334 at 4037
$\AA$.

$z_{abs}=$2.0330-  \markcite{sbs1988} SBS88 identify both C IV and Si IV 
doublets for this redshift.  We detect Ly$\alpha$ at 3687  $\AA$ and
Si III $\lambda$1206 at 3659 $\AA$.  Our red spectrum shows marginal evidence
for the features listed by \markcite{sbs1988} SBS88. 

\subsection{Q 0400+258 \hspace{0.5in} $z_{em}=$2.108}

No previously published absorption line spectrum of this QSO was found in
our searches.  Unfortunately, the low signal-to-noise of the blue portion of
our spectrum (3208 $\AA$ - 3659 $\AA$) prevents us from identifying any lines
in the Lyman alpha forest.  We find only one significant line at 3752 $\AA$
from which we cannot identify any metal line systems.  

\subsection{Q 0747+610 \hspace{0.5in} $z_{em}=$2.491}

In their catalog of QSO absorption lines, \markcite{j1991} Junkkarinen et al. 
(1991) note two metal line systems found for this object by \markcite{a1979}
Afanasjev et al. (1979).  These systems were identified at $z_{abs}=$1.986
and $z_{abs}=$2.210.  \markcite{y1991} York et al. (1991) give both of these
systems a B rating in their reference catalog of heavy element systems in QSO
spectra.  According to their explanation of their rating system, this
B rating indicates that either a C IV or Mg II doublet was identified for
these systems with the correct doublet ratio, but that no other lines but
Lyman alpha were detected.  However, \markcite{j1991} Junkkarinen et al.
note that for the $z_{abs}=$1.986 system, N V, Si II, C II , Si IV, and Al II
lines were detected in addition to H I and C IV; and for the $z_{abs}=$2.210 
system, Si II$^{*}$, N V, C II, Si IV, and Al III lines were detected in 
addition to H I and C IV.
\markcite{ss1992} SS92 (cf. Section~\ref{sec-q0027}) do not 
confirm either of these metal line system 
redshifts.  Instead, they find three others at $z_{abs}=$1.1282, 
$z_{abs}=$2.0076, and $z_{abs}=$2.4865.

We confirm the $z_{abs}=$1.986 system of \markcite{a1979}  Afanasjev et al. 
(1979) with our identification of Ly$\alpha$ at 3629 $\AA$, a possible N I 
$\lambda$1135 at 3389 $\AA$, 
a possible Si II $\lambda$1190 line at 3554 $\AA$, Si II $\lambda$1193 at
3562 $\AA$, and
N I $\lambda$1200 at 3582 $\AA$.
We also confirm their $z_{abs}=$2.210 system with our detection
of Ly$\alpha$ at 3903 $\AA$, a possible N II $\lambda$1083 line at 3480 $\AA$,
Si III $\lambda$1206 at 3874 $\AA$, Si II $\lambda$1260 at 4047 $\AA$,
and O I $\lambda$1302 at 4180 $\AA$. 

We do not find any lines at the position of the $z_{abs}=$1.1282
system of SS92 which they identify by a weak Mg II doublet.  
We identify a
metal line system at $z_{abs}=$2.0071, in accordance with the $z_{abs}=$2.0076
system found by these authors.  At this redshift, we find a strong Ly$\alpha$
line at 3656 $\AA$, Si II $\lambda$1190 and $\lambda$1193 at 3580 $\AA$ and
3589 $\AA$, Si III $\lambda$1206 at 3629 $\AA$, Si II $\lambda$1260 and
$\lambda$1304 at 3791 $\AA$ and 3923 $\AA$, O I $\lambda$1302 at 3915 $\AA$,
and possible C II $\lambda$1334 absorption at 4014 $\AA$.  It is clear that
some of these Si II lines are blends given their relative strengths.
Our confirmation of the $z_{abs}=$2.4865 system of
SS92 is not as strong.  We find Ly$\alpha$ and Ly$\beta$  at 4237 $\AA$ and 3575
$\AA$ respectively for this redshift.  But we do not detect any other species
with any confidence.

The absorption line spectrum of this object is a rich one.  We find a total
of 145 significant lines and we find twelve 
metal line systems in addition to the
ones discussed above.  As is the case for all of our objects,  it is unlikely 
that all of these systems are real since SS92 do not report any lines from their
red spectrum at these redshifts.  However, we have kept all the 
systems that cannot be definitively ruled out on the basis of our data.  For
all redshifts below 1.742, the Ly$\alpha$ line falls outside the spectral
range of our data.  The values of these redshifts are based upon the strongest
line that was detected for each system.

$z_{abs}=$1.4102-  This system is based upon a C IV doublet at 3731 $\AA$
and 3738 $\AA$ and a Si IV doublet at 3359 $\AA$ and 3381 $\AA$.  We also
find Al II $\lambda$1670 at 4028 $\AA$.

$z_{abs}=$1.4529-  The value for this redshift is based upon a C IV
doublet at 3798 $\AA$ and 3804 $\AA$.  In addition, we detect 
Si IV $\lambda$1393 at 3419 $\AA$ and Si IV $\lambda$1402 at 3441 $\AA$
(though it must be a blend if it is present otherwise the Si IV doublet 
ratio is less than one), Si II $\lambda$1526 at 3745 $\AA$, and Al II
$\lambda$1670 at 4098 $\AA$.

$z_{abs}=$1.5986-  For this system, we identify Si II $\lambda$1304 at
3389 $\AA$, possible C II $\lambda$1334 absorption at 3466 $\AA$, a possible
Si IV $\lambda$1393 line at 3621 $\AA$ (no $\lambda$1402 is found), Si II
$\lambda$1526 at 3967 $\AA$, and a possible, weak C IV $\lambda$1548 line at
4023 $\AA$. A weak feature is 
present at the position of C IV $\lambda$1550, but it is not identified as
a significant (3 $\sigma$) line.

$z_{abs}=$1.6822-  This redshift is based upon Si II $\lambda$1260. 
We also find a possible blended N V $\lambda$1242 line at 3333 $\AA$ 
($\lambda$1238  
is out of the wavelength range of our line list), O I $\lambda$1302 
at 3492 $\AA$, Si II
$\lambda$1304 at 3498 $\AA$, C II $\lambda$1334 at 3580 $\AA$, and a rather
doubtful Si IV doublet at 3738 $\AA$ and 3761 $\AA$. 

$z_{abs}=$1.7324-  This system, based on a possible C IV doublet at
4230 $\AA$ and 4237 $\AA$, is a relatively tentative one due to the
inconsistent doublet ratios of this pair and of a possible Si IV doublet 
at 3808 $\AA$ and 3833 $\AA$.
We also find O I $\lambda$1302 at 3558 $\AA$. 

$z_{abs}=$1.8123-  For this system, we find Ly$\alpha$ at 3419 $\AA$.  In
addition, we find a possible Si II $\lambda$1193 line at 3357 $\AA$, Si III
$\lambda$1206 at 3393 $\AA$, a possible blended Si II $\lambda$1260 line
at 3544 $\AA$, O I $\lambda$1302 at 3662 $\AA$, and C II $\lambda$1334
at 3753 $\AA$.

$z_{abs}=$1.8728-  This system consists of Ly$\alpha$ at 3492 $\AA$, Si II 
$\lambda$1190 and $\lambda$1193 at 3419 $\AA$ and 3428 $\AA$, N I 
$\lambda$1200 at 3447 $\AA$, Si III 
$\lambda$1206 at 3466 $\AA$, a possible N V $\lambda$1238 line at 3558 $\AA$,
Si II $\lambda$1260 at 3621 $\AA$, C II $\lambda$1334 at 3833 $\AA$ and a 
possible
Si IV $\lambda$1393 at 4003 $\AA$ (no $\lambda$1402 component is found.) 

$z_{abs}=$2.0070,2.0093-  We find a metal line system of two components at these
redshifts. 
The first component shows Ly$\alpha$ at 3656 $\AA$, Si II $\lambda$1190 and 
$\lambda$1193 at 3580 $\AA$ and 3589 $\AA$, a possible Si III $\lambda$1206
line at 3618 $\AA$, Si II $\lambda$1260 and $\lambda$1304 at 3791 $\AA$ and
3923 $\AA$, C II $\lambda$1334 at 4014 $\AA$, and Si IV $\lambda$1393
at 4192 $\AA$.  The $\lambda$1402 component of this doublet is blended
with the same line corresponding to the other system at $z_{abs}=$2.009.
The second component consists of Ly$\alpha$ at 3658 $\AA$,
a possible N V $\lambda$1238 line at 3728 $\AA$ (no $\lambda$1242 line is
found), Si II $\lambda$1260 at 3792 $\AA$, C II $\lambda$1334 at 4015 $\AA$,
and a tentative Si IV doublet at 4194 $\AA$ and 4221 $\AA$ (with a doublet
ratio less than one due to blending.)

$z_{abs}=$2.0476-  This system is composed of Ly$\alpha$ at 3705 $\AA$,
N I $\lambda$1200 at 3656 $\AA$ (blended with Ly $\alpha$ at
$z_{abs}=$2.007 if present), Si III $\lambda$1206 at 3677  $\AA$, 
O I $\lambda$1302 at 3967 $\AA$, and Si IV $\lambda$1393 at 4247 $\AA$. 

$z_{abs}=$2.1391-  At this redshift, we identify Ly$\alpha$ at 3816 $\AA$,
N I $\lambda$1135 and $\lambda$1200 at 3562 $\AA$ and 3767 $\AA$, and a
possible N V doublet at 3889 $\AA$ and 3901 $\AA$ for which the $\lambda$1242
component must be blended as it is stronger than both the $\lambda$1238
component of the doublet and Ly$\alpha$.

$z_{abs}=$2.1724-  This system consists of Ly$\alpha$ at 3856 $\AA$, 
tentative N I $\lambda$1135 and $\lambda$1200 lines at 3601 $\AA$ and
3808 $\AA$, Si III $\lambda$1206 at
3828 $\AA$, and a possible C II $\lambda$1334 line at 4235 $\AA$. 

$z_{abs}=$2.2849-  This system is identified on the basis of strong Ly$\alpha$
absorption at 3993 $\AA$, Ly$\beta$ at 3369 $\AA$, O VI $\lambda$1031 and
$\lambda$1037 at 3389 $\AA$ and 3408 $\AA$, possible Si II $\lambda$1190
and $\lambda$1193 lines at 3777 $\AA$ and 3786 $\AA$,
and a possible Si III $\lambda$1206 line at 3964 $\AA$. 

\subsection{Q 0836 + 710 \hspace{0.5in} $z_{em}=$2.218}

\markcite{sk1993} Stickel \&  K\"{u}hr (1993) report an absorption feature 
in their spectrum of this object at 5360 $\AA$ which they identify as the
Mg II doublet at $z_{abs}=$0.914.  We find
Al III $\lambda$1854 at 3550 $\AA$.  Also, we have a red spectrum of this
object in the vicinity of Mg II emission. 
This spectrum 
does show the Mg II doublet at 
5359 $\AA$ and 5372 $\AA$, giving a redshift of 0.916. 

We find several other redshift systems in our data: 

$z_{abs}=$1.4256- This system is a double-component C IV absorber 
with the Si IV doublet at 
3380 $\AA$ and
3403 $\AA$, the C IV doublet at 3755 $\AA$ and 3762 $\AA$, a possible
Si II $\lambda$1526 at 3702 $\AA$, and Fe II $\lambda$1608 at 3902 $\AA$
Two components in each line are
evident in the spectrum, with the second, weaker component at $z_{abs}=$1.4249
which, unlike the first component, shows Al II $\lambda$1670 absorption, at 
4051 $\AA$.

$z_{abs}=$1.6681- At this redshift, we detect absorption from 
Ly$\alpha$ at 3243 $\AA$, 
C II $\lambda$1334 at 3561 $\AA$,  and a Si IV doublet at 3719 $\AA$ and
3742 $\AA$ (though its implied doublet ratio is less than one.)  There is
no Mg II absorption in our red spectrum.

$z_{abs}=$1.7331-  This system  consists of Ly$\alpha$ at 3322 $\AA$, 
O I $\lambda$1302 at
3558 $\AA$, the N V doublet at 3386 $\AA$ and 3397 $\AA$, and a possible Si IV 
$\lambda$1393 line at 3809 $\AA$.  The expected position of the Mg II doublet 
falls on a poorly subtracted sky line in the red spectrum.

We find a two-component associated absorption system at $z_{abs}=$2.1800 
consisting of only Ly$\alpha$ (3866 $\AA$) and Ly$\beta$ (3261 $\AA$ and
3263 $\AA$.)

The absorption features at 3964 $\AA$, 3970 $\AA$, 3975 $\AA$, and 3983 $\AA$ 
are identified as traps in the CCD. 

\subsection{Q 0848+153 \hspace{0.5in} $z_{em}=$2.014}

\markcite{ysb1982} YSB82 (cf. Section~\ref{sec-q0226}) find eight
absorption lines blueward of Ly$\alpha$ emission in their spectrum of 
this QSO.   They do not identify any of them.   \markcite{sbs1988} SBS88 (cf. 
Section~\ref{sec-q0150}) detect only one line in their spectrum of this
object (one which is not found by YSB88.)  \markcite{ss1992} SS92 (cf. 
Section~\ref{sec-q0027}) find four absorption features in their red 
spectrum and identify three of them as a Mg II doublet and Fe II $\lambda$2600
at $z_{abs}=$1.0254.  Neither we nor YSB82 nor SBS88 observed the region
of the spectrum necessary to confirm the C IV doublet for this system; but
we do identify Fe II $\lambda$1608 at 3259 $\AA$.
We find no other lines at this redshift or any other 
metal line systems from our data.  We do note that lines 8 and 11
in our line list match the position of the Si IV doublet at $z_{abs}=$1.5738
well, although we cannot call this a true metal line system based on our
criteria.

\subsection{Q 0936+368 \hspace{0.5in} $z_{em}=$2.025}

We have found no previously published spectrum of this object.  Due to
low signal-to-noise in the blue region of our spectrum (3200-3400 $\AA$) the
spectrum was truncated at roughly 3400 $\AA$ for the purposes of the line list.
The absorption features at 3942 $\AA$, 3948 $\AA$, and 3955 $\AA$ are
traps in the CCD.

The only system found is a C IV doublet at 4001 $\AA$ and
4006 $\AA$ and C II $\lambda$1334 at 3448 $\AA$ from a system  
at $z_{abs}=$1.5841.

\subsection{Q 0952+335 \hspace{0.5in} $z_{em}=$2.504}

Our spectrum of this object shows a 
damped Lyman alpha system at 3765 $\AA$ with an observed equivalent width
of  30.97 $\AA$.  The absorption features at 4277 $\AA$, 4282 $\AA$, 4286 $\AA$,
and 4290 $\AA$ are traps in the CCD.
We find ten possible metal line systems:

$z_{abs}=$0.5393- This system consists of several Fe II lines ($\lambda$2344
at 3609 $\AA$, $\lambda$2374 at 3655 $\AA$, $\lambda$2382 at 3668 $\AA$, 
$\lambda$2586 at 3981 $\AA$, and $\lambda$2600 at 4002 $\AA$) and a
possible Mg II doublet at 4304 $\AA$ and 4314 $\AA$. However, these Mg II lines
are weaker than all of the Fe II lines identified, contrary to what is
expected; and the relative strengths of the Fe II lines are also not
entirely consistent with the expected values.  Although the possibility of
blending keeps us from ruling out this system altogether, it is a 
tentative one.  

$z_{abs}=$1.5362-  This redshift is   based upon a C IV $\lambda$1548
line at 3927 $\AA$.  The expected position of C IV $\lambda$1550 for this
redshift falls on a bad column in the data.  We also detect a Si IV doublet
at 3535 $\AA$ and 3558 $\AA$, Si II $\lambda$1526 at 3872 $\AA$, Fe II
$\lambda$1608 at 4079 $\AA$, and Al II $\lambda$1670 at 4237 $\AA$.

$z_{abs}=$2.0399-  For this system, we find Ly$\alpha$ at 3695 $\AA$, 
Si III $\lambda$1206 at 3668 $\AA$, C II $\lambda$1334 at 4055 $\AA$,  and
the Si IV doublet at 4237 $\AA$ and 4265 $\AA$.  Also,
the position of the N V doublet falls within the damped Lyman
alpha line at 3763 $\AA$. 

$z_{abs}=$2.0555-  Ly$\alpha$ for this system is found at 3714 $\AA$.
We also identify Fe II $\lambda$1145 absorption at 3498 $\AA$, 
possible Si III $\lambda$1206 absorption at 3687 $\AA$, possible Si II 
$\lambda$1260 and $\lambda$1304 absorption at 3850 $\AA$ and 3985 $\AA$, 
and a Si IV doublet at 4258 $\AA$ and 4286 $\AA$. 

$z_{abs}=$2.0965-  This system is the damped Ly$\alpha$ absorber noted
above.
The metal lines found at this redshift include Si II
$\lambda$1190 and $\lambda$1193 at 3685 $\AA$ and 3695 $\AA$ (possible), a
N I $\lambda$1200 line at 3714 $\AA$, Si III $\lambda$1206 
at 3735 $\AA$, Si II
$\lambda$1260 at 3903 $\AA$, C II $\lambda$1334 at 4130 $\AA$, and
Si IV $\lambda\lambda$1393,1402 at 4314 $\AA$ and 4342 $\AA$. 

$z_{abs}=$2.1670-  For this system, we find Ly$\alpha$  at 3850 $\AA$,  Fe II
$\lambda$1143 and $\lambda$1145 at 3620 $\AA$ and 3626 $\AA$, N I $\lambda$1200
at 3801 $\AA$, and Si III $\lambda$1206 at 3820 $\AA$.

$z_{abs}=$2.1850-  This system consists of Ly$\alpha$ at 3872 $\AA$, Si II
$\lambda$1193 at 3801 $\AA$, N I $\lambda$1200 at 3820 $\AA$,
Si II $\lambda$1260 and $\lambda$1304 at 4014 $\AA$ and 4153 $\AA$, and 
O I $\lambda$1302 at 4147 $\AA$.

$z_{abs}=$2.2102-  At this redshift, we detect Ly$\alpha$ at 3903 $\AA$,
Si II $\lambda$1190 and $\lambda$1193 at 3820 $\AA$ and 3830 $\AA$, Si III
$\lambda$1206 at 3872 $\AA$, a possible N V $\lambda$1238 line at 3976 $\AA$
(no $\lambda$1402 component is found), a blended Si II $\lambda$1260 line 
at 4046 $\AA$,
and O I $\lambda$1302 at 4180 $\AA$.  The expected position of Si II 
$\lambda$1304 falls on a bad column in the data.

$z_{abs}=$2.2924-  For this system, we identify Ly$\alpha$ at 4002 $\AA$,
N II $\lambda$1083 at 3569 $\AA$, 
Si III $\lambda$1206 at 3972 $\AA$,
and the N V doublet at 4079 $\AA$ and 4092 $\AA$.

$z_{abs}=$2.3189-  This system consists of Ly$\alpha$ at 4035 $\AA$, Fe II
$\lambda$1143 and $\lambda$1145 at 3795 $\AA$ and 3801 $\AA$, Si II $\lambda$
1193 at 3959 $\AA$, and Si II $\lambda$1260 at 4183 $\AA$.

\subsection{Q 0955+472 \hspace{0.5in} $z_{em}=$2.482}

We note the presence of associated absorption in the spectrum of
this radio loud QSO, at
4203 $\AA$, 4206 $\AA$, 4219 $\AA$, and 4241 $\AA$, separated from the
position of the Lyman alpha emission by 2121 km s$^{-1}$, 1910 km s$^{-1}$,
990 km s$^{-1}$, and -539 km s$^{-1}$, respectively.  
We do not find metal
line systems consistent with these redshifts, but we do find Ly$\beta$
absorption in our spectrum for the first, third, and fourth systems listed
above at 3547 $\AA$, 3561 $\AA$, and 3579 $\AA$.  The Ly$\beta$ line for
the second system appears to be blended with Ly$\beta$ for the first
system at 3549 $\AA$, but is not identified as a significant line by our
line-finding program.  The metal line systems we find are as follows:

$z_{abs}=$1.7251-  This system is identified on the basis of a possible
C IV doublet at 4219 $\AA$ and 4225 $\AA$.  The other metal 
lines detected are O I $\lambda$1302 at 3547 $\AA$ and Si II $\lambda$1304
at 3554 $\AA$.  This system is relatively insecure.

$z_{abs}=$2.2849-  For this system, we find Ly$\alpha$ at 3993 $\AA$, N II
$\lambda$1083 at 3561 $\AA$, 
blended N I $\lambda$1200 absorption at 3943 $\AA$,
Si III $\lambda$1206 at 3963 $\AA$, and a possible N V
doublet for which the $\lambda$1238 component is blended with the
Lyman alpha complex at 4071 $\AA$, and the $\lambda$1242 component is
detected at 4082 $\AA$.

$z_{abs}=$2.3453,2.3481-  Ly$\alpha$ for this system is part of the Lyman
alpha complex at 4067 $\AA$.  Other lines detected include a possible, blended 
N I $\lambda$1135 line and N I $\lambda$1200 at 3796 $\AA$ and 4014 $\AA$, 
Fe II $\lambda$1145 at 3830 $\AA$, Si II $\lambda$1190 and $\lambda$1193 at
3984 $\AA$ and 3993 $\AA$, Si III $\lambda$1206 at 4038 $\AA$, and a possible
N V doublet at 4144 $\AA$ and 4156 $\AA$.

$z_{abs}=$2.4087-  This system consists of Ly$\alpha$ at 4144 $\AA$, 
N I $\lambda$1135 and $\lambda$1200 at 3869 $\AA$ and 4090 $\AA$, 
and Si III $\lambda$1206 at 4112 $\AA$.
Despite the fact that the putative N I $\lambda$1135 
line shows good redshift agreement with this system, it is treated as
a possible identification because the stronger line of the same species, 
N I $\lambda$1200, shows poorer agreement.

The absorption features at 4277 $\AA$, 4282 $\AA$, 4286 $\AA$, and 4290 $\AA$
are traps in the CCD.

\subsection{Q 0956+122 \hspace{0.5in} $z_{em}=$3.308}

Sargent et al. (1989) obtained a spectrum of this object with 4 $\AA$
resolution from 3150 $\AA$ to 4700 $\AA$ and 6 $\AA$ resolution from
4600 $\AA$ to 7000 $\AA$.  They
find weak C IV systems at $z_{abs}=$2.9145 and
$z_{abs}=$3.2230.  We find only Ly$\alpha$ at $z_{abs}=$2.9156.
The system at $z_{abs}=$3.2230 is identified as a Lyman
limit system by Steidel (1990) from a higher resolution ($\sim$1.1 $\AA$) 
spectrum.  He identifies C IV and Si IV doublets, Si III $\lambda$1206,
C III $\lambda$977 and several Lyman series lines.
We confirm this system with our
detection of Ly$\alpha$ at 5134 $\AA$, N I $\lambda$1200 at 5069 $\AA$,
and Si III $\lambda$1206 at 5095 $\AA$.
Songaila \& Cowie (1996)
identify this system as a partial Lyman limit system at $z_{abs}=$3.2216.
Sargent et al. (1989)  also find a Lyman limit system with no corresponding 
heavy element lines at $z_{abs}=$3.096.  We identify strong Ly$\alpha$ 
absorption at this redshift as well as a possible Si II $\lambda$1260 line 
at 5162 $\AA$.  Both of these lines are found in the spectrum of Steidel (1990),
but they are not attributed to a Lyman limit system.
Instead, Steidel (1990) finds another Lyman limit system at 
$z_{abs}=$3.11.  We detect strong Lyman alpha absorption at this
redshift as well as Si III $\lambda$1206.   The position of Si II $\lambda$1260
falls on a trap in the CCD.
Several other metal line systems were also found by this author:

$z_{abs}=$0.0456-  Our spectrum does not extend far enough into the red
to allow us to confirm the Na I $\lambda\lambda$5891, 5897 lines tentatively 
identified at this redshift.

$z_{abs}=$2.3104-  Steidel (1990) tentatively identifies Ly$\alpha$, 
C IV $\lambda$1548 and Al II $\lambda$1670 at this redshift.  We find
C II $\lambda$1334 at 4418 $\AA$, a double-component Si IV doublet
at 4614 $\AA$ and 4636 $\AA$, Si II $\lambda$1526 at 5054 $\AA$, and
C IV $\lambda$1548 and $\lambda$1550 (blended with Ly$\alpha$ at
$z_{abs}=$3.223) at 5125 $\AA$ and 5134 $\AA$, 
Hu et al. (1995) identify this double-component Si IV doublet as well
in a high resolution ($\sim$0.13 $\AA$) spectrum taken with the HIRES 
Spectrograph on the Keck Telescope.  
The $\lambda$1393 line is seen at $z_{abs}=$2.3104 and
$z_{abs}=$2.3109.
 
$z_{abs}=$2.7169-  Steidel (1990) finds C II $\lambda$1334, a C IV doublet, and
Al II $\lambda$1670 at this redshift.
We confirm this system with the detection of Ly$\alpha$ at
4519 $\AA$, N I $\lambda$1200 at 4461 $\AA$, Si III $\lambda$1206
at 4484 $\AA$, the N V doublet at 4604 $\AA$ and 4618 $\AA$, and C II
$\lambda$1334 at 4959 $\AA$.  The position of O I $\lambda$1302 falls on
a bad region in the spectrum.
  
$z_{abs}=$2.7261-  Steidel (1990) finds a weak C IV doublet at this redshift.
We do not find Ly$\alpha$ corresponding to this redshift.

$z_{abs}=$2.8320-  Steidel (1990) finds a weak C IV doublet at this redshift
as well as Ly$\beta$, Si II $\lambda$1260, and C II $\lambda$1334.  We
identify Ly$\alpha$ at 4659 $\AA$, N I $\lambda$1200 at 4599 $\AA$, Si II
$\lambda$1260 at 4830 $\AA$, a possible Si II $\lambda$1304 line
at 5002 $\AA$ (blended with Ly$\alpha$ at $z=$3.1145), O I $\lambda$1302
4990 $\AA$, and a possible C II $\lambda$1334 line at 5118 $\AA$.
 
$z_{abs}=$3.1045-  Steidel (1990) identifies Ly$\alpha$, C III $\lambda$977,
and the C IV doublet for this secure system.  We confirm strong Ly$\alpha$
absorption at 4990 $\AA$ and find a Si II $\lambda$1260 line at 5172 $\AA$.
 
$z_{abs}=$3.1530-  Steidel (1990) finds a weak C IV doublet, a Si IV doublet,
Si II $\lambda$1190 and $\lambda$1193, Si III $\lambda$1206, and several
Lyman series lines.  We detect Ly$\alpha$ at 5048 $\AA$, a possible N I 
$\lambda$1200 line at 4980 $\AA$ (blended with Ly$\alpha$ at $z_{abs}=$3.0963),
a tentative Si III $\lambda$1206 line at 5012 $\AA$, and the N V doublet
at 5144 $\AA$ and 5157 $\AA$.  We detect the features identified by
Steidel (1990) as Si II $\lambda$1190 and $\lambda$1193, but since our
spectrum shows no feature at the position of Si II $\lambda$1260, we do
not confirm those identifications. 

We identify several other possible metal line systems from our spectrum:

$z_{abs}=$2.8342-  This system is separated by 172 km s$^{-1}$ from the
system found by Steidel (1990) at $z_{abs}=$2.8320.  Ly$\alpha$ is
detected at 4661 $\AA$, the N V doublet at 4750 $\AA$ and 4765 $\AA$, and
Si II $\lambda$1260 at 4832 $\AA$.  The Si II $\lambda$1304 and C II
$\lambda$1334 identified with the $z_{abs}=$2.8320 system are more likely
associated with this system.

$z_{abs}=$3.0490-  This system consists of Ly$\alpha$ at 4922 $\AA$, possible
Fe II $\lambda$1143 and $\lambda$1145 lines at 4630 $\AA$ and 4636 $\AA$, 
Si III $\lambda$1206 at 4884 $\AA$ and Si II $\lambda$1260 at 5103 $\AA$. 
Steidel (1990) finds no line which would correspond to Fe II $\lambda$1608
at $\sim$6510 $\AA$ or C IV at $\sim$6270 $\AA$.

$z_{abs}=$3.0528-  At this redshift, we identify Ly$\alpha$ at 4927 $\AA$,
N I $\lambda$1135 and $\lambda$1200 at 4599 $\AA$ and 4862 $\AA$, Si III
$\lambda$1206 at 4890 $\AA$, and the N V doublet at 5021 $\AA$ and 5036 $\AA$.
There is a line in the Steidel (1990) line list at 6274 $\AA$, which would
correspond to C IV $\lambda$1548 at this redshift, but none at 6285 $\AA$,
which would correspond to C IV $\lambda$1550.

$z_{abs}=$3.1321-  This system is composed of Ly$\alpha$ at 5023 $\AA$,
a possible N II $\lambda$1083 line at 4480 $\AA$, N I $\lambda$1135 and
$\lambda$1200 at 4689 $\AA$ and 4959 $\AA$, a possible N V doublet,
both components of which are blended with other lines (see line list),
at 5118 $\AA$ and 5134 $\AA$, and Si II $\lambda$1260 at 5208 $\AA$.
No C IV is detected by Steidel (1990).

$z_{abs}=$3.1975-  At this redshift, we detect Ly$\alpha$ at 5103 $\AA$, N I
$\lambda$1200 at 5036 $\AA$, Si III $\lambda$1206 at 5065 $\AA$, and a
possible N V doublet at 5200 $\AA$ and 5217 $\AA$.  A feature at 6497 $\AA$
in the line list of Steidel (1990) would correspond to C IV $\lambda$1548
at this redshift, but no $\lambda$1550 component is present.

$z_{abs}=$3.2461-  This system consists of Ly$\alpha$ at 5162 $\AA$, Fe II
$\lambda$1143 and $\lambda$1145 at 4855 $\AA$ and 4862 $\AA$, N I 
$\lambda$1135 and $\lambda$1200 at 4753 $\AA$ and 5095 $\AA$, and Si III
$\lambda$1206 at 5122 $\AA$.  Steidel (1990) finds no C IV doublet
or Fe II $\lambda$1608 at this redshift.

$z_{abs}=$3.2774-  At this redshift, we identify Ly$\alpha$ at 5200 $\AA$,
N II $\lambda$1083 at 4636 $\AA$, N I $\lambda$1135 and $\lambda$1200 at
4855 $\AA$ and 5134 $\AA$, and Si III $\lambda$1206 at 5162 $\AA$.
Steidel (1990) finds no C IV at this redshift. 

The absorption features at 5176 $\AA$, 5181 $\AA$, 5185 $\AA$, and 5189 $\AA$
are identified as traps in the CCD.

\subsection{Q 1009+299 \hspace{0.5in} $z_{em}=$2.633}

There are no previously published absorption line spectra of this object.
From our data, we find eight candidate metal line systems including a complex
of associated absorption near the quasar redshift:

$z_{abs}=$1.8484-  This system is identified by the C IV doublet
at 4410 $\AA$ and 4418 $\AA$, the $\lambda$1550 component of which is blended
with Ly$\alpha$ at $z_{abs}=$2.6339.  Other lines found include 
O I $\lambda$1302
at 3709 $\AA$, Si II $\lambda$1304 and $\lambda$1526 at 3715 $\AA$ and 4349
$\AA$.

$z_{abs}=$2.2611-  For this system, we identify Ly$\alpha$ at 3964 $\AA$,
Si III $\lambda$1206 at 3934 $\AA$, Si II $\lambda$1260 at 4110 $\AA$,
O I $\lambda$1302 at 4246 $\AA$, and a possible C II $\lambda$1334 line
at 4353 $\AA$.  The expected positions of Fe II $\lambda$1143 and 
$\lambda$1145 fall on bad columns in the data.

$z_{abs}=$2.3582-  This system is comprised of Ly$\alpha$ at 4082 $\AA$,
N I $\lambda$1200 at 4030 $\AA$, Si III $\lambda$1206 at 4052 $\AA$, 
Si II $\lambda$1260
at 4232 $\AA$, and O I $\lambda$1302 at 4373 $\AA$.

$z_{abs}=$2.3809-  At this redshift, we detect Ly$\alpha$ at 4110 $\AA$,
N II $\lambda$1083 at 3665 $\AA$,  N I $\lambda$1200 at 4056 $\AA$, Si III
$\lambda$1206 at 4079 $\AA$, a possible N V $\lambda$1242 line at 4201 $\AA$
(the expected position of the $\lambda$1238 component falls on a bad region
in the spectrum), and O I $\lambda$1302 at 4403 $\AA$.

$z_{abs}=$2.4068-  For this system, we identify very strong, weakly damped  
damped Ly$\alpha$ absorption at 4141 $\AA$,
Si II $\lambda$1190 at 4056 $\AA$ (the position of Si II $\lambda$1193 falls
on a bad region in the spectrum),N I $\lambda$1200 
at 4087 $\AA$, Si III
$\lambda$1206 at 4110 $\AA$, Si II $\lambda$1260 at 4294 $\AA$, and O I 
$\lambda$1302 at 4436 $\AA$.  The position of Si II $\lambda$1304 falls
on a bad column. 

$z_{abs}=$2.5236-  At this redshift, we identify Ly$\alpha$ at 4283 $\AA$,
N I $\lambda$1135 at 3998 $\AA$, Si II $\lambda$1193 at 4205 $\AA$, N I
$\lambda$1200 at 4227 $\AA$, and Si III $\lambda$1206 at 4252 $\AA$.  The
expected position of Si II $\lambda$1260 falls on bad columns in the data.

$z_{abs}=$2.5531-  This system consists of Ly$\alpha$ at 4319 $\AA$, 
Ly$\beta$ at 3645 $\AA$, O VI
$\lambda$1031 and $\lambda$1037 at 3667 $\AA$ and 3686 $\AA$, N II 
$\lambda$1083 at 3851 $\AA$, Fe II $\lambda$1143 at 4061 $\AA$ (the position
of $\lambda$1145 falls on bad columns in the spectrum), and Si II $\lambda$1193
at 4240 $\AA$.

$z_{abs}=$2.6158-  For this associated absorber, 
Ly$\alpha$ is found at 4396 $\AA$, Ly$\beta$ at 3709 $\AA$,
C II $\lambda$1036 at 3746 $\AA$, possible Fe II $\lambda$1143 and
Fe II $\lambda$1145 blended with Ly$\alpha$ at $z_{abs}=$2.40677 at 
4134 $\AA$ and 4141 $\AA$, and 
Si III $\lambda$1206 at 4362 $\AA$.

The absorption features at 4412 $\AA$, 4418 $\AA$, 4422 $\AA$, and 4425 $\AA$
are identified as traps in the CCD.

\subsection{Q 1207+399 \hspace{0.5in} $z_{em}=$2.459}

According to our literature searches, there is no previously published
absorption line spectrum of this QSO.  From our data, we find two metal line
systems:

$z_{abs}=$2.1116-  At this redshift, we detect a blended Ly$\alpha$ line
at 3781 $\AA$, Si III $\lambda$1206 at 3765 $\AA$, Si II $\lambda$1260
at 3922 $\AA$, 
C II $\lambda$1334 at 4152 $\AA$, Si IV $\lambda\lambda$1393,1402
at 4337 $\AA$ and 4365 $\AA$, and a possible C IV $\lambda$1548 line at
4816 $\AA$.  The expected position of C IV $\lambda$1550 for this redshift
falls just outside our spectral range.

$z_{abs}=$2.1561-  At this redshift, we find Ly$\alpha$ at 3837 $\AA$,
Ly$\beta$ at 3238 $\AA$, Si III $\lambda$1206 at 3808 $\AA$, 
the N V doublet at 3907 $\AA$ and 3922 $\AA$, Si II
$\lambda$1260 at 3977 $\AA$, and C II $\lambda$1334 at 4212 $\AA$.

The absorption features present
at 4576 $\AA$, 4587 $\AA$, 4595 $\AA$ and 4603 $\AA$ are traps in the CCD.

\subsection{Q 1210+175 \hspace{0.5in} $z_{em}=$2.564}

This QSO was observed by \markcite{foltz1987} Foltz et al. (1987) who noted 
a possible damped Lyman alpha system in their spectrum at roughly 3500 $\AA$.
According to Wolfe et al. (1995) this system is a confirmed damped Ly$\alpha$
absorber with an equivalent width of 11.3 $\AA$. 
Ly$\alpha$ for this candidate is not within our spectral range for 
this object.  However,
we find five metal line systems from our data, one of which is consistent
with this damped system.

$z_{abs}=$1.8917-  This system is the damped Ly$\alpha$ absorber discussed
above.   Ly$\alpha$ at this redshift is outside our spectral range, but we
do detect the Si IV doublet
at 4030 $\AA$ and 4056 $\AA$.  Other lines detected include Si II 
$\lambda$1260, $\lambda$1304, and $\lambda$1526 at 3645 $\AA$, 3772 $\AA$,
and 4414 $\AA$, O I $\lambda$1302 at 3765 $\AA$, and C II $\lambda$1334
at 3859 $\AA$.

$z_{abs}=$2.0548-  At this redshift, 
we identify Ly$\alpha$ at 3713 $\AA$, Si II $\lambda$1193, $\lambda$1260,
and $\lambda$1304 at 3645 $\AA$, 3850 $\AA$, and 3985 $\AA$, and C II 
$\lambda$1334 at 4076 $\AA$.  The Si II $\lambda$1304 line must be 
a blend if it is present.

$z_{abs}=$2.1240-  For this system, we detect Ly$\alpha$ at 3798 $\AA$,
Si III $\lambda$1206 at 3768 $\AA$, a N V doublet at 3868 $\AA$
and 3881 $\AA$, a possible Si II $\lambda$1260 line at 3938 $\AA$, and
C II $\lambda$1334 at 4169 $\AA$.  

$z_{abs}=$2.1974-  For this system, we identify Ly$\alpha$ at 3887 $\AA$,
N I $\lambda$1200 at 3837 $\AA$, Si III $\lambda$1206 
at 3859 $\AA$, Si II $\lambda$1260 at 4030 $\AA$, 
O I $\lambda$1302 at 4164 $\AA$, and C II $\lambda$1334 at 4266 $\AA$.

$z_{abs}=$2.5786-  This system consists of Ly$\alpha$ at 4350 $\AA$, 
Ly$\beta$ at 3671 $\AA$, and O VI $\lambda$1031 and $\lambda$1037 at
3693 $\AA$ and 3714 $\AA$.  Both O VI lines are stronger than Ly$\alpha$
and Ly$\beta$ indicating either that they are blends or that the
line of sight through this absorber intersects regions dominated by
highly ionized gas.  The latter interpretation is likely because the
redshift of this absorber is larger than the QSO emission redshift,
indicating that this absorbing material must be infalling gas associated
with the QSO itself.

\subsection{Q 1231+294 \hspace{0.5in} $z_{em}=$2.018}

\markcite{tdw19899} Thompson et al. (1989) measure an emission redshift
of $z_{em}=$2.011 $\pm$ 0.001 for this QSO from [O IV]+Si IV 
$\lambda\lambda$1397-1406 and C III] $\lambda$1909 emission lines.  Our
spectrum of Ly$\alpha$ emission gives a redshift of $\sim$2.018.  

We find two metal line systems from our absorption line spectrum.

$z_{abs}=$1.4780-  This system consists of the C IV doublet at 3836 $\AA$
and 3843 $\AA$ and the Si IV doublet at 3454 $\AA$ and 3477 $\AA$.

$z_{abs}=$1.8755-  For this system, we identify Ly$\alpha$  at 3496 $\AA$,
possible N I $\lambda$1135 and $\lambda$1200 lines at 3264 $\AA$ and 3450 $\AA$,
a possible Fe II $\lambda$1145 line at 3292 $\AA$, and O I $\lambda$1302
at 3745 $\AA$, blended with C IV $\lambda$1550 at $z_{abs}=$1.4145.

Lastly, we identify a C IV doublet at $z_{abs}=$1.4145 and a C IV doublet at
$z_{abs}=$1.1672 along
with Al II $\lambda$1670 at 3621 $\AA$, though we detect no other lines
at these redshifts.
The absorption features at 3937 $\AA$, 3942 $\AA$, 3946 $\AA$, and 3950 $\AA$
are traps in the CCD.  The feature at 3722 $\AA$ is spurious as well,
and it most likely a cosmic ray.

\subsection{Q 1323-107 \hspace{0.5in} $z_{em}=$2.360}

The only previously published spectrum found for this object is a spectrum
including Ly$\alpha$ and C IV emission from \markcite{k1981} Kunth et al. 
(1981).  They find an emission redshift of 2.360 for the QSO.   We find
four candidate metal line systems from our absorption line spectrum:

$z_{abs}=$1.4244-  This system is based upon the Si IV doublet at
3379 $\AA$ and 3401 $\AA$.  At this redshift, we also detect C II
$\lambda$1334 at 3235 $\AA$ and Si II $\lambda$1526 at 3701 $\AA$.  No
C IV doublet is detected.

$z_{abs}=$1.4727-  This system is identified by the C IV doublet at
3828 $\AA$ and 3835 $\AA$.  Other lines detected include O I $\lambda$1302
at 3220 $\AA$ and C II $\lambda$1334
at 3300 $\AA$. 

$z_{abs}=$1.4922-  This system is based upon the C IV doublet at 3858
$\AA$ and 3864 $\AA$.  Due to the large uncertainty in the position of
the line center for the $\lambda$1550 component, the redshifts of the
doublet components agree to within $<$1$\sigma$.  We also detect
O I $\lambda$1302 at 3246 $\AA$, possible Si II $\lambda$1304 and
$\lambda$1526 lines at 3250 $\AA$ and 3803 $\AA$ respectively, and Fe II
$\lambda$1608 at 4008 $\AA$.  Our red spectrum of this object (see Paper II)
actually
extends slightly blueward of 1.0 $\AA$ resolution blue spectrum and shows a 
possible Si II
$\lambda$1260 line at 3144 $\AA$.

$z_{abs}=$1.8415-  This system consists of Ly$\alpha$ at 3454 $\AA$, N I
$\lambda$1135 and $\lambda$1200 at 3225 $\AA$ and 3409 $\AA$, a possible O I
$\lambda$1302 line at 3701 $\AA$, Si II $\lambda$1260 and $\lambda$1304 at
3582 $\AA$, and 3706 $\AA$ respectively, and C II $\lambda$1334 at 3792 $\AA$.
The Si II $\lambda$1260 line must be blended because its equivalent width is
larger than that of Ly$\alpha$.  The N I and Si II line matches have been
retained despite poor redshift agreement between the two lines of the same
species due to the fact the errors in the line centers of lines 32 (N I 
$\lambda$1200) and 64 (Si II $\lambda$1260) are large enough for these 
redshifts to agree to within $\sim$3$\sigma$.  Our red spectrum shows 
no lines redward of Ly$\alpha$ for this system.

\subsection{Q 1329+412 \hspace{0.5in} $z_{em}=$1.934}

\markcite{sbs1988} SBS88 (cf. Section~\ref{sec-q0150}) 
find six absorption line systems in their spectrum
of this object. \markcite{ss1992} SS92
(cf. Section~\ref{sec-q0027}) confirm two of these systems and find 
another.  These are the systems these authors report and the additional
information gained from our spectrum:

$z_{abs}=$0.5009-  \markcite{sbs1988} SBS88 regard this system as probable from
their identification of the Mg II doublet.  The only search lines that fall in
our spectral range for this redshift are Fe II $\lambda$2344-$\lambda$2600.
We find none of these.

$z_{abs}=$1.2821-  This system is identified by \markcite{ss1992} SS92 from a
strong Mg II doublet.  The spectrum of \markcite{sbs1988} SBS88 did not 
cover the region of C IV absorption, but ours does and we find no significant
lines that would correspond to the C IV doublet at this redshift. 

$z_{abs}=$1.4716-  This system is identified by \markcite{sbs1988} SBS88
on the basis  of an $``$unambiguous" C IV doublet.  \markcite{ss1992} SS92
find no Mg II absorption at this redshift.  We confirm the C IV doublet
identification of \markcite{sbs1988} SBS88 and also find a tentative O I
$\lambda$1302 line at 3217 $\AA$.

$z_{abs}=$1.6010-  \markcite{sbs1988} SBS88 find a strong C IV doublet
at this redshift which they note is likely to be blended with another
C IV doublet at a nearby redshift.   \markcite{ss1992} SS92 identify
the Mg II doublet at this redshift.  Our spectrum shows the strong C IV
doublet found by \markcite{sbs1988} SBS88 in addition to Si II $\lambda$1260
at 3279 $\AA$, C II $\lambda$1334 at 3471 $\AA$, and the Si IV doublet
at 3625 $\AA$ and 3648 $\AA$.  In addition, we find that the position of
the C IV $\lambda$1548 for $z_{abs}=$1.5980 corresponds to a significant
line in our spectrum while the $\lambda$1550 component at this redshift
appears to be strongly blended with C IV $\lambda$1548 at $z_{abs}=$1.6007.

$z_{abs}=$1.8359-  This system is identified 
by \markcite{sbs1988} SBS88
by the C IV doublet and confirmed by \markcite{ss1992} SS92 who find the Mg II
doublet at $z_{abs}=$1.8355.  We detect Ly$\alpha$ at 3447 $\AA$, a possible 
Fe II $\lambda$1145 line at 3246 $\AA$, possible Si II $\lambda$1193 and 
$\lambda$1260 lines at 3384 $\AA$ and 3575 $\AA$, and C II $\lambda$1334 
at 3785 $\AA$.

$z_{abs}=$1.8401-  This system is identified by 
\markcite{sbs1988} SBS88 on the basis of a C IV doublet.  \markcite{ss1992} SS92
do not detect Mg II. We do detect a strong
Ly$\alpha$ line consistent with this redshift at 3453 $\AA$.

$z_{abs}=$1.9406-  \markcite{sbs1988} SBS88 identify this system on the basis
of the C IV doublet.  \markcite{ss1992} SS92 do not observe the spectral region
encompassing Mg II at this redshift; but we detect 
Ly$\alpha$ at 3575 $\AA$ and a N V doublet
at 3643 $\AA$ and 3654 $\AA$.  This system,  having a redshift larger than the
QSO emission redshift, is probably associated with the QSO. 

The absorption features at 3969 $\AA$, 3974 $\AA$, 3979 $\AA$, and 3983 $\AA$
are traps in the CCD.

We detect a possible C IV doublet redward of Ly$\alpha$ emission but 
blueward of the spectral range of \markcite{ss1992} SS92 at a redshift
of 1.35285.  The components are detected at 3643 $\AA$ and 3648 $\AA$
along with Fe II $\lambda$1608 at 3785 $\AA$.  The $\lambda$1548 component
of the doublet coincides with N V $\lambda$1238 at $z_{abs}=$1.9404;
and the $\lambda$1550 component coincides with the Si II $\lambda$1402 for 
the well-established system at $z_{abs}=$1.6010 described above. 
No other lines are found.   Also, we find another possible C IV doublet
in the Ly$\alpha$ forest at $z_{abs}=$1.2480; but no other lines
are detected at this redshift either.

Lastly, \markcite{lwt1995} Lanzetta, Wolfe, \& Turnshek (1995) report
a damped Lyman alpha system at $z_{abs}=$0.5193 in the IUE spectrum of 
\markcite{lts1993} Lanzetta, Turnshek, \& Sandoval (1993).   Again,
the only lines in our spectral range for this redshift are Fe II 
$\lambda$2344-$\lambda$2600.
We detect only the strongest of these lines, Fe II $\lambda$2382 at 3621 $\AA$.

\subsection{Q 1337+285 \hspace{0.5in} $z_{em}=$2.541}

Our literature search yielded no previously published optical spectrum
of this QSO.  From our spectrum, we detect two possible heavy metal
absorption systems:

$z_{abs}=$2.5081-  This relatively secure system consists of Ly$\alpha$ 
at 4265 $\AA$, Ly$\beta$
at 3598 $\AA$, possible O VI $\lambda$1031 and $\lambda$1037 lines
at 3619 $\AA$ and 3640 
$\AA$, C II $\lambda$1036 at 3636 $\AA$, N II $\lambda$1083 at 3803 $\AA$,
a possible Fe II $\lambda$1145 line at 4017 $\AA$, 
and Si II $\lambda$1190 and $\lambda$1193 at 4176 $\AA$ and 4186 $\AA$.

$z_{abs}=$2.5228-  For this  system, we  detect Ly$\alpha$ at
4283 $\AA$, Ly$\beta$ at 3614 $\AA$, O VI $\lambda$1031 at 3636 $\AA$,
possible N I $\lambda$1135 and $\lambda$1200 lines at 3998 $\AA$ and 
4226 $\AA$, Fe II
$\lambda$1143 and $\lambda$1145 at 4027 $\AA$ and 4033 $\AA$, 
and Si III $\lambda$1206 at 4249 $\AA$. 

The absorption features at 4339 $\AA$,  4344 $\AA$, 4347 $\AA$, 
and 4352 $\AA$ are traps
in the CCD.

\subsection{Q 1346-036 \hspace{0.5in} $z_{em}=$2.362}

The spectrum of this QSO blueward of Ly$\alpha$ emission has been studied
by \markcite{ysb1983} YSB83 (cf. Section~\ref{sec-q0226}).  
We confirm all the absorption features seen by these authors  
with the exception of the line they detect at 3844 $\AA$ which falls on
a bad region in our spectrum.  They find no metal line absorbers from their
data, but suggest a possible Mg II doublet at  $z_{abs}=$0.4453.  We detect
this tentative doublet at 4043 $\AA$ and 4054 $\AA$ ($z_{abs}=$0.4458) but
find no Fe II absorption at this redshift.  We detect the 4051 $\AA$ line
reported by \markcite{ysb1983} YSB83, but identify Mg II $\lambda$2803
with the absorption feature at 4054 $\AA$  for better redshift agreement.

\markcite{ltw1987} LTW87 (cf. Section~\ref{sec-q0226})
report no absorption features
in their red (6250 $\AA$ - 8350 $\AA$) spectrum of this object.  And
\markcite{wolfe1986} W86 (cf. Section~\ref{sec-q0037}) find no 
damped Lyman alpha candidates
in their 3200 $\AA$ - 5200 $\AA$ spectrum.

The only additional identifications we make for this object are are two 
Ly$\alpha$-Ly$\beta$ pairs at 3965 $\AA$ and 3345 $\AA$ ($z_{abs}=$2.2616)
and at 4028 $\AA$ and 3450 $\AA$ ($z_{abs}=$2.3630).
For the $z_{abs}=$2.2616 pair, the Ly$\beta$ line is stronger than 
Ly$\alpha$ and must be a blend; also, our red spectrum (see Paper II)
shows the C IV doublet
for this system at 5050 $\AA$ and 5058 $\AA$.  
The $z_{abs}=$2.3630 redshift is
larger than the QSO emission redshift indicating that it must be 
associated with the QSO, although not an associated absorber per se, as it
shows no metal lines Our red spectrum does not show the C IV doublet
at this redshift.

\subsection{Q 1358+115 \hspace{0.5in} $z_{em}=$2.589}

\markcite{wolfe1986}  W86 (cf. Section~\ref{sec-q0037})
find several 4$\sigma$ 
absorption features in their 10 $\AA$ resolution spectrum of this object.
We confirm these lines with the exception of the features they report at
3573 $\AA$, 3874 $\AA$, and 4092 $\AA$.   We also confirm the feature they
report at  4074 $\AA$ having less than 4$\sigma$ significance.

We find six possible metal line systems from our data:

$z_{abs}=$0.5084-  This system is a Mg II absorber for which the doublet
is detected at 4218 $\AA$ and 4228 $\AA$.  The $\lambda$2803 component
of the doublet is blended with Ly$\alpha$ at $z_{abs}=$2.4778. 
We also detect Fe II $\lambda$2382 and $\lambda$2600
at 3593 $\AA$ and 3922 $\AA$ and Mg I $\lambda$2853 at 4303 $\AA$.

$z_{abs}=$2.4158-  This system is composed of a strong Ly$\alpha$ line 
at 4152 $\AA$, possible
Si II $\lambda$1190 and $\lambda$1193 lines at 4065 $\AA$ and 4075 $\AA$, 
a N I possible $\lambda$1200 line at 4098 $\AA$, and
Si III $\lambda$1206  at 4121 $\AA$.  The
expected position of N I $\lambda$1135 for this redshift falls on a bad column
in the data.

$z_{abs}=$2.5559-  For this system, we find Ly$\alpha$ at 4323 $\AA$, 
Ly$\beta$ at 3647 $\AA$, Si II $\lambda$1190 and $\lambda$1193 at 
4234 $\AA$ and 4243 $\AA$, N I $\lambda$1200 at 4266 $\AA$, and Si III 
$\lambda$1206 at 4290 $\AA$.

$z_{abs}=$2.5630-  This system is composed of Ly$\alpha$ at 4331 $\AA$,
Ly$\beta$ at 3655 $\AA$ and
Si II $\lambda$1190 and $\lambda$1193 at 4241 $\AA$ and 
4251 $\AA$.  The Si II $\lambda$1193 line is blended with Ly$\alpha$ at
$z_{abs}=$2.4968.

$z_{abs}=$2.5763-  At this redshift, we find an associated absorber showing
Ly$\alpha$ at 4348 $\AA$, Ly$\beta$ blended with the feature
at 3672 $\AA$ (Ly$\beta$ at $z_{abs}=$2.5799),
O VI $\lambda$1031 and $\lambda$1037 at 3689 $\AA$ and 3709 $\AA$.

$z_{abs}=$2.57996-  This system is another associated absorber
for which we identify Ly$\alpha$ at
4353 $\AA$, Ly$\beta$ at 3672 $\AA$, and O VI $\lambda$1031 and $\lambda$1037
at 3694 $\AA$ and 3714 $\AA$.

\subsection{Q 1406+492 \hspace{0.5in} $z_{em}=$2.161}

Literature searches yielded no previously published absorption spectrum
of this QSO.  From our data, we find two possible heavy element absorption
systems:

$z_{abs}=$1.4330-  This redshift is based upon the C IV doublet at
3767 $\AA$ and 3773 $\AA$.  We also detect the Si IV doublet at
3391 $\AA$ and 3411 $\AA$.  However, the Si IV $\lambda$1402 line
must be a blend (possibly with Si IV $\lambda$1393 at $z_{abs}=$1.4474) 
due to its equivalent width relative to the $\lambda$1393
component and its poor redshift agreement with it.

$z_{abs}=$1.4470-  This system is another C IV absorber for which the C IV
doublet is found at 3788 $\AA$ and 3795 $\AA$.  We also find C II $\lambda$1334
at 3266 $\AA$, the Si IV doublet for which the $\lambda$1393 component
lies at 3411 $\AA$ and the $\lambda$1402 component is blended with the
feature at 3435 $\AA$, Si II $\lambda$1526 at 3736 $\AA$, and Fe II $\lambda$1608
at 3936 $\AA$.

A C IV doublet is found at
at $z_{abs}=$1.5253; and
we find a Ly$\alpha$, Ly$\beta$ pair due to an
absorber at $z_{abs}=$2.1540.
The absorption features present at 3962 $\AA$, 3967 $\AA$, 3968 $\AA$, 
3974 $\AA$, 3978 $\AA$, and 3981 $\AA$ are traps in the CCD.

\subsection{Q 1408+009 \hspace{0.5in} $z_{em}=$2.260}

According to a literature search, this is the first published spectrum 
of this  object.  Five possible metal line systems are found:

$z_{abs}=$1.3158-  This system is identified by a Si IV doublet at
3228 $\AA$ and 3248 $\AA$ as well as Si II $\lambda$1526 absorption at
3535 $\AA$ and a Fe II $\lambda$1608 line at 3725 $\AA$.

$z_{abs}=$1.5190-   This system is a C IV absorber with $\lambda$1548
identified at 3900 $\AA$ and $\lambda$1550 at 3906 $\AA$.  Also found
are C II $\lambda$1334 at 3363 $\AA$ and Si II $\lambda$1526 at 3843 $\AA$.

$z_{abs}=$1.6929-  This system consists of Ly$\alpha$ at 3274 $\AA$, Si III
$\lambda$1206 at 3248 $\AA$, Si II $\lambda$1260 at 3394 $\AA$, and O I 
$\lambda$1302 at 3506 $\AA$.  Despite the fact that this Ly$\alpha$
line is relatively strong (EW$_{0}=$1.153 $\AA$) all of the other lines
identified are stronger, creating the need to invoke the possibility of
blending for all of them.  For this reason, this system is considered 
uncertain.

$z_{abs}=$1.9956-  At this redshift, we detect Ly$\alpha$ at 3642 $\AA$,
N I $\lambda$1200 at 3595 $\AA$, 
Si II $\lambda$1260 and
$\lambda$1304 at 3774 $\AA$ and 3906 $\AA$, and O I $\lambda$1302
at 3900 $\AA$.

$z_{abs}=$2.1991-  For this system, we identify Ly$\alpha$ at 3889 $\AA$,
a blended Ly$\beta$ line at 3282 $\AA$, Si III $\lambda$1206 at 3859 $\AA$,
and Si II $\lambda$1260 at 4032 $\AA$.

The absorption features at 4575 $\AA$, 4586 $\AA$, 4602 $\AA$ are traps in the
CCD.

\subsection{Q 1421+330 \hspace{0.5in} $z_{em}=$1.905}

The rest-UV absorption spectrum of this object has been studied by many
authors. \markcite{wwpt1979} Weymann et al. (1979) find C IV
in their 2.5 $\AA$ resolution spectrum at $z_{abs}=$1.462, but not the 
expected Si IV and C II absorption.   This redshift
is confirmed by \markcite{k1992} Koratkar et al. (1992) and by our data.  We
find Si IV at 3433 $\AA$ and 3455 $\AA$ and C IV at 3813 $\AA$ and 3820 $\AA$.

\markcite{u1984} Uomoto (1984) finds several tentative systems in his red
spectrum of this QSO:

$z_{abs}=$0.2249-  \markcite{u1984} Uomoto (1984) detects a Mg II doublet
at this reshift.  Our spectrum shows this identification to be unlikely
given the implied velocity separation of the doublet lines ($\sim$310
km s$^{-1}$) if they are associated with the features at 3428 $\AA$ and
3433 $\AA$ in our data.  

$z_{abs}=$0.3236-  \markcite{u1984} Uomoto (1984) finds a Mg II doublet at
this redshift.  Our spectrum does not show these lines, nor any others 
at this redshift. 

$z_{abs}=$0.9030-  \markcite{u1984} Uomoto (1984) finds several Fe II
lines at this redshift, $\lambda$2344, $\lambda$2374, $\lambda$2382,
$\lambda$2586, and $\lambda$2600.  Also, Mn II $\lambda$2594 $\AA$ and
a Mg II doublet are detected.  This Mg II doublet is confirmed by 
\markcite{ss92} SS92 (cf. Section~\ref{sec-q0027}.)  Our spectrum
shows Al III $\lambda$1854 and $\lambda$1862 absorption at 3530 $\AA$ and
3544 $\AA$. 

$z_{abs}=$1.1732- \markcite{u1984} Uomoto (1984) finds a Mg II doublet
at this redshift which is confirmed by \markcite{ss92} SS92.  We find
Si II $\lambda$1526 at 3318 $\AA$; but no C IV or Al III.

$z_{abs}=$1.2252-  \markcite{u1984} Uomoto (1984) finds a C IV doublet
at this redshift.  We detect absorption at the position of the $\lambda$1548
component, but none at the position of $\lambda$1550.  

\markcite{foltz1986} Foltz et al. (1986) find four additional systems in
their 1 $\AA$ resolution spectrum covering 3820 $\AA$ to 4035 $\AA$:

$z_{abs}=$0.4565-  A Mg II doublet is detected
at this redshift.  These lines should fall at the very red edge of
our spectrum.  While there are some possible features present, we are not
able to confirm this system.

$z_{abs}=$1.5847-  A C IV doublet is detected
at this redshift.   We detect O I $\lambda$1302 at 3368 $\AA$, C IV $\lambda$1548
at 4001 $\AA$, and an apparent absorption feature, but no significant line, at
the position of C IV $\lambda$1550.

$z_{abs}=$1.7177-  \markcite{foltz1986} Foltz et al. (1986) find a C IV 
doublet and Al II $\lambda$1670 at this redshift.  We confirm this system
with our detections of Ly$\alpha$ at 3304 $\AA$, Si III $\lambda$1206 at
3279 $\AA$, and O I $\lambda$1302 at 3539 $\AA$.

$z_{abs}=$1.7590-  \markcite{foltz1986} Foltz et al. (1986) detect a C IV
doublet at this redshift.  We detect a  Ly$\alpha$ line consistent with this
system at 3355 $\AA$.

\markcite{c1989} Caulet (1989) detects C IV at four redshifts including
$z_{abs}=$1.7171 and $z_{abs}=$1.4621 (see above).  The other systems
detected are $z_{abs}=$1.7010 and $z_{abs}=$1.7755 for which we detect
no Ly$\alpha$ absorption.

Lastly, \markcite{lwt1995} Lanzetta et al. (1995) report a possible 
Lyman limit absorber in their IUE spectrum at $z_{LLS}=$1.4798.  We find
possible absorption features at the positions of O I $\lambda$1302,
Si II $\lambda$1304, and  Si II $\lambda$1526 for this redshift.  
These features are not 
identified as 3$\sigma$ lines by FINDSL, however.   We do not
detect C IV, Si IV, or C II.

The absorption features at 3967 $\AA$, 3972 $\AA$, and 3980 $\AA$ are
traps in the CCD.

\subsection{Q 1422+231 \hspace{0.5in} $z_{em}=$3.623}

This object is a gravitationally lensed quasar (Bechtold \& Yee 1995, 
hereafter BY95.) Therefore, due to uncertainties in the amplification by the
lensing, it will only be used for the analysis of the Ly$\alpha$ forest 
statistics and not in the proximity effect analysis in Paper II.

Bechtold \& Yee (1995) obtained a spectrum of this object from  4818 $\AA$
to 5684 $\AA$
with 1.8 $\AA$ resolution using the Subarcsecond Imaging Spectrograph on the
Canada-France-Hawaii Telescope.  A red spectrum from 6246 $\AA$
to 7179 $\AA$ with 2.0 $\AA$ resolution was also obtained in order to
identify metal line systems using the 
Red Channel Spectrograph on the Multiple Mirror Telescope.  The systems
identified by these authors are as follows:

$z_{abs}=$3.091-  This system is identified by a strong C IV doublet. 
We detect a marginally consistent doublet at 6323 $\AA$ and 6331 $\AA$
in our red spectrum (see Paper II).
BY95 also find Ly$\alpha$, Si II $\lambda$1193, N I $\lambda$1200, Si II 
$\lambda$1260, and O I $\lambda$1302.  We confirm the Ly$\alpha$ feature
at 4973 $\AA$ and find features at 4882 $\AA$, 4907 $\AA$, 5157 $\AA$,  
and 5328 $\AA$, in marginal agreement with the other lines found by these
authors.  No Si II $\lambda$1190 is detected by us or BY95.  The O I 
$\lambda$1302 line, if present, is blended with Ly$\alpha$ at $z_{abs}=$3.3830.

$z_{abs}=$3.382-  This system is also based upon a strong C IV doublet
seen in the red spectrum of BY95.  These authors also identify Ly$\alpha$ and
Si II $\lambda$1260 blended with a double-component Ly$\alpha$ line at 5519
$\AA$.  We confirm their C IV doublet from our red spectrum; and in our
Ly$\alpha$ forest spectrum, we 
detect a strong Ly$\alpha$ line consistent with this redshift 
at 5328 $\AA$, but do not confirm a Si II $\lambda$1260 line corresponding to
the one found by BY95.

$z_{abs}=$3.515-  This system is based upon a weak C IV doublet for which
BY95 also identify Ly$\alpha$, Si II $\lambda$1190 and $\lambda$1193,
and Si III $\lambda$1206.  We confirm the Ly$\alpha$ line at 5489 $\AA$; we
find N I $\lambda$1200 at 5418 $\AA$; but we do not find features
corresponding to the Si II and Si III lines above.  We do detect weak features 
at the correct position of C IV for this system in our red spectrum.

$z_{abs}=$3.536,3.538-  These systems are identified by strong C IV doublets
by BY95.  We confirm these in our red spectrum,  but the two components 
are not resolved.
These authors also identify Ly$\alpha$ and Si III $\lambda$1206
for both components.   We confirm these features, Ly$\alpha$ at 5513 $\AA$
and 5517 $\AA$, and Si III at 5471 $\AA$ and 5475 $\AA$; and we make an
additional identification of N I $\lambda$1200 at 5445 $\AA$.  
Songaila \& Cowie (1996) identify a strong redshift system at $z_{abs}=$3.5353
in their high resolution ($\sim$0.15 $\AA$) spectrum taken with HIRES on the
Keck Telescope.  They are able to derive column densities for several
species, including C II, C IV, S II, Si III, Si IV, and N V (upper limit).

$z_{abs}=$3.587-  BY95 find a weak C IV doublet at this redshift, and
we confirm this detection in our red spectrum.
They also identify Ly$\alpha$ and Si II $\lambda$1193.  We confirm 
these features
at 5577 $\AA$ and 5475 $\AA$ and make the additional identifications of
N II $\lambda$1083 at 4973 (blended with Ly$\alpha$ at $z_{abs}=$3.0906), 
Si III $\lambda$1206 at 5534 $\AA$ and N I $\lambda$1200 at 5504 $\AA$.
Songaila \& Cowie (1996) find a strong system at $z_{abs}=$3.5862
and derive column densities or upper limits for C II, C III, C IV, Si II,
Si III, Si IV, and N V.

$z_{abs}=$3.624-  BY95 find another weak C IV doublet at this redshift,
along with Ly$\alpha$, Si II $\lambda$1190 and $\lambda$1193, and Si III
$\lambda$1206.  We detect the weak C IV absorption in our red spectrum.  
In our  Ly$\alpha$ forest spectrum, we
confirm Ly$\alpha$ at 5621 $\AA$ and the Si II lines
at 5504 $\AA$ and 5578 $\AA$; but we find no Si III line.

Songaila and Cowie identify a third strong redshift system from their
data at $z_{abs}=$3.4464 for which they derive column densities for
C IV, Si III, and Si IV and upper limits on the column densities for
C II, C III, and Si II.  We detect a strong C IV doublet in our red spectrum;
but in the Ly$\alpha$ forest, we identify only a strong Ly$\alpha$ line at 5407 
$\AA$ corresponding to this system.  These authors also identify a  
partial Lyman limit system at $z_{abs}=$3.3809 showing C IV, Si IV. and C II.
Our spectrum shows Ly$\alpha$ at 5324 $\AA$ and a possible
Si II $\lambda$1260 line at 5522 $\AA$.

Lastly, we make two more metal line system identifications based upon
strong Ly$\alpha$ absorption in our spectrum:

$z_{abs}=$3.3460-  This system is composed of Ly$\alpha$ at 5283 $\AA$,
possible Si II $\lambda$1190 and $\lambda$1193 lines at 5174 $\AA$ and 5187
$\AA$, Si III $\lambda$1206 at 5244 $\AA$, and Si II $\lambda$1260 at
5477 $\AA$.  

$z_{abs}=$3.4945-  At this redshift, we identify Ly$\alpha$ at 5464 $\AA$,
Fe II $\lambda$1143 and $\lambda$1145 at 5137 $\AA$ and  5146 $\AA$,
possible Si II $\lambda$1190 and $\lambda$1193 lines at 5348 $\AA$ and
5363 $\AA$, and N I $\lambda$1200 at 5392 $\AA$.

\subsection{Q 1435+638 \hspace{0.5in} $z_{em}=$2.066}

The absorption line spectrum of this QSO has been studied by several
authors.  \markcite{sbs1988} SBS88 (cf. Section~\ref{sec-q0150}) report
four C IV systems:

$z_{abs}=$1.4590-  \markcite{sbs1988} SBS88 find a weak, possible C IV 
doublet at this redshift.  We confirm this identification, but note that
these lines are more likely O I $\lambda$1302 and Si II $\lambda$1304
at $z_{abs}=$1.9233.  We find no other lines at this redshift.

$z_{abs}=$1.4792-  \markcite{sbs1988} SBS88 find a second weak, possible
C IV doublet at this redshift.  Our spectrum shows only the $\lambda$1548
component at 3837 $\AA$.  There is a weak absorption feature at the 
position of the $\lambda$1550 component, but no significant line is 
identified.  No other lines are found at this redshift.

$z_{abs}=$1.5925-  \markcite{sbs1988} SBS88 find a probable C IV doublet
at this redshift.  We identify O I $\lambda$1302 at 3376 $\AA$ and find
a possible, weak absorption feature (but no 3$\sigma$ line) at the 
position of C II $\lambda$1334.

$z_{abs}=$1.9235-  \markcite{sbs1988} SBS88 regard this C IV doublet as
certain.  They also find C II $\lambda$1334 and a possible Si IV 
$\lambda$1393 line.  We detect a strong Ly$\alpha$ line for this redshift
at 3554 $\AA$, Si II $\lambda$1260 at 3685 $\AA$, O I $\lambda$1302
at 3808 $\AA$, Si II $\lambda$1304 at 3813 $\AA$, C II $\lambda$1334 at
3901 $\AA$.  In addition, \markcite{ss1992} SS92 find a strong Mg II doublet
at this redshift.

\markcite{lwt1995} Lanzetta et al. (1995) report no damped Lyman alpha
candidates in their ultraviolet spectrum.
The absorption features at 3824 $\AA$ and 3829 $\AA$ are traps in the CCD.

\subsection{Q 1604+290 \hspace{0.5in} $z_{em}=$1.962}

A literature search yielded no previously published absorption line
spectrum of this QSO.  Our spectrum shows no significant absorption
lines.  However, the signal-to-
noise of the data blueward of Lyman alpha  is poor ($\leq$2 over the range
3200-3500 $\AA$) and the spectrum is truncated blueward of 3493 $\AA$. 
The apparent absorption features redward of Ly$\alpha$ emission are
identified as traps in the CCD.

\subsection{Q 1715+535 \hspace{0.5in} $z_{em}=$1.932}

The Lyman alpha forest spectrum of this QSO has been studied by several 
authors.  \markcite{sbs1988} SBS88 (cf. Section~\ref{sec-q0150}) find
three systems from their 3750-4930 $\AA$ spectrum:

$z_{abs}=$0.3673-  \markcite{sbs1988} SBS88 identify a Mg II doublet
at this redshift.  The $\lambda$2796 component falls on a series
of traps in the CCD at 3824 $\AA$
in our spectrum; and we do not detect the $\lambda$2803 component.
Mg I $\lambda$2853 coincides with a feature at 3902 $\AA$; but 
we find no Fe II lines to corroborate this Mg II system, which is therefore
still regarded as uncertain.
\markcite{nm1992} Nelson \& Malkan (1992)  find no candidates for this
system in their photometric search for [O II] emission from 
Mg II absorption systems.
They do note a galaxy at a redshift of 0.449, but we detect no Fe II
at this redshift.

$z_{abs}=$1.6330-  \markcite{sbs1988} SBS88 detect a C IV doublet and Si
II $\lambda$1526 at this redshift.  Our spectrum shows C II $\lambda$1334 
at 3512 $\AA$, and a Si IV doublet at 3669 $\AA$ an 3692 $\AA$.  The IUE
spectrum of \markcite{lts1993} Lanzetta et al. (1993) appears to show
an absorption feature at $\sim$3200 $\AA$, which would coincide with 
Ly$\alpha$.

$z_{abs}=$1.7587-  \markcite{sbs1988} SBS88 report a C IV doublet
at this redshift as well.  We confirm this system with our detections of
Ly$\alpha$ at 3354 $\AA$ and a possible Si II $\lambda$1260 line at
3476 $\AA$.  We find possible weak absorption features at the positions of the
Si IV doublet.

\markcite{sbs1988} SBS88 also report a possible Galactic Ca II 
$\lambda$3935 line.  We confirm
the detection of this line at 3934 $\AA$.
\markcite{ss1992} SS92 (cf. Section~\ref{sec-q0027}) detect no lines in
their 5950-8040 $\AA$ and 5130-8950 $\AA$ spectra of this object.

In addition to the absorption line systems discussed above, we find
four other systems from our spectrum:

$z_{abs}=$1.3412-  At this redshift, we detect a  C IV doublet at 3635 $\AA$
and 3631 $\AA$ and Al II $\lambda$1670 at 3911 $\AA$.  We note the presence
of a weak absorption feature (but no 3$\sigma$ line) at the expected position
of Si II $\lambda$1526.  Although we find only three lines for this system,
the IUE spectrum of \markcite{lts1993} Lanzetta et al. (1993) appears to
show a feature at $\sim$2850 $\AA$ which could be identified with Ly$\alpha$
at this redshift.

$z_{abs}=$1.4711-  This system consists of a possible C IV $\lambda$1548
line at 3826 $\AA$ (no $\lambda$1550 absorption is detected), Si II 
$\lambda$1526 at 3772 $\AA$, and C II $\lambda$1334 at 3297 $\AA$.  We
find weak features, but no significant lines at the positions of O I 
$\lambda$1302 and Si II $\lambda$1304.  Only three line detections for
this system are regarded as acceptable as well given a possible absorption
line in the IUE spectrum of \markcite{lts1993} Lanzetta et al. (1993)
at $\sim$3005 $\AA$ which can be regarded as Ly$\alpha$ at this redshift.

$z_{abs}=$1.8746-  For this system, we find Ly$\alpha$ at 3494 $\AA$, possible 
Si II $\lambda$1193 and $\lambda$1260 lines at 3431 $\AA$ and 3625 $\AA$,
N I $\lambda$1200 at 3449 $\AA$, and possible Si III $\lambda$1206 absorption
at 3467 $\AA$.  \markcite{ss1992} SS92 find no Mg II $\lambda$2796 at 
this redshift.

$z_{abs}=$1.8963-  At this redshift, we detect Ly$\alpha$ at 3521 $\AA$,
N I $\lambda$1200 at 3476 $\AA$, a blended Si III $\lambda$1206 line 
at 3494 $\AA$, and O I $\lambda$1302 at 3772 $\AA$.

The apparent absorption features at 3818 $\AA$, 3821 $\AA$, 3824 $\AA$, 
and 3826 $\AA$ are identified as traps in the CCD.

\subsection{Q 2134+004 \hspace{0.5in} $z_{em}=$1.941}

We find no previously published spectrum for this object.  Our data
show 19 absorption lines.  Line \#17 is tentatively identified at
Mg II $\lambda$2796 at $z_{abs}=$0.3654.  The $\lambda$2803 component
of the doublet as well as Mg I 2853 coincide with absorption features
which are not identified as 3$\sigma$ lines by FINDSL.  No Fe II
lines are found at this redshift.

\subsection{Q 2251+244 \hspace{0.5in} $z_{em}=$2.359}

\markcite{c1976} Carswell et al. (1976) report one metal line system 
in their spectrum (3250-5200 $\AA$) of this object.  This system is
an associated absorber at $z_{abs}=$2.3638; and these authors identify
Ly$\alpha$, Ly$\beta$, C III $\lambda$977, O VI $\lambda$1031, N I
$\lambda$1200, N V $\lambda$1238, Si IV $\lambda$1393, and C IV $\lambda$1548.
This system is confirmed by \markcite{btt90} Barthel et al. (1990)
from their 5.0 $\AA$ resolution spectrum over the range 3870-7730 $\AA$ 
who detect Ly$\alpha$, N V $\lambda$1238, Si IV $\lambda$1393, and C IV
$\lambda$1548 as well.  The N V, Si IV, and C IV doublets are also
confirmed by \markcite{abe1994} Aldcroft et al. (1994).  We confirm
this system as well with our identifications of Ly$\alpha$ at 4088 $\AA$,
Ly$\beta$ at 3450 $\AA$ and O VI $\lambda$1031 and a blended $\lambda$1037
line at 3470 $\AA$ and 3490 $\AA$ respectively.  In addition, our
red spectrum of this object (see Paper II) shows the N V doublet at
4167 $\AA$ and 4181 $\AA$, the Si IV doublet at 4688 $\AA$ and 4718 $\AA$,
and the C IV doublet at 5205 $\AA$ and 5214 $\AA$.  The region of the 
spectrum blueward of $\sim$3290 $\AA$ has been removed from our analysis
due to low signal-to-noise ($\lesssim$2.) 

\markcite{btt90} Barthel et al. (1990) report four other systems:

$z_{abs}=$1.7495-  At this redshift, the authors detect a C IV doublet.
We detect only a possible N V $\lambda$1242 line at 3416 $\AA$.  Since
no lines are detected at shorter wavelengths in our spectrum, no $\lambda$1238
component is identified.  \markcite{abe1994} Aldcroft et al. (1994)
confirm this system.  Our red spectrum does show a possible C IV doublet 
associated with this system at 4257 $\AA$ and 4264 $\AA$.

$z_{abs}=$1.0901-  For this system, the authors identify Fe II $\lambda$2382,
a Mg II doublet and Mg I $\lambda$2852.  Our spectrum shows only a possible
Al II $\lambda$1670 line at 3490 $\AA$.  The only other lines that fall within
the range of our line list are Al III $\lambda$1854 and $\lambda$1862, but 
these are not found.  \markcite{abe1994} Aldcroft et al. (1994) confirm
this system.  Our red spectrum confirms the Mg II doublet identification
made by \markcite{btt90} Barthel et al. (1990) (5842 $\AA$ and 5857 $\AA$) but
not the Fe II or the Mg I identifications.

$z_{abs}=$2.1554-  The authors find C II $\lambda$1334 and a C IV doublet
at this redshift.  \markcite{abe1994} Aldcroft et al. (1994) confirm this
and also detect Si IV $\lambda$1393.  In our spectrum, we identify Ly$\alpha$
for this system at 3835 $\AA$, Si II $\lambda$1193 at 3765 $\AA$ (the position
of $\lambda$1190 falls on a bad column in  the data), N I $\lambda$1200
at 3786 $\AA$, a possible Si III $\lambda$1206 line at 3807 $\AA$, and Si II
$\lambda$1260 at 3976 $\AA$.  In addition, our red spectrum shows C II 
$\lambda$1334 at 4122 $\AA$, Si II $\lambda$1526 at 4816 $\AA$,
the C IV doublet at 4885 $\AA$  and 4893 $\AA$, Al II $\lambda$1670 at
5272 $\AA$, and a possible Al III $\lambda$1854 line at 5272 $\AA$.  (No
Al III $\lambda$1862 line is found.)

$z_{abs}=$2.3524-  \markcite{btt90} Barthel et al. (1990) identify C IV
and N V doublets at this redshift which are confirmed by \markcite{abe1994}
Aldcroft et al. (1994).  In our spectrum, we identify Ly$\alpha$ at 4074 $\AA$,
Ly$\beta$ at 3438 $\AA$, O VI $\lambda$1031 and $\lambda$1037 at 3459 $\AA$
and 3478 $\AA$, and a possible Fe II $\lambda$1145 line at 3838 $\AA$.
We also confirm the N V and C IV doublet found by \markcite{btt90} 
Barthel et al. (1990) from our red spectrum, though the N V doublet we
identify has a doublet ratio less than one.  In addition, our red spectrum
shows O I $\lambda$1302 at 4367 $\AA$ and C II $\lambda$1334 at 4475 $\AA$.

We also identify a number of other systems from our data:

$z_{abs}=$1.8993- This system is composed of Ly$\alpha$ at 3525 $\AA$,
possible Si II $\lambda$1190 and $\lambda$1193 absorption at 3450 $\AA$ and
3459 $\AA$, N I $\lambda$1200 at 3478 $\AA$, O I $\lambda$1302 at 3775 $\AA$,
and Si II $\lambda$1260 3653 $\AA$.  In addition, our red spectrum shows
Si II $\lambda$1526 at 4427 $\AA$.

$z_{abs}=$2.0336-  At this redshift, we identify Ly$\alpha$ at 3688 $\AA$,
possible, blended Si II $\lambda$1193 and $\lambda$1260 lines at 3620 $\AA$ and 
3824 $\AA$, Si III $\lambda$1206 at 3660 $\AA$, and O I $\lambda$1302
at 3949 $\AA$.  Our red spectrum shows Si II $\lambda$1526 at 4631 $\AA$,
a possible, blended Fe II $\lambda$1608 line at 4879 $\AA$,  and 
Al II $\lambda$1670 at 5068 $\AA$.

$z_{abs}=$2.0570-  For this system, we find  Ly$\alpha$ at 3716 $\AA$,
N I $\lambda$1135 (blended) and $\lambda$1200 at 3470 $\AA$ and 3668 $\AA$,
a blended Si III $\lambda$1206 line at 3688 $\AA$, and the N V doublet at 
3786 $\AA$ and 3799 $\AA$.  Also, our red spectrum shows the Si IV doublet
at 4262 $\AA$ and 4289 $\AA$ as well as Fe II $\lambda$1608 at 4915 $\AA$.

$z_{abs}=$2.1052-  This system consists of Ly$\alpha$ at 3775 $\AA$, 
Si III $\lambda$1206 at 3746 $\AA$, and Si II $\lambda$1260 and $\lambda$1304
at 3913 $\AA$ and 4050 $\AA$.  Our red spectrum shows a possible,
blended Al II $\lambda$1670 line at 5188 $\AA$.

$z_{abs}=$2.3158-  This system is composed of Ly$\alpha$ at 4031 $\AA$, a
possible O VI $\lambda$1037 line at 3440 $\AA$ ($\lambda$1031 is outside
the range of the line list), N I $\lambda$1200 at 3979 $\AA$, and Si III 
$\lambda$1206 at 4001 $\AA$.  Our red spectrum extends slightly blueward
of the higher resolution Ly$\alpha$ forest spectrum and shows some evidence for
Ly$\beta$ at 3401 $\AA$ and O VI $\lambda$1031 at 3422 $\AA$, as well as
Si II $\lambda$1260, $\lambda$1304, and $\lambda$1526 at 4181 $\AA$, 
4327 $\AA$, and 5065 A$\AA$, O I $\lambda$1302 at 4319 $\AA$, and C II
$\lambda$1334 at 4427 $\AA$.

\subsection{Q 2254+024 \hspace{0.5in} $z_{em}=$2.090}

The radio properties and the UV emission lines of this object have been
widely studied. \markcite{ss1992} SS92 (cf. Section~\ref{sec-q0027})
find no absorption lines in their red spectra (5128-8947 $\AA$).  Due
to the poor signal-to-noise ($\leq$2) of the blue region of our spectrum,
only the portion redward of 3450 $\AA$ was used for our line list.  
We find 25 absorption lines
but no metal line systems according to our criteria.  Only two possible
identifications are made:   a C IV doublet at $z_{abs}=$1.4751 for which
the $\lambda$1550 component must actually blended with the feature at
3837 $\AA$; and a Si IV doublet at $z_{abs}=$1.7323 for which the 
corresponding Ly$\alpha$ line
falls in the low signal-to-noise region of the data and is not seen.
However, our red spectrum (see Paper II) does lend some confirmation 
to the possible 
$z_{abs}=$1.7323  system as it shows this Si IV doublet as well as 
Si II $\lambda$1526 at 4171 $\AA$, a C IV doublet at 4233 $\AA$ and 4238 $\AA$,
and Al II $\lambda$1670 at 4564 $\AA$.  No lines redward of Ly$\alpha$ are
confirmed for the $z_{abs}=$1.4751 system.

\subsection{Q 2310+385 \hspace{0.5in} $z_{em}=$2.181}

No previously published spectrum of this QSO was found.
Due to poor signal-to-noise blueward of 3571 $\AA$, only the portion of the
spectrum redward of this wavelength was used for the purposes of our line list.  
Fifteen significant 
absorption lines were found, but none of these could be identified with any 
heavy element absorption systems.  Three identifications of doublets
redward of Ly$\alpha$ emission
could be made:  C IV doublets at $z_{abs}=$1.4998 and 
$z_{abs}=$1.5036; and a Mg II doublet at $z_{abs}=$0.3840.

\subsection{Q 2320+079 \hspace{0.5in} $z_{em}=$2.088}

We found no previously published spectrum of this object.
We find a double component damped Ly$\alpha$ complex in our spectrum 
at 3712 $\AA$ and 3715 $\AA$.  Each of these components shows
Si II $\lambda$1193 (3645 $\AA$) and Si III $\lambda$1206 (3685 $\AA$ and
3687 $\AA$.)  Si II $\lambda$1190 is present, but not identified as a 
3$\sigma$ line by FINDSL.  The feature at 3553 $\AA$ is most likely
a cosmic ray.

\subsection{Q 2329-020 \hspace{0.5in} $z_{em}=$1.896}

No previously published spectra of this QSO were found.
We find 17 significant absorption lines in our spectrum.  We make a number
of identifications of doublets redward of Ly$\alpha$ emission.
Two C IV doublets are seen at $z_{abs}=$1.2902
and $z_{abs}=$1.2922, a separation of $\sim$260 km s$^{-1}$.  
The 
second doublet also appears to have weak features of Si II $\lambda$1526
and Al II $\lambda$1670 associated with it.  A strong C IV doublet is also 
detected 
at $z_{abs}=$1.3339 along with weak features at the positions of Si II
$\lambda$1526 and Al II $\lambda$1670.  This QSO also
shows associated Ly$\alpha$ absorption at 3509 $\AA$, 3513 $\AA$,
3521 $\AA$, and 3531 $\AA$, but no metals lines are found at this redshifts.

\subsection{Data from the Literature} 

\begin{figure}
\epsscale{1.00}
\plotone{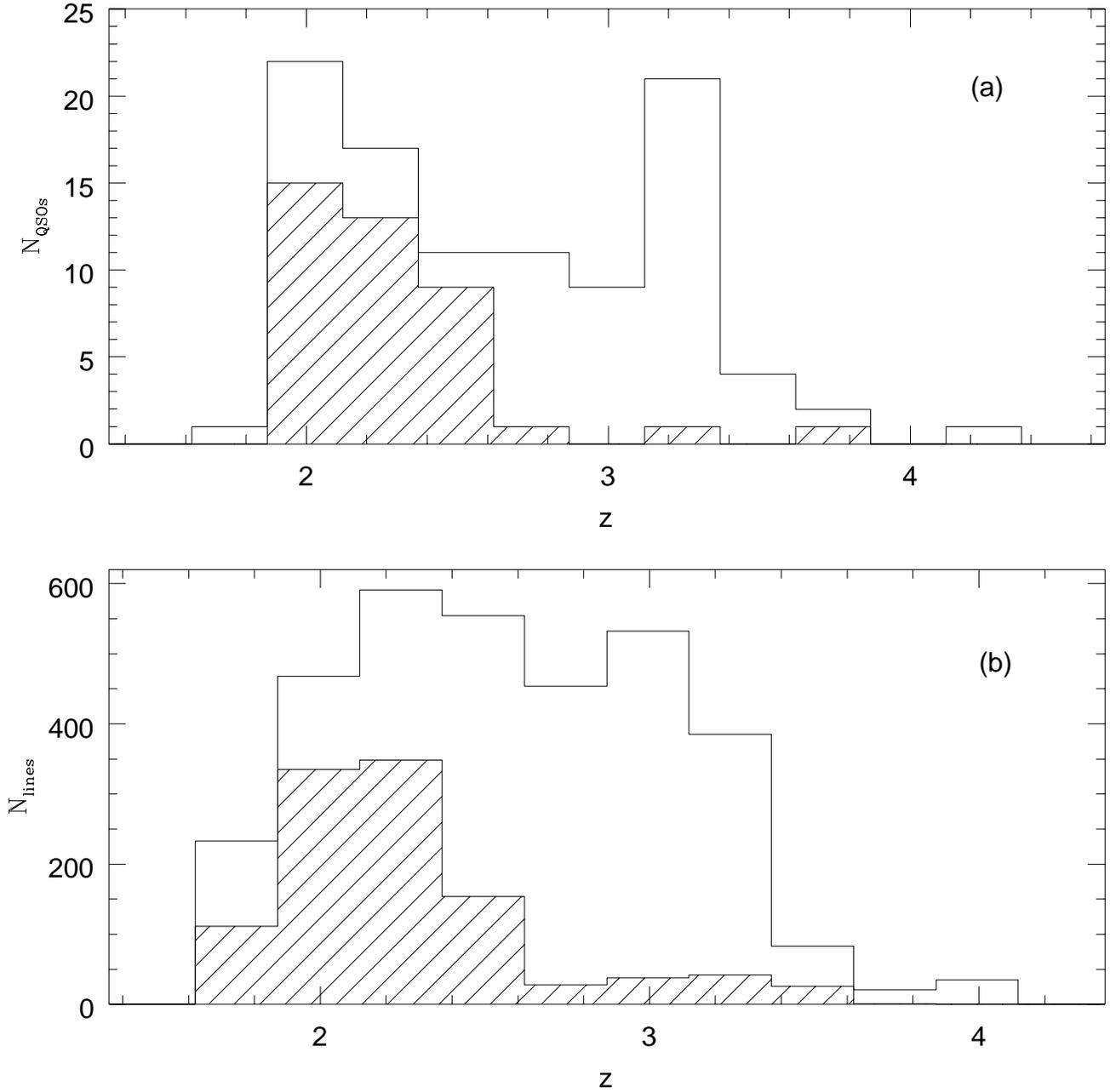}
\caption{
(a) Histogram of 99 QSO redshifts, includes QSOs presented in
this paper (shaded region) and objects
from the literature; (b) Histogram of 3356 absorption line redshifts 
from QSOs presented in this paper (shaded region) and objects from
the literature, using 
a variable equivalent width threshold, includes all lines between each QSO's
Ly$\beta$ and Ly$\alpha$ emission lines \label{fig:zhist} }
\end{figure}

Spectra that met three basic criteria were gathered from the literature. In all
cases, the
errors were published, the resolution was equal to or better than 200 km 
s$^{-1}$, and no broad absorption line features were present, which would 
indicate the
presence of material intrinsic to the QSO (see Table 5 of B94).
Table~\ref{table-lit} is a 
list of the objects 
chosen to supplement the sample and the reference for each.  
Figure~\ref{fig:zhist} shows histograms 
of the distribution of QSO redshifts and absorption line redshifts 
for the total sample. 

The line list for the QSO 1603+383 was provided by 
Dobrzycki, Engels, \& Hagen  (1999)
prior to publication.  This object has a B magnitude of
$15.9$.

\section{Results and Discussion}
\label{sec-results}

The number of Ly$\alpha$ lines per unit redshift per unit equivalent
width can be parametrized as follows:
\begin{equation}
\frac{\partial^{2} {\cal N}}{\partial z \partial W}= \frac{A_{0}}{W^{*}}(1+z)
^{\gamma}\exp\left(-\frac{W}{W_*}\right).
\end{equation}
Integrating this equation over equivalent width with a constant 
threshold equivalent width throughout each spectrum gives
\begin{equation}
\frac{d{\cal N}}{dz}= {\cal A}_{0}(1+z)^{\gamma} \label{eq:dndz} . 
\end{equation}
To solve for the parameters $\gamma$ and $W^{*}$, we use a maximum 
likelihood technique which allows for an equivalent width threshold
that varies with wavelength.  We also derive these parameters using
various fixed threshold values; and in this case, the procedure reduces
to the method described in the Appendix of MHPB,
using corrected expressions for their equations (A8) and (A2a).  
However, the variable
threshold information is still used in the fixed threshold case,
as regions of the spectrum for which the threshold lies above the 
fixed value, ie. where not all significant lines could be detected even if
they were present, are excluded.

The solutions for the statistics $\gamma$ and $W^{*}$ are listed in 
Table~\ref{table-paramtab}. 
Each sample excludes regions of the spectra within $\Delta$z of 0.15 of the 
QSO emission redshift, chosen to eliminate any effects on the line density 
due to proximity to the QSO.
A variable equivalent width threshold
gives a value of $1.23 \pm 0.16$ for $\gamma$.  This is 
lower than the value of $2.75 \pm 0.29$ found by LWT for
for a fixed equivalent width threshold of 0.36 $\AA$ over the range
$1.7 < z < 3.8$, and the value of 
$1.89 \pm 0.28$ found by B94 for a fixed threshold of 0.32 $\AA$ over
the range $1.6 < z < 4.1$.  Using a fixed threshold of 0.32 $\AA$, the
value of $\gamma$ derived from our data is $1.88 \pm 0.22$, in good
agreement with that of B94. In Table~\ref{table-paramtab},
no error is quoted for ${\cal A}_{0}$ because it is strongly correlated
with the error in $\gamma$.

We calculate the Kolmogorov-Smirnov (KS) probability
that a power law number density
distribution given by
Equ.~\ref{eq:dndz} for each of these values of $\gamma$ is a good
representation of the data (cf. Appendix of MHPB).
A high probability (P$_{KS}$) that the maximum deviation from the 
cumulative number
distribution could occur by chance if the data set is drawn from an assumed
parent distribution indicates that the
choice of parent distribution is justified.
These results are included in Table~\ref{table-paramtab}.
The total sample and each of the subsamples
is described well by a single power law, as illustrated by the
high KS probabilities obtained.  
The KS probability obtained from our data set with a
fixed equivalent width threshold of 0.32 $\AA$ and the LWT $\gamma$ value
of 2.75 is 0.0020, while the B94 value of 1.85 gives 0.97, as it
is in good agreement with our maximum likelihood result. 

The errors in $\gamma$ and $W^{*}$ are calculated 
by our software by
fitting a parabola to the peak of the logarithm of the likelihood function, 
using the fact that the likelihood function itself 
should be distributed as a Gaussian  in
$\gamma$ and $W^{*}$ near its maximum value.   In order to avoid any 
assumptions about the distribution of the statistics of interest, 
a resampling technique was used to independently
estimate the distribution.
Jackknife samples (Babu \& Feigelson 1996, Efron 1982) 
of our original data set were constructed, 100 in all,
each with one QSO from the original sample removed.  We used the same
program
to calculate $\gamma$ and $W^{*}$ for each jackknife sample, for the
case of $W_{thr}$=0.32 $\AA$.   The goal is to understand how the values
of these statistics found by our software vary with random variations 
in the data.
The weighted mean of all the jackknife values for $\gamma$ is 1.91 and
for $W^{*}$ it is 0.309 $\AA$.
Since we cannot treat each of the 100 values of these statistics as 
independent measurements of $\gamma$ and $W^{*}$, 
the jackknife errors show how
well the error calculated by the software estimates the true distribution of
the statistics calculated.  The jackknife results for $\sigma_{\gamma}$
and $\sigma_{W^{*}}$ are 0.26  and 0.011  respectively.
The fact that the jackknife errors are $\sim$20\% larger than the
error calculated by our software may reflect the fact that the jackknife
estimate of the variance tends to be conservative (Efron 1982) or it
may indicate the
the presence of additional sources of random error. 
In any case, the jackknife results do agree with the total data
set result to well within the errors.

The two
questions we now ask are whether the number densities of
strong and weak lines evolve differently
with redshift and whether there is a difference in $\gamma$ for low and high
redshift subsamples, ie. does $\gamma$ evolve over the history of the universe
after the observed break at z $\sim$ 1.7?
In this context, strong lines will refer to lines with rest equivalent
widths greater than 0.32 $\AA$ and weak lines will refer to those with
rest equivalent widths between 0.16 $\AA$ and 0.32 $\AA$.
The total absorption line sample was divided into low and high 
redshift subsamples  at an absorption redshift of 2.5, giving
1084 and 995 lines in each subset, respectively.  
For the remainder of this paper, 
the low and high redshift subsamples will refer to 
Lyman $\alpha$ forest absorption lines with redshifts above and below 
2.5, respectively.

\begin{figure}
\epsscale{1.00}
\plotone{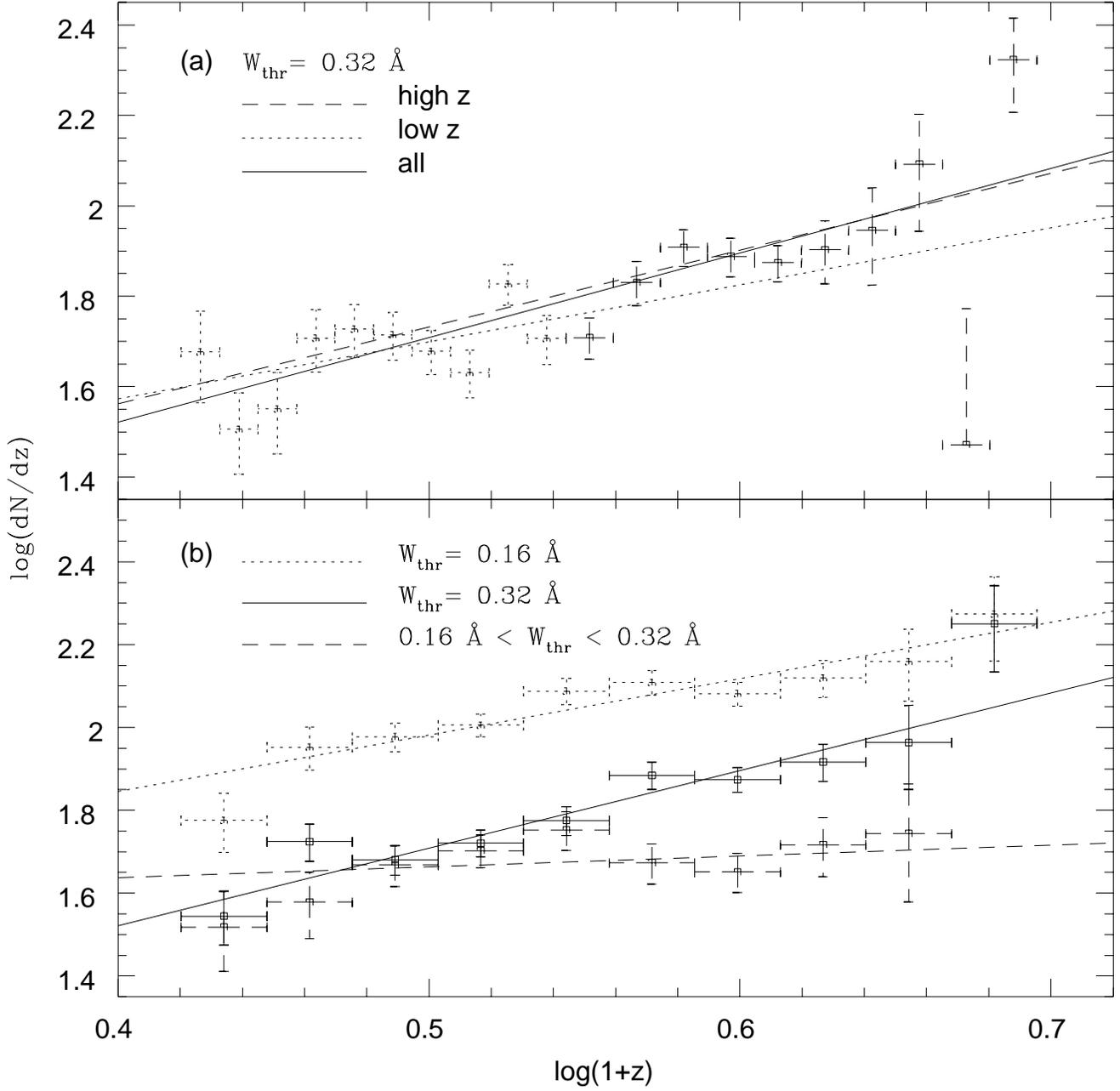}
\caption{
(a) log$(d{\cal N}/dz)$ vs. log$(1+z)$ for z$<$2.5
(dotted line),
z$>$2.5 (dashed line),
and all lines (solid line) each using a fixed threshold of 0.32 $\AA$;
(b) log$(d{\cal N}/dz)$ vs. log$(1+z)$ for different equivalent width
thresholds:
W $>$ 0.16 $\AA$ (dotted line); W $>$ 0.32 $\AA$ (solid line);
0.16 $\AA <$ W $<$ 0.32 $\AA$ (dashed line) \label{fig:dndz} }
\end{figure}

Figure~\ref{fig:dndz} is a set of 
plots of log$(d{\cal N}/dz)$ versus log$(1+z)$ for the
various subsamples of our data set which are binned solely for
display purposes.  The straight lines are derived from the parameters
given in Table~\ref{table-paramtab}. Figure 4a shows the low and high
redshift subsamples and the solutions for each along with the solution
for the total sample.  Each of these are generated 
with a fixed equivalent width 
threshold of 0.32 $\AA$.  Figure 4b shows the results for strong (W $>$
0.32  $\AA$) and
weak lines (0.16 $<$ W $<$ 0.32 $\AA$) considered separately.
Column 8 of Table~\ref{table-paramtab} lists the KS probabilities for each 
case considered.

No log$(d{\cal N}/dz)$ versus log$(1+z)$ plots are shown and no KS
probabilities are quoted for any case in which a variable threshold was
used.  This is because the distribution in redshift is now related to the
equivalent width of each line.   The separation of these two distributions,
which is possible in the case of a constant threshold, is not possible; and
the formalism of MHPB can no longer be applied.  Nevertheless, since the
implementation of a variable threshold allows the most efficient use of the
data, we consider these values of $\gamma$ to be reliable, especially in light
of the reasonable KS probabilities in the constant threshold cases.

Considering the moderate resolution and signal-to-noise of our data, it
is worth investigating how well we are recovering the true parameters
describing the line distribution.  Recall from the discussion of the
simulations in Section~\ref{sec-absdata}
that our 5$\sigma$ line lists are 55\% complete due to blending. 
To address this point, we generated more sets of artificial spectra
based on the 56 objects in our data set for which we have detailed spectral
information
in the way described in Section~\ref{sec-lineid}.   
The redshift of each QSO in these sets is equal to that 
of one of the 39 new MMT spectra presented in this paper 
or to that 
of one of the 17 spectra presented in Dobrzycki and Bechtold (1996). 
In order to investigate how signal-to-noise impacts this analysis,
we created
three sets of these 56 artificial spectra with the resolution of the data,
$\sim$1 $\AA$, one set having
signal-to-noise ratios half that of the data (median S/N $\sim$ 5),
another having signal-to-noise ratios equal to that of the data
(median S/N $\sim$ 10), and
another having twice
the signal-to-noise of the data (median S/N $\sim$ 20).  
The input parameters used were
$\gamma$=1.88, $\beta$=1.46, N$_{lower}$=1x10$^{13}$ cm$^{-2}$,
N$_{upper}$=1x10$^{16}$ cm$^{-2}$, $<$b$>$=28 km s$^{-1}$, and $\sigma_{b}$=
10 km s$^{-1}$.

In the low S/N simulation, FINDSL spuriously identified one
simulated line, out of 1722 lines above threshold, as two separate lines,
both of 5$\sigma$ significance or greater.  This did not occur in either
the data S/N simulation, or in the high S/N simulation,
so we remain confident that the Lyman $\alpha$ lines in our line lists
are real absorption features.

We also generated set of synthetic spectra with higher resolution than the data.
Two sets were made with resolution $\Delta\lambda\sim$ 0.7 $\AA$, one
with the same signal-to-noise as the data, and 
another with median S/N $\sim$ 20.
Finally, a Keck/HIRES data set was simulated by generating spectra
with $\Delta\lambda\sim$ 0.2 $\AA$ and median S/N $\sim$ 40.

The simulation line lists
were analyzed in the same way as the data
to determine the value of
$\gamma$ input into the FINDSL analysis.  
This $\gamma$ is not necessarily equal to the simulation input
$\gamma$, 1.88, because, in generating the artificial spectra, the
simulation software does not fix the redshift and equivalent width distributions
by the input parameters, but rather draws line redshifts and equivalent widths
from a distribution given by Equation~\ref{eq:dndz}.
FINDSL line lists were then generated and
$\gamma$ was calculated again using these line
lists.
This was done for both the variable threshold and the case of an
equivalent width threshold of 0.32 $\AA$ for all redshifts, and at 
high and low redshifts separately.  
The two values of $\gamma$ for each case are compared with each other
in order to determine how well the redshift distribution in the FINDSL 
line lists reflects the distribution output by the simulations.
The results are listed in Table~\ref{table-sim}.
The simulation resolution and median signal-to-noise ratio
are given in the first two columns; the redshift range and the threshold
used for the $\gamma$ solution are given in columns (3) and (4);
and the values of $\gamma$ derived from the simulation line lists
($\gamma_{\rm simulation}$) and from the FINDSL line lists 
($\gamma_{\rm FINDSL}$) are given
in columns (5) and (6), respectively.  
DB96 discuss this simulation software in detail and use it 
to investigate the column density distribution of Lyman $\alpha$  
lines.  Their data set, a subset of ours, encompassed a limited redshift
path and was therefore insensitive to a determination of $\gamma$ from the
simulations.
Presumably, if 
we ran the large number of simulations for which this software was  
designed, we would recover $\gamma$=1.88 in column (5) of Table~\ref{table-sim}; 
but since we 
are merely trying to determine the reliability of our methods for identifying
significant lines, we will leave this for future work.
These Monte-Carlo simulations create line lists by distributing lines
according to the input
value of $\gamma$, which is independent of redshift and equivalent width.  It
is for this reason, and because we have created a relatively small number 
of synthetic spectra in order to simulate our data set, that we do not take the
values of $\gamma$ derived either from the simulation line lists or from
the FINDSL line lists to truly reflect the redshift distribution of Lyman
$\alpha$ lines.  We use these simulations 
only to investigate how well our techniques for 
identifying significant lines and calculating 
$\gamma$ recovers the value input into the FINDSL analysis.  

\begin{figure}
\epsscale{1.00}
\plotone{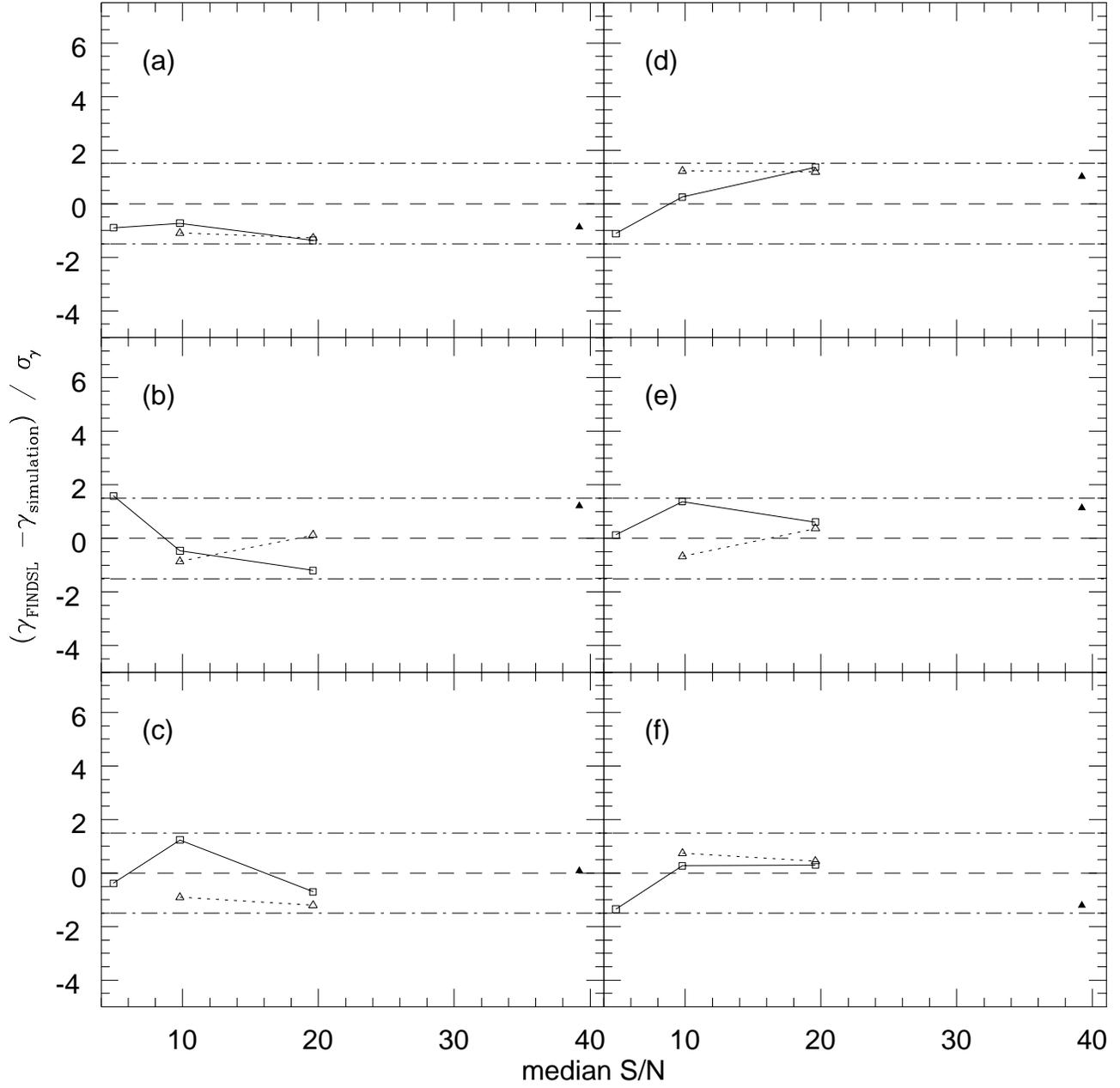}
\caption{
( $\gamma_{\rm FINDSL}$-$\gamma_{\rm simulation}$)/$\sigma_{\gamma}$  
vs. median signal-to-noise,
open squares and solid line- data resolution, $\sim 1 \AA$;
open triangles and dotted line- $\Delta\lambda\sim$ 0.7 $\AA$;
filled triangles- $\Delta\lambda\sim$  0.2 $\AA$:
(a) variable threshold;
(b) variable threshold, z$<$2.5;
(c) variable threshold, z$>$2.5;
(d) W $>$ 0.32 $\AA$;
(e) W $>$ 0.32 $\AA$, z$<$2.5;
(f) W $>$ 0.32 $\AA$, z$>$2.5 \label{fig:gamma} }
\end{figure}

Figure~\ref{fig:gamma} also demonstrates these results.  It shows
the number of sigma difference between the 
output (FINDSL line lists) and input (simulation line lists)
values of $\gamma$, (a)-(c) for the variable
threshold case and (d)-(f) for the W$_{thr}$=0.32 $\AA$ case.  
The square points and solid lines indicate the results for the 
simulations  at the resolution of the data in this paper, $\Delta\lambda\sim$ 1
$\AA$.  The open triangles and dotted lines show the results for
the simulations at higher resolution,  $\Delta\lambda\sim$ 0.7 $\AA$; and
the filled triangle shows the result for the Keck/HIRES simulation,
$\Delta\lambda\sim$ 0.2 $\AA$.

\begin{figure}
\epsscale{1.00}
\plotone{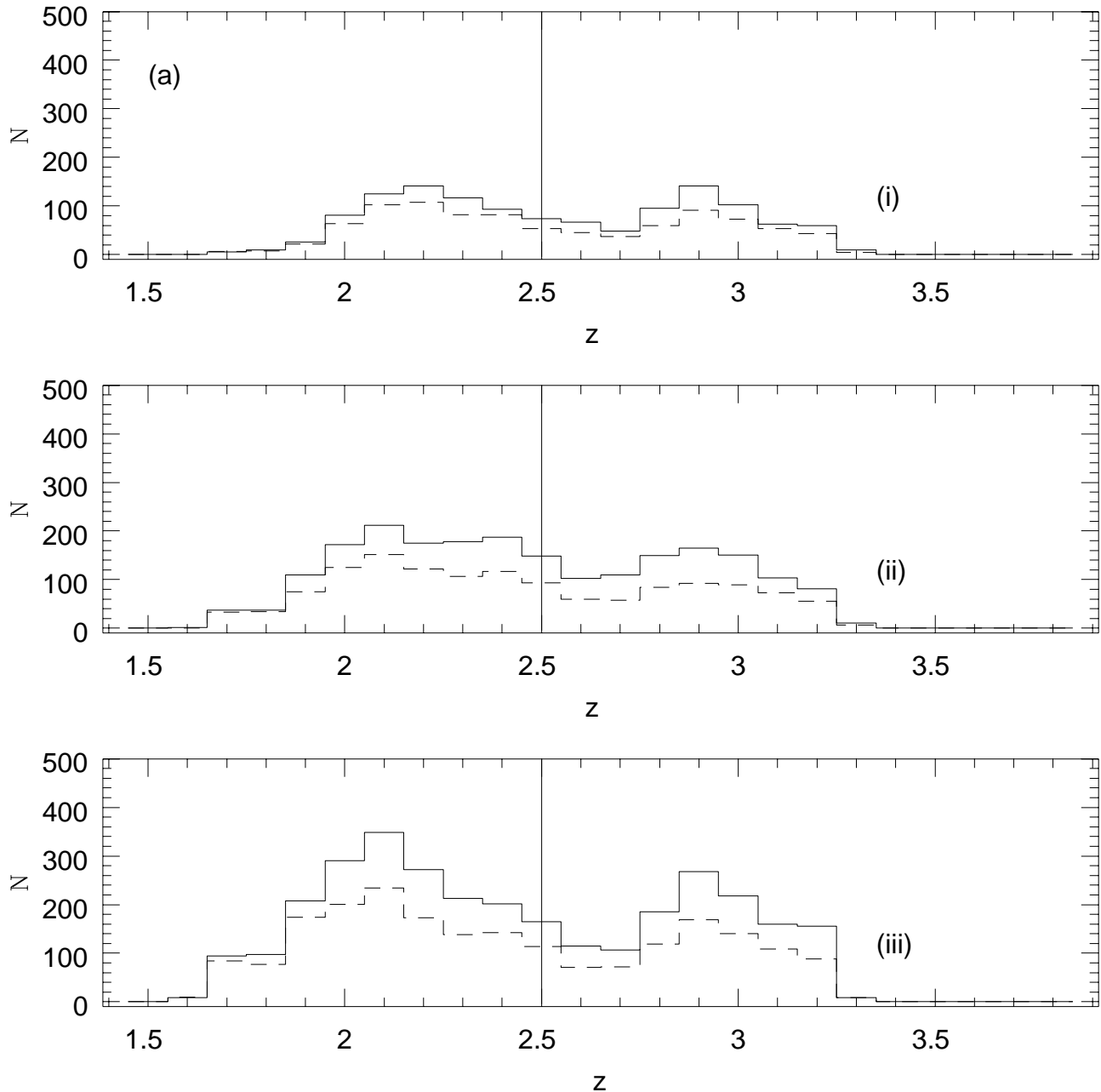}
\caption{
Histograms of the line distribution used to solve 
for the simulation input $\gamma$ (solid line) and FINDSL output 
$\gamma$ (dashed line); vertical line at z=2.5 marks the
division between low z and high z solutions for $\gamma$ in 
Fig.~\ref{fig:gamma}: (a) variable threshold, $\Delta\lambda\sim$ 1 $\AA$- 
(i) median S/N $\sim$ 5, 
(ii) median S/N $\sim$ 10,
(iii) median S/N $\sim$ 20;
(b) variable threshold, $\Delta\lambda\sim$ 0.7 $\AA$- 
(i) median S/N $\sim$ 10,
(ii) median S/N $\sim$ 20,
(iii) $\Delta\lambda\sim$ 0.2 $\AA$, median S/N $\sim$ 40; 
(c) constant threshold, $\Delta\lambda\sim$ 1 $\AA$ -
(i) median S/N $\sim$ 5,
(ii) median S/N $\sim$ 10,
(iii) median S/N $\sim$ 20;
(d) constant threshold, $\Delta\lambda\sim$ 0.7 $\AA$-
(i) median S/N $\sim$ 10,
(ii) median S/N $\sim$ 20,
(iii) $\Delta\lambda\sim$ 0.2 $\AA$, median S/N $\sim$ 40 \label{fig:hist}}
\end{figure}

Histograms of the line distributions used in the  
input (simulation line lists) and output (FINDSL line lists)
$\gamma$ solutions are shown in Figure~\ref{fig:hist}(a-d).
Also, plots of 
log$(d{\cal N}/dz)$ versus log$(1+z)$ analogous to those in 
Figure~\ref{fig:dndz} for the data resolution, data S/N, constant threshold
simulations are shown in Figure~\ref{fig:dndzsim}.  As in
Figures~\ref{fig:hist}(a-d), the solid lines
correspond to the maximum likelihood solution for $\gamma$ and ${\cal A}_{0}$
for the simulation line lists and the dashed lines correspond to the solution
for the FINDSL line lists.  These figures demonstrate that the process of
simulation lines above threshold being $``$blended out" with other features
in the final FINDSL line lists
dominates over lines below threshold being $``$blended in" by blending
with other features below threshold in all cases.  
Overall, therefore, the FINDSL line lists suffer from a net loss of lines due 
to the blending out of significant features.

However, this blending has not significantly affected the value of $\gamma$.  
The only case for which the simulation and subsequent FINDSL solutions for
$\gamma$ differ by more than 1.5$\sigma$, indicated by the dashed-dotted lines in
Figure~\ref{fig:gamma}, is the constant threshold solution for the lowest S/N simulation at
low redshift, the leftmost point in Figure~\ref{fig:gamma}(b).  It should be noted that
some visual inspection of the simulation spectra was necessary to achieve
this overall agreement between the simulation and the FINDSL $\gamma$'s.  
This examination was commensurate with that done on the data, especially during the
course of the metal line identifications, so no significant bias is introduced 
into the simulation analysis by doing this.
The FINDSL program tended to miss some weak lines in the high redshift
spectra due to crowding of features.  Some lines were also missed  by FINDSL 
at low redshift, where the signal-to-noise is lowest.  The equivalent width
thresholds used in the solution for $\gamma$ required that the weakest lines
at low S/N be left out of the simulation line list solution, so missing them
with FINDSL had little effect.  However, in some cases, FINDSL either failed to find lines
above threshold at low S/N or failed to fit them with the proper equivalent width.
These omissions did adversely affect the agreement between the $\gamma$ solutions, 
as these lines were
included in the solution using simulation line lists.
Upon inspection of the simulated spectra, all of these lines were identified and
the simulation and FINDSL line list solutions for $\gamma$ were brought into
agreement.

For the total sample and the high redshift subsample, including weak lines 
in the maximum likelihood solution tends to make $\gamma$ more shallow.  
Both our data and  high resolution work (Cristiani et
al. 1995, Giallongo et al. 1996) indicate that the tendency for $\gamma$ 
to change in either direction 
when weaker lines are included is not a significant one.
Decreasing the column density cutoff from log(N$_{HI}$)=13.8 to 13.3
at z$\sim$3,
Cristiani et al. (1995) find that $\gamma$ increases from 1.86 to 2.17; but
this is a change of less than 1$\sigma$.  Giallongo et al. (1996) find that 
decreasing the column density cutoff from log(N$_{HI}$)=14 to 13.3, again
at z$\sim$3,
decreases $\gamma$ from 2.7 to 2.49, $\sim$1$\sigma$. 
However, using only weak lines for our total sample gives a 
$\gamma$ of
$0.26\pm0.33$, a value consistent with no evolution for $q_{0}= 0.5$;
while using
all lines with rest equivalent width greater than 0.32 $\AA$ gives
a value 4$\sigma$ larger, 1.88.  In the case of the high redshift subsample,
this difference is 2.6$\sigma$.  Weak lines being blended out in the
crowded, high redshift regions of the spectra is undoubtedly contributing
to this effect. 
In our simulations, lines with rest equivalent widths between 0.16 $\AA$  
and 0.32 $\AA$ yield a $\gamma$ of 2.25$\pm$0.40 for the simulation output 
lines, while the FINDSL line lists give a significantly lower value of
1.30$\pm$0.49. The plots of log$(d{\cal N}/dz)$ versus log$(1+z)$ 
analogous to Figure~\ref{fig:dndzsim} for these weak lines 
are shown in Figure~\ref{fig:dndzsimweak}; and this solution for
all redshifts is shown in panel (a). Recall that
the input simulation redshift distribution is independent of the line 
width.  By contrast, the
simulation line list and FINDSL line list values of $\gamma$ for lines 
with equivalent widths greater than 0.32 $\AA$  are 1.62$\pm$0.27 and 
1.70$\pm$0.30, respectively.  
This indicates that though we can be confident that we are recovering
the true $\gamma$ for lines with equivalent widths greater than 0.32 $\AA$,
weak lines blended out at high redshift
in our data may indeed produce this flattening of $\gamma$ seen when
weak lines are included in the solution. 

For the  low redshift subsample, the  weak lines
give a steeper $\gamma$, but this difference is
not statistically significant.
The Weymann et al. (1998) results at z $<$ 1.7 suggest the opposite, that
lines of higher rest equivalent width yield larger values of $\gamma$.
These authors find a difference in the
evolution rates for Ly$\alpha$ absorbers with and without identified
associated metal lines. Their interpretation of this is
that it can be attributed to a difference in the rate of
evolution of lines of different strengths.
This scenario is supported by the higher redshift results of Kim et al. (1997).
Their high resolution data 
suggest that there is a break in the column density distribution of Ly$\alpha$
lines at log(N$_{HI}$) $\geq$ 14.8 and z $\sim$ 3.3
and that this break occurs at lower column
densities and becomes more pronounced as redshift decreases.  These results imply
that weak lines should show a flatter $\gamma$ at all redshifts and that
the difference in the rate of evolution between strong and weak lines
should be more significant at redshifts less than 2.5 than at redshifts 
greater than 2.5.  

The $\gamma$'s 
derived from the simulation and FINDSL line lists
for the spectra generated at the data resolution and signal-to-noise
listed in Table~\ref{table-sim}
are generally in good agreement with one another for strong and weak 
lines at low and high
redshift, noting however, the large uncertainties for the weak 
line $\gamma$'s. 
The FINDSL $\gamma$'s for weak lines for all redshifts and at
low redshift
are systematically lower than the simulation $\gamma$'s, due to 
blending out of weak features preferentially at high redshift.
The high redshift solution does not suffer from this as lines are
evenly blended out at all redshifts greater than 2.5, as 
demonstrated in 
Figure~\ref{fig:dndzsimweak}(a-c).
In any case, this comparison indicates that there is no tendency 
for blending to work
to artificially produce the trend noted above, namely that the $\gamma$ for
weak lines is steeper than the $\gamma$ for strong lines at low redshift,
contrary to the results of other authors.

For a variable threshold at high redshifts, 
$\gamma$ flattens by 1.5$\sigma$ compared to
the value found for low redshift lines, to $0.64 \pm 0.47$ for z $>$ 2.5
from $1.57 \pm 0.42$ at $z < 2.5$.
Again, the difference is not statistically significant; but a trend exists
in that the maximum likelihood
values of $\gamma$  found for the low redshift subsample
are larger than those found for the high redshift
subsample in all cases in which weak lines are included, while for
strong lines, $\gamma$ increases from low to high redshift.
The agreement between the $\gamma$'s derived from the 
simulation line lists and
the FINDSL line lists indicates that, at the resolution and signal-to-noise
of the data, this trend is not artificially imposed by blending. 

Equivalently, one can investigate the distribution in equivalent width
as a function of redshift.  The value of the parameter $W^{*}$ increases
from low to high redshift from 0.282 $\AA$ to 0.330 $\AA$ in the
case of a constant 0.32 $\AA$ threshold, a difference of $\sim$3$\sigma$
in the sense that the distribution is more shallow at high redshift.  Both
of these results imply that there exist more weak lines relative to strong
ones at low redshift than at high redshift.
Given the
discussion above, it is likely that at least some of this difference
can be attributed to increased blending of weak lines at high redshifts.
Nevertheless, the Kim et al. (1997) analysis supports this interpretation,
as do the results of the hydrodynamic simulations of Dav\'{e} et al. 1999.
These authors find that $W^{*}$ does indeed increase with redshift from z=0
to z=3 due to the onset of structure formation.  
The values of $W^{*}$ they derive from
their simulated spectra at high resolution are smaller than
those measured in this paper or at low redshift by Weymann et al. (1998).
They find, however, that the effects of blending in even low redshift, moderate 
resolution spectra, comparable to the FOS data,  
can raise the measured values to those found by Weymann et al. (1998).

This effect is demonstrated by Figure~\ref{fig:ewhist}, a histogram
of the rest equivalent width distribution of lines in the simulation and FINDSL
line lists for the data resolution ($\sim$1 $\AA$) simulations with median
signal-to-noise ratios of 5, 10, and 20 in the variable threshold case.
As expected, the number of lines blended out is largest
at low equivalent width, flattening out the overall distribution and
in turn raising the value of $W^{*}$ derived.

If a fixed equivalent width threshold of 0.32 $\AA$ is used
(rows 9, 11, 15, and 17 in Table~\ref{table-paramtab}), weak lines are thrown
out and the distribution in redshift is flatter at low redshift than at high 
redshift, though not significantly so:
$\gamma=1.30 \pm 0.60$ for z $<$ 2.5, versus $1.69 \pm 0.60$ for z $>$ 2.5,
a difference of less than 0.5$\sigma$.  
Interestingly, Stengler-Larrea et al. (1995) find  
$\gamma= 1.50 \pm 0.39$ for Lyman limit absorbers between 
z=0.32 and z=4.11, in reasonable agreement with our values of $\gamma$ using 
W$_{thr}$=0.32 $\AA$ for both the low and high redshift subsamples. The 
total sample of lines with W $>$ 0.32 $\AA$ gives a somewhat larger 
value of $\gamma$, 1.88 $\pm$ 0.22, but including the low redshift 
data of Bahcall et
al. (1993) yields a value of 1.70 $\pm$ 0.19, consistent with the result for
Lyman limit systems.
It has been proposed that Ly$\alpha$ absorbers with log(N$_{HI}$) $\gtrsim$
14, the value of the break in the column density distribution,   
are associated with the outer halos of galaxies responsible for 
Lyman limit systems and damped
Lyman $\alpha$ systems
(Giallongo et al. 1996, Lanzetta et al. 1995, 1996, Chen et al. 1998).
This column density
is approximately equivalent to the equivalent width threshold of 0.32
$\AA$ used in this study; and the agreement between our values of $\gamma$ and
that for Lyman limit systems lends some credence to this scenario. 

\acknowledgements
We extend thanks to the staff of the Multiple Mirror Telescope Observatory
for their assistance with the observations, to T. Aldcroft for 
use of his program FINDSL, 
and to K. -V. Tran for assistance with data reduction and continuum fits.
We also thank S. Morris for a helpful referee report.
J. S. acknowledges the support of the National Science Foundation
Graduate Research Fellowship and the Zonta Foundation Amelia Earhart
Fellowship. 
J. B.  acknowledges support from AST-9058510 and AST-9617060 of the
National Science Foundation.
A. D.  acknowledges support from NASA Contract No. NAS8-39073 (ASC).
This research has made use of the NASA/IPAC Extragalactic Database (NED) 
which is operated by the Jet Propulsion Laboratory, California Institute
of Technology, under contract with the National Aeronautics and Space
Administration. 

\pagebreak

\pagebreak

\clearpage
\clearpage

\clearpage
\clearpage

\clearpage
\clearpage

\clearpage
\clearpage

\clearpage
\clearpage

\clearpage
\begin{figure}
\epsscale{1.00}
\plotone{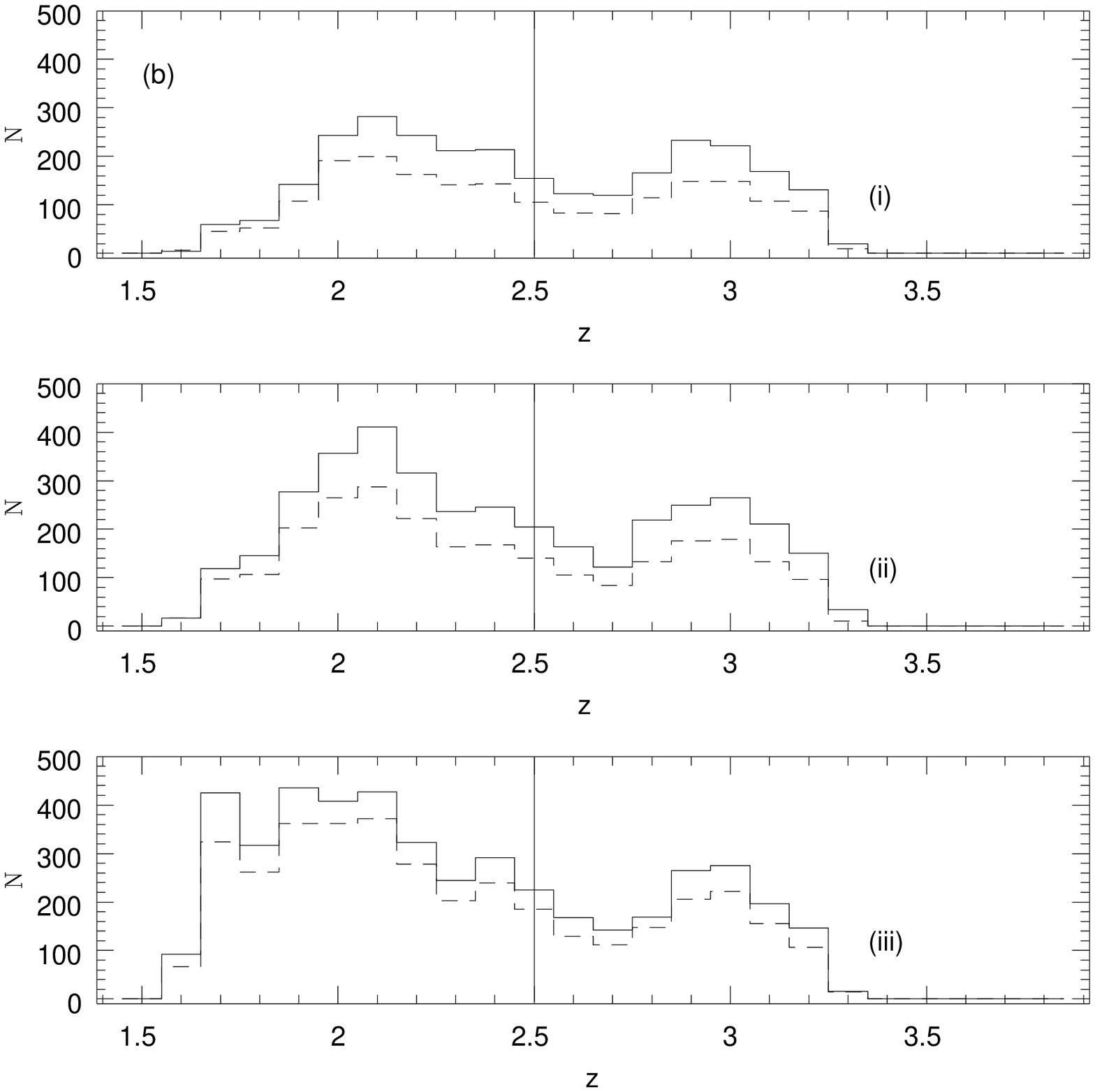}
\end{figure}
\clearpage

\clearpage
\begin{figure}
\epsscale{1.00}
\plotone{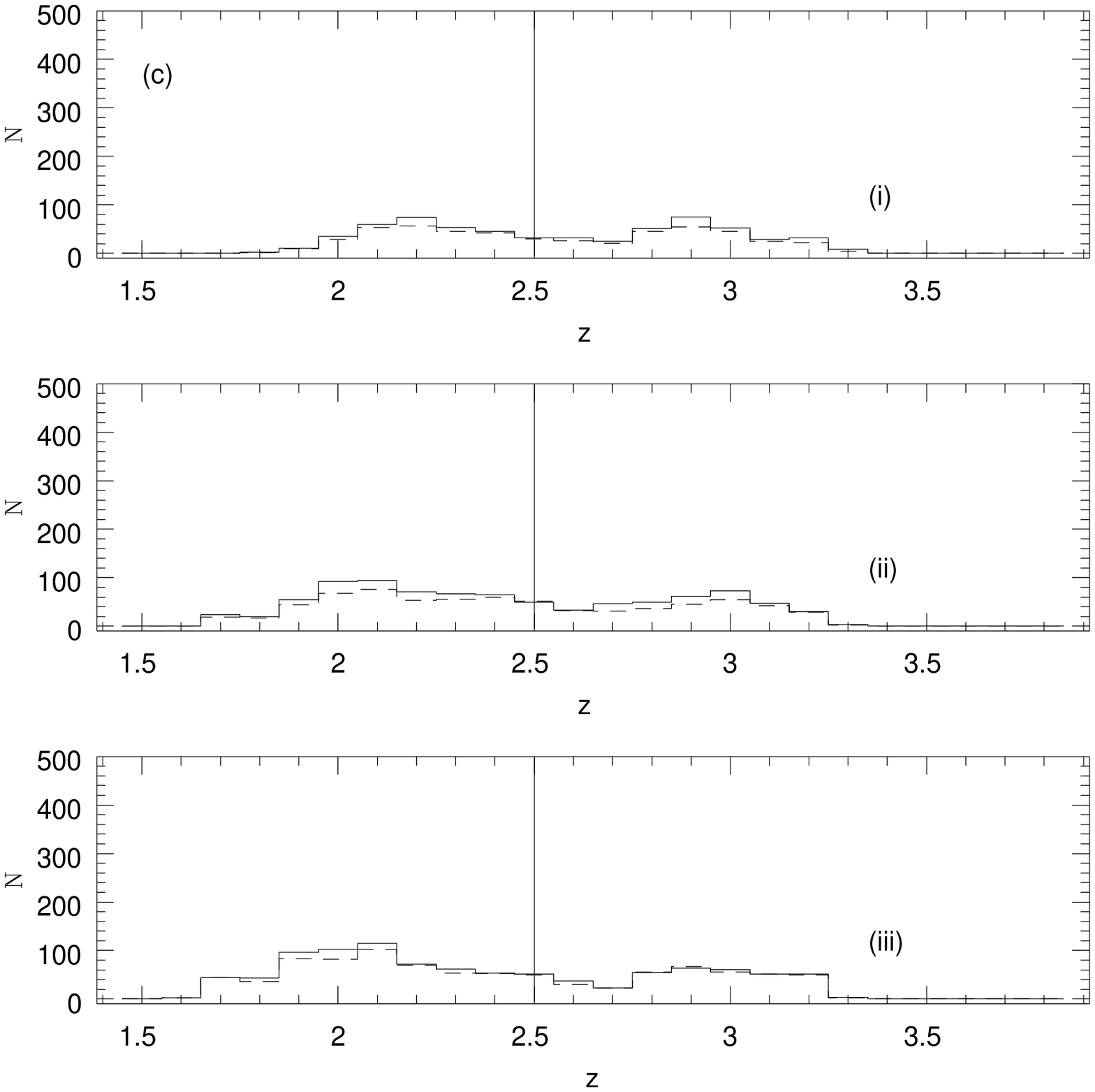}
\end{figure}
\clearpage

\clearpage
\begin{figure}
\epsscale{1.00}
\plotone{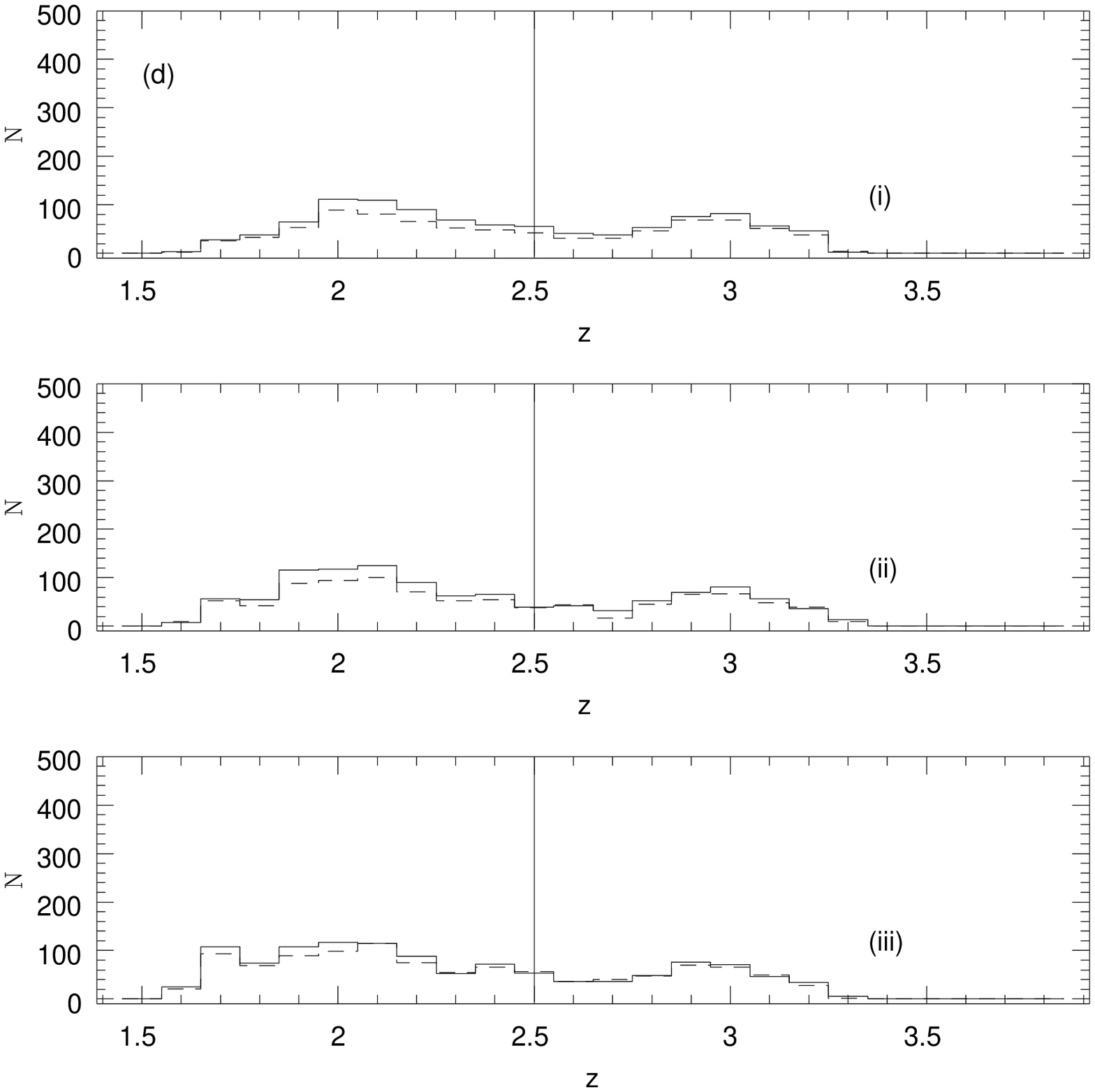}
\end{figure}
\clearpage

\clearpage
\begin{figure}
\epsscale{1.00}
\plotone{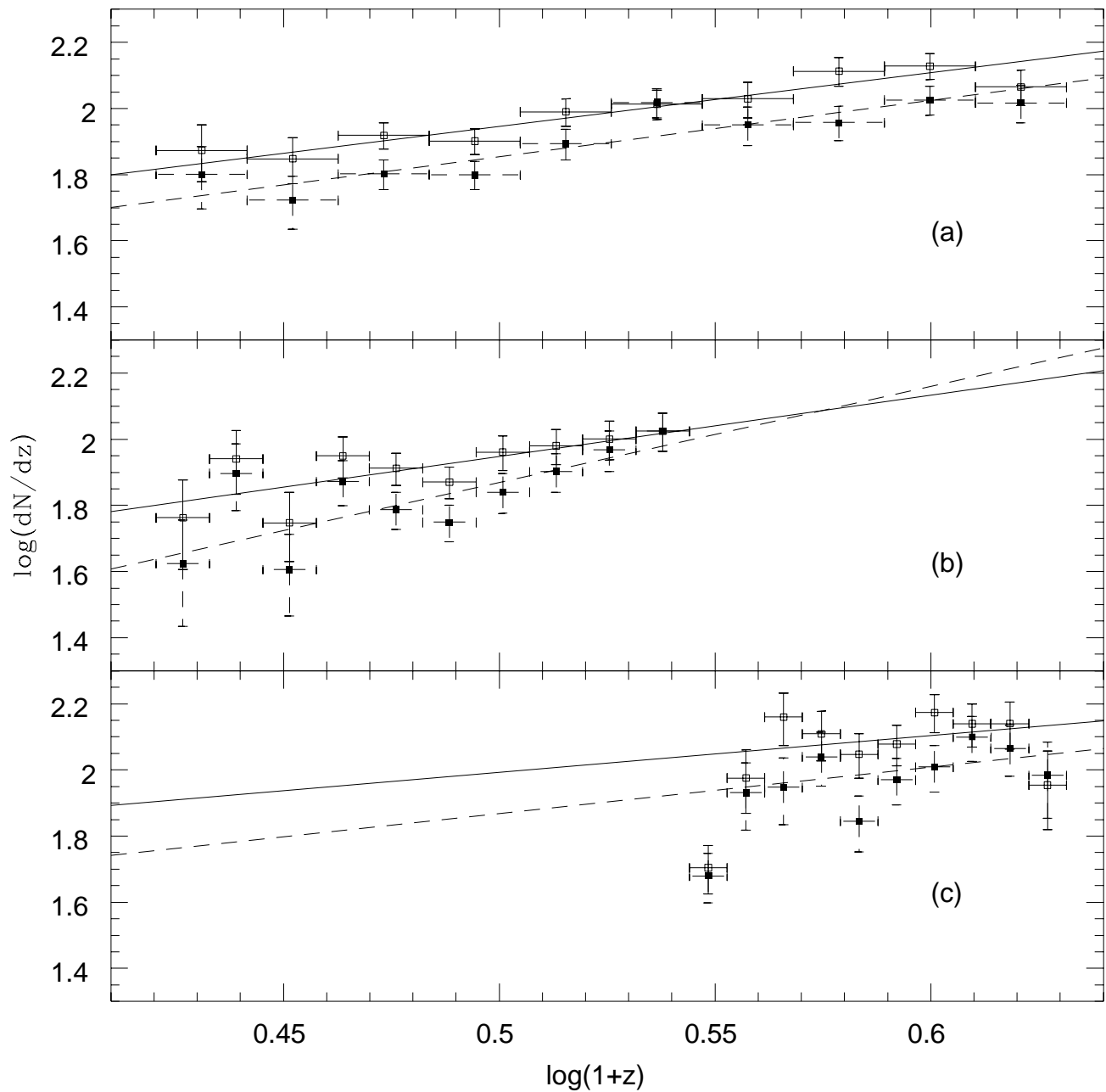}
\caption
{log$(d{\cal N}/dz)$ vs. log$(1+z)$ for
the data resolution, data S/N
simulation line lists (solid line) and FINDSL
line lists (dashed line), for lines with W $>$ 0.32 $\AA$;
(a) all z; (b) z$<$2.5; (c) z$>$2.5
\label{fig:dndzsim}}
\end{figure}
\clearpage

\clearpage
\begin{figure}
\epsscale{1.00}
\plotone{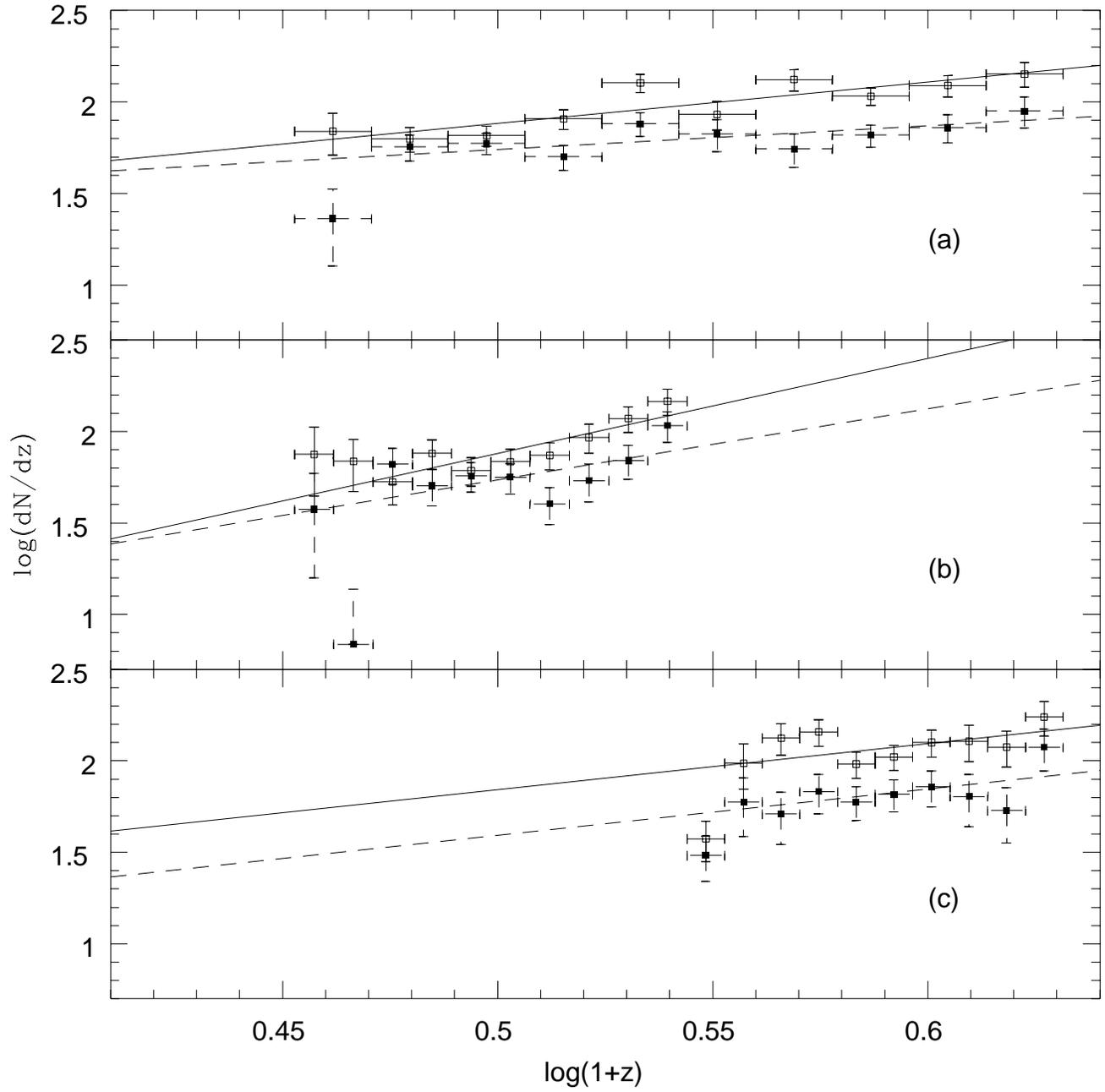}
\caption
{Same as Figure~\ref{fig:dndzsim}, but for lines
with 0.16 $\AA <$ W $<$ 0.32 $\AA$. 
\label{fig:dndzsimweak}}
\end{figure}
\clearpage

\clearpage
\begin{figure}
\epsscale{1.00}
\plotone{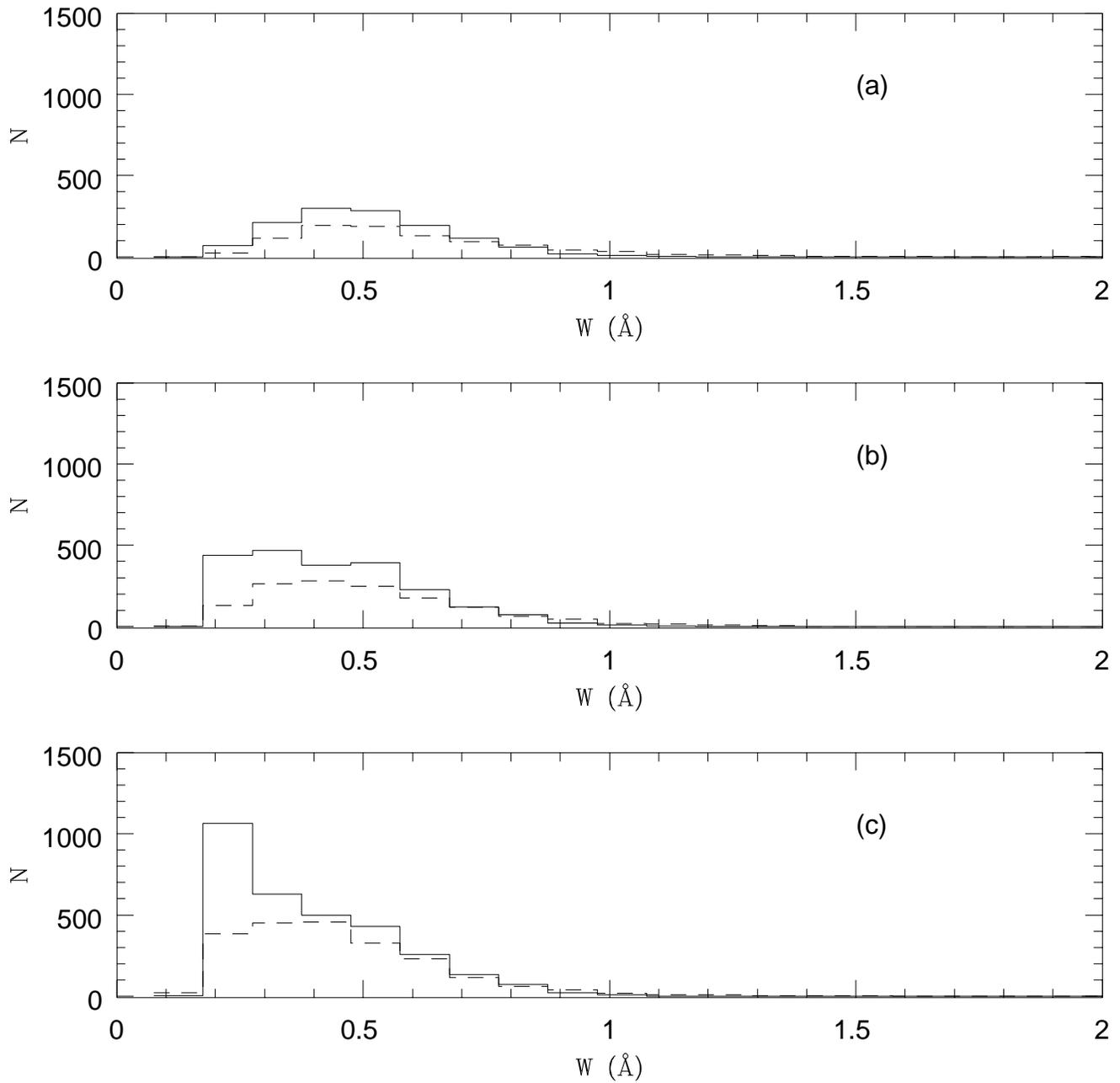}
\caption
{Rest equivalent width distribution of lines
in the data resolution, data S/N simulation line lists 
(solid line) and in the
FINDSL line lists (dashed line), variable threshold case: 
(a) median S/N $\sim$ 5; (b) median S/N $\sim$ 10; 
(c) median S/N $\sim$ 20 \label{fig:ewhist} }
\end{figure}
\clearpage

\begin{deluxetable}{llccccccc}
\tablefontsize{\footnotesize}
\tablecolumns{9}
\tablewidth{42pc}
\tablecaption{Summary of QSO Absorption Line Spectra 
Observations \label{table-mmtobj} }
\tablehead{
\colhead{QSO} & \colhead{Alternate} & \colhead{$z_{em}$} & \colhead{Ref.} &
\colhead{m$_V$} & \colhead{Instr.} & \colhead{Date} & \colhead{Total Exposure}&
\colhead{Wavelength} \\
\colhead{} & \colhead{Name} & \colhead{} & \colhead{(a)}  &
\colhead{(b)} & \colhead{(c)}  &
\colhead{} & \colhead{(seconds)} & \colhead{($\AA$)}  }
\startdata
 0006+020 &        & 2.34  &1  & 17.5  &  2 & 15Nov93 & 7200 & 3200-4088 \nl
 0027+018 & UM 247 & 2.31  &2  & 18.9  &  1 & 25Oct92 & 3600 & 3136-4118 \nl
 0037-018 & UM 264 & 2.34  &1  & 18.0  &  2 & 7Jan94  & 7200 & 3205-4109 \nl
 0049+007 & UM 287 & 2.27  &3  & 17.8  &  1 & 23-25Oct92 & 9600 & 3150-4111 \nl
 0123+257 & PKS    & 2.37  &3  & 17.5  &  2 & 16Nov93 & 9000 & 3198-4094 \nl
 0150-202 & UM 675 & 2.14  &4  & 17.1  &  1 & 24-25Oct92 & 6000 & 3173-4126 \nl
 0153+744 & S5     & 2.34  &3  & 16.0  &  2 & 15Nov93 & 3600 & 3192-4088 \nl
 0226-038 & PKS    & 2.07  &3  & 16.9  &  2 & 16Nov93 & 3000 & 3198-4095 \nl
 0348+061 &        & 2.05  &4  & 17.6  &  1 & 25Oct92 & 2400 & 3130-4112 \nl
 0400+258 & B2     & 2.10  &1  & 18.0  &  2 &7Jan94   & 3000 & 3209- 4121\nl
 0747+613 &        & 2.49  &1  & 17.5  &  1 & 25Oct92 & 3600 & 3323-4269 \nl
 0836+710 & S5     & 2.21  &3  & 16.5  &  2 & 15Nov93 & 3600 & 3192-4088 \nl
 0848+155 &        & 2.01  &4  & 17.7  &  2 & 15Nov93 & 3600 & 3192-4088 \nl
 0936+368 & CSO 233 & 2.02 &1  & 17.0  &  2 & 4Apr94  & 3600 & 3176-4058 \nl
 0952+338 & CSO 239 & 2.50 &1  & 17.0  &  2 & 7Jan94  & 5400 & 3486-4389 \nl
 0955+472 & PC      & 2.48 &1  & 17.7  &  2 & 7Jan94  & 3600 & 3486-4389 \nl
 0956+122 &         & 3.30 &1  & 17.5  &  2 & 7Jan94  & 3600 & 4394-5293 \nl
 1009+299 & CSO 38  & 2.63 &1  & 16.0  &  2 & 7Jan94  & 3600 & 3622-4525 \nl
 1207+399 &         & 2.45 &3  & 17.5  &  3 & 5Apr94  & 900   &3201-4824 \nl
\tablebreak
 1210+175 &         & 2.56 &1  & 17.4  &  2 & 4June94 & 3600 & 3572-4453 \nl
 1231+294 & CSO 151 & 2.01 &1  & 16.0  &  2 & 12Mar94 & 1800 & 3172-4053 \nl
 1323-107 & POX188  & 2.36 &5  & 17.0  &  2 & 4June94 & 5400 & 3200-4087 \nl
 1329+412 & PG      & 1.93 &1  & 16.3  &  2 & 3June94 & 1800 & 3202-4087 \nl
 1337+285 &         & 2.54 &1  & 17.1  &  2 & 3June94 & 3600 & 3574-4455 \nl
 1346-036 &         & 2.36 &3  & 17.2  &  2 & 18Jul93 & 3600 & 3275-4155 \nl
 1358+115 &         & 2.58 &1  & 16.5  &  2 & 18Jul93 & 3600 & 3547-4424 \nl
 1406+492 & CSO 609 & 2.16 &1  & 17.0  &  2 & 3-4June94 & 3400 & 3201-4085 \nl
 1408+009 & UM 645  & 2.26 &3  & 18.0  &  3 & 5Apr94 &900 &3200-4807\nl
 1421+330 & MKN 679 & 1.90 &4  & 16.7  &  2 & 4June94  & 1800 & 3200-4084 \nl
 1422+231 &         & 3.62 &3  & 16.5  &  2 & 16-17Jul93 & 1800 & 4853-5716 \nl
 1435+638 &         & 2.06 &3  & 15.0  &  2 & 16-17Jul93 & 7200 & 3100-3942 \nl
 1603+383 & HS      & 2.51 &6  & 16.9  &  4 & 12-13Apr97 & 3300 & 3532-5045 \nl
 1604+290 & KP 63   & 1.96 &1  & 17.0  &  2 & 18Jul93 & 3600 & 3100-3943 \nl
 1715+535 & PG      & 1.93 &4  & 16.3  &  2 & 16-17Jul93 & 9000 & 3100-3938 \nl
 2134+004 & PKS     & 1.94 &1  & 17.5  &  1 & 24-25Oct92 & 7200 & 3173-4125 \nl
 2251+244 & PKS     & 2.35 &3  & 17.8  &  2 & 16Nov93 & 12000 & 3200-4093 \nl
 2254+022 & PKS     & 2.09 &4  & 17.0  &  2 & 16-17Jul93 & 7200 & 3100-3936 \nl
 2310+385 & UT      & 2.18 &3  & 17.5  &  1 & 25Oct92  & 1200 & 3200-4118 \nl
\tablebreak
 2320+079 & PKS     & 2.08 &1  & 17.5  &  2 & 17Jul93 & 5400 & 3160-3940 \nl
 2329-020 & UM 164  & 1.89 &1  & 17.0  &  2 & 18Jul93 & 3600 & 3060-3943 \nl
\tablenotetext{\footnotesize{a}}{\footnotesize{(1) this paper, from Ly$\alpha$ emission; (2) Baker et al.
1994; (3)Scott et al. (2000) and references therein; 
(4) Steidel \& Sargent 1991; (5) Hewitt \& 
Burbidge 1993; (6) Dobrzycki, Engels, \& Hagen  (1999)} }
\tablenotetext{\footnotesize{b}}{\footnotesize{as listed in Hewitt \& Burbidge 1993, with the
exception of 1603+383 for which V was calculated from the flux-
calibrated spectrum (unpublished)}}
\tablenotetext{\footnotesize{c}}{\footnotesize{Instrument Set-up:
(1) Big Blue Reticon, 832 l mm$^{-1}$ 2$^{nd}$ order, 
1$\arcsec$x3$\arcsec$ slit; 
(2) 3Kx1K CCD, 832 l mm$^{-1}$ 2$^{nd}$ order, 1$\arcsec$x180$\arcsec$ slit;
(3) 3Kx1K CCD, 800 lmm$^{-1}$  1$^{st}$ order, 1$\arcsec$x180$\arcsec$ slit;
(4) 3Kx1K CCD, 1200 lmm$^{-1}$  1$^{st}$ order, 1$\arcsec$x3$\arcsec$ slit}}
\enddata
\end{deluxetable}

\begin{deluxetable}{lcc}
\tablefontsize{\footnotesize}
\tablewidth{25pc}
\tablecaption{QSO Spectra from the Literature \label{table-lit} }
\tablehead{
\colhead{QSO} & \colhead{$z_{em}$} & \colhead{Reference}  }
\startdata
0000-263 &4.111 &1  \nl
0001+087 &3.243 &1  \nl 
0002+051 &1.899 &2    \nl
0002-422 &2.763 &3,4    \nl
0014+813 &3.384 &1,5 \nl
0029+073 &3.294 &1  \nl
0058+019 &1.959 &6    \nl
0100+130 &2.690 &4    \nl
0114-089 &3.205 &1,5  \nl
0119-046 &1.937 &7    \nl
0142-100 &2.727 &6    \nl
0237-233 &2.222 &6    \nl
0256-000 &3.374 &1,5  \nl
0301-005 &3.223 &1  \nl
0302-003 &3.286 &1,5  \nl
0334-204 &3.126 &1  \nl
0421+019 &2.051 &2  \nl
0424-131 &2.166 &6    \nl
0453-423 &2.656 &3,4    \nl
0636+680 &3.174 &1,5    \nl
0731+653 &3.033 &1   \nl      
0831+128 &2.739 &1,5   \nl 
0837+109 &3.326 &6     \nl
0848+163 &1.925 &6     \nl
0905+151 &3.173 &1   \nl
0913+072 &2.784 &1,5   \nl
0938+119 &3.192 &1   \nl 
1017+280 &1.928 &6     \nl
1033+137 &3.092 &1   \nl 
1115+080 &1.725 &2     \nl
1159+124 &3.502 &6     \nl
1206+119 &3.108 &1,5  \nl
1208+101 &3.822  &1   \nl
1215+333 &2.606  &1,5   \nl  
1225-017 &2.831  &1,5  \nl 
1225+317 &2.200  &4     \nl
1247+267 &2.039  &6     \nl
1315+472 &2.590  &1,5  \nl 
1334-005 &2.842  &1,5   \nl 
1400+114 &3.177  &1   \nl 
1402+044 &3.206  &1   \nl 
1410+096 &3.313  &1   \nl 
1442+101 &3.554  &1   \nl 
1451+123 &3.251  &1   \nl 
1511+091 &2.878  &6     \nl
1512+132 &3.120  &1   \nl 
1548+092 &2.748  &1,5   \nl 
1601+182 &3.227  &1   \nl 
1602+178 &2.989  &1   \nl 
1607+183 &3.134  &1,5  \nl 
1614+051 &3.216  &1   \nl
1623+269 &2.526  &1,5  \nl 
1700+642 &2.744  &1,5  \nl
1738+350 &3.239  &1   \nl 
1946+770 &3.020  &5  \nl
2126-158 &3.280  &4     \nl
2233+131 &3.295  &1   \nl 
2233+136 &3.209  &1   \nl 
2311-036 &3.041  &1   \nl 
\tablenotetext{\footnotesize{a}}{\footnotesize{References: (1)Bechtold 1994; (2) Young, Sargent, \&
Boksenberg 1982a; (3)Sargent et al. 1979; (4) Sargent et al. 1980;  
(5) Dobrzycki \& Bechtold 1996; (6) Sargent, Boksenberg, \& Steidel 1988;
(7) Sargent, Young, \& Boksenberg 1982}}
\enddata
\end{deluxetable}

\begin{deluxetable}{lccllccl}
\tablefontsize{\footnotesize}
\tablewidth{40pc}
\tablecaption{Maximum Likelihood Estimations of $\gamma$,$W^{*}$, 
and ${\cal A}_{0}$ 
 \label{table-paramtab} }
\tablehead{
\colhead{ } & \colhead{Sample} & \colhead{No. lines} & \colhead{$W$ limit} 
  & \colhead{$\gamma$} 
  & \colhead{W$^{*}$ ($\AA$)}  &\colhead{${\cal A}_{0}$}  &\colhead{P$_{KS}$}  
   \\
\colhead{ } & \colhead{(a)} & \colhead{} & \colhead{} 
 & \colhead{(b)}
 & \colhead{(b)} & \colhead{(b)} & \colhead{} }
\startdata
1........ &1 &2079 &variable     &1.23$\pm$0.16 &0.313$\pm$0.006 &-    &-  \nl
2........ &1 &1295 &$W>0.16 \AA$ &1.35$\pm$0.21 &0.300$\pm$0.008 &20.1 &0.46 \nl
3........ &1 &1131 &$W>0.32 \AA$ &1.88$\pm$0.22 &0.307$\pm$0.009 &5.78 &0.98 \nl
4........ &1 &1208 &$0.16<W<1.00 \AA$ &1.11$\pm$0.22 &0.238$\pm$0.006 &25.4 
          &0.53  \nl
5........ &1 &1007 &$0.32<W<1.00 \AA$ &1.59$\pm$0.24 &0.226$\pm$0.007 &7.47 
          &0.96  \nl
6........ &1 &555  &$0.16<W<0.32 \AA$ &0.26$\pm$0.33 &0.075$\pm$0.003 &34.1 
          &0.26  \nl
7........ &2 &1084 &variable     &1.57$\pm$0.42 &0.284$\pm$0.008 &-    &-  \nl
8........ &2 &605  &$W>0.16 \AA$ &2.42$\pm$0.62 &0.257$\pm$0.010 &5.86 &0.72 \nl
9........ &2 &534  &$W>0.32 \AA$ &1.30$\pm$0.60 &0.282$\pm$0.012 &11.1 &0.93 \nl
10....... &2 &578  &$0.16<W<1.00 \AA$ &2.26$\pm$0.63 &0.218$\pm$0.009 &6.77 
          &0.53 \nl
11....... &2 &491  &$0.32<W<1.00 \AA$ &1.07$\pm$0.63 &0.229$\pm$0.010 &13.2
          &0.78 \nl
12....... &2 &298  &$0.16<W<0.32 \AA$ &2.47$\pm$0.88 &0.073$\pm$0.004 &2.72 
          &0.93 \nl
13....... &3 &995  &variable     &0.64$\pm$0.47 &0.348$\pm$0.010 &-    &-  \nl 
14....... &3 &690  &$W>0.16 \AA$ &0.46$\pm$0.55 &0.338$\pm$0.012 &67.9 &0.87 \nl
15....... &3 &597  &$W>0.32 \AA$ &1.69$\pm$0.60 &0.330$\pm$0.013 &7.62 &0.83 \nl
16....... &3 &630  &$0.16<W<1.00 \AA$ &-0.05$\pm$0.58 &0.256$\pm$0.010 &125.
          &0.98  \nl
17....... &3 &516  &$0.32<W<1.00 \AA$ &1.26$\pm$0.65  &0.223$\pm$0.009 &11.8
          &0.92  \nl
18....... &3 &257  &$0.16<W<0.32 \AA$ &-1.22$\pm$0.94 &0.077$\pm$0.004 &251.
          &0.86  \nl
\tablenotetext{\footnotesize{a}}{\footnotesize{(1)entire sample; (2) low redshift subsample; (3) high 
 redshift subsample}}
\tablenotetext{\footnotesize{b}}{\footnotesize{see Equ.~\ref{eq:dndz}}}
\enddata 
\end{deluxetable}

\begin{deluxetable}{cccccc}
\tablefontsize{\footnotesize}
\tablecolumns{6}
\tablewidth{42pc}
\tablecaption{Simulation Results for $\gamma$
\label{table-sim} }
\tablehead{
\colhead{$\Delta\lambda$ ($\AA$)} & \colhead{median S/N} &
\colhead{z range} & \colhead{$W$ limit} &
\colhead{$\gamma_{\rm simulation}$} &
\colhead{$\gamma_{\rm FINDSL}$} \\ 
\colhead{(1)} & \colhead{(2)} &
\colhead{(3)} & \colhead{(4)} &
\colhead{(5)} &
\colhead{(6)} }
\startdata
1.0 &4.9 &all z &variable &1.99$\pm$0.25 &1.71$\pm$0.30 \\ 
1.0 &4.9 &z $<$ 2.5 &variable &1.47$\pm$0.74 &2.82$\pm$0.84 \\ 
1.0 &4.9 &z $>$ 2.5 &variable &2.25$\pm$0.76 &1.88$\pm$0.98 \\ 
1.0 &4.9 &all z &0.32 $\AA$ &2.33$\pm$0.37 &1.86$\pm$0.42 \\
1.0 &4.9 &z $<$ 2.5 &0.32 $\AA$ &2.24$\pm$1.19 &2.42$\pm$1.32 \\
1.0 &4.9 &z $>$ 2.5 &0.32 $\AA$ &3.12$\pm$1.12 &1.38$\pm$1.28 \\
1.0 &9.8 &all z &variable &1.63$\pm$0.18 &1.47$\pm$0.22 \\ 
1.0 &9.8 &z $<$ 2.5 &variable &2.61$\pm$0.47 &2.36$\pm$0.54 \\ 
1.0 &9.8 &z $>$ 2.5 &variable &0.36$\pm$0.61 &1.35$\pm$0.79 \\ 
1.0 &9.8 &all z &0.32 $\AA$ &1.62$\pm$0.27 &1.70$\pm$0.30 \\
1.0 &9.8 &z $<$ 2.5 &0.32 $\AA$ &1.85$\pm$0.68 &2.90$\pm$0.76 \\
1.0 &9.8 &z $>$ 2.5 &0.32 $\AA$ &1.10$\pm$0.97 &1.40$\pm$1.08 \\
1.0 &9.8 &all z &0.16 $<$ W $<$ 0.32 $\AA$ &2.25$\pm$0.40 &1.30$\pm$0.49\\
1.0 &9.8 &z $<$ 2.5 &0.16 $<$ W $<$ 0.32 $\AA$ &5.18$\pm$1.26 &3.88$\pm$1.50 \\
1.0 &9.8 &z $>$ 2.5 &0.16 $<$ W $<$ 0.32 $\AA$ &2.51$\pm$1.23 &2.52$\pm$1.68 \\
1.0 &19.6 &all z &variable &1.91$\pm$0.14 &1.67$\pm$0.17 \\ 
1.0 &19.6 &z $<$ 2.5 &variable &1.82$\pm$0.34 &1.33$\pm$0.40 \\ 
1.0 &19.6 &z $>$ 2.5 &variable &3.34$\pm$0.53 &2.87$\pm$0.66 \\ 
1.0 &19.6 &all z &0.32 $\AA$ &1.79$\pm$0.24 &2.14$\pm$0.25 \\
1.0 &19.6 &z $<$ 2.5 &0.32 $\AA$ &1.48$\pm$0.57 &1.86$\pm$0.61 \\
1.0 &19.6 &z $>$ 2.5 &0.32 $\AA$ &3.61$\pm$1.00 &3.92$\pm$1.02 \\
0.7 &9.8 &all z &variable &1.77$\pm$0.15 &1.56$\pm$0.18 \\ 
0.7 &9.8 &z $<$ 2.5 &variable &1.86$\pm$0.39 &1.46$\pm$0.46 \\ 
0.7 &9.8 &z $>$ 2.5 &variable &2.34$\pm$0.54 &1.73$\pm$0.67 \\ 
0.7 &9.8 &all z &0.32 $\AA$ &2.11$\pm$0.24 &2.44$\pm$0.27 \\
0.7 &9.8 &z $<$ 2.5 &0.32 $\AA$ &2.04$\pm$0.61 &1.58$\pm$0.69 \\
0.7 &9.8 &z $>$ 2.5 &0.32 $\AA$ &2.38$\pm$0.92 &3.12$\pm$1.01 \\
0.7 &19.6 &all z &variable &1.76$\pm$0.12 &1.56$\pm$0.15 \\ 
0.7 &19.6 &z $<$ 2.5 &variable &1.98$\pm$0.30 &2.02$\pm$0.35 \\ 
0.7 &19.6 &z $>$ 2.5 &variable &2.42$\pm$0.49 &1.68$\pm$0.61 \\ 
0.7 &19.6 &all z &0.32 $\AA$ &1.90$\pm$0.22 &2.19$\pm$0.24 \\
0.7 &19.6 &z $<$ 2.5 &0.32 $\AA$ &1.69$\pm$0.52 &1.90$\pm$0.57 \\
0.7 &19.6 &z $>$ 2.5 &0.32 $\AA$ &3.56$\pm$0.96 &4.01$\pm$1.03 \\
0.2 &39.2 &all z &variable &1.41$\pm$0.10 &1.31$\pm$0.12 \\ 
0.2 &39.2 &z $<$ 2.5 &variable &1.22$\pm$0.23 &1.54$\pm$0.25 \\ 
0.2 &39.2 &z $>$ 2.5 &variable &0.72$\pm$0.48 &0.77$\pm$0.55 \\ 
0.2 &39.2 &all z &0.32 $\AA$ &1.68$\pm$0.21 &1.91$\pm$0.21 \\
0.2 &39.2 &z $<$ 2.5 &0.32 $\AA$ &1.69$\pm$0.45 &2.24$\pm$0.48 \\
0.2 &39.2 &z $>$ 2.5 &0.32 $\AA$ &0.86$\pm$0.96 &-0.33$\pm$0.98 \\
\enddata
\end{deluxetable}
\end{document}